\documentclass{article}%
\usepackage{amssymb}
\usepackage{amsmath}
\usepackage{makeidx}
\usepackage{amsfonts}
\usepackage{graphicx}%
\begin{document}

\author{Hartmut Wachter\thanks{E-Mail: Hartmut.Wachter@gmx.de}\\An der Schafscheuer 56\\D-91781 Wei\ss enburg, Federal Republic of Germany}
\title{Momentum and Position Representations for the $q$-de\-formed Euclidean Quantum Space}
\maketitle
\date{}

\begin{abstract}
We summarize some basics about mathematical tools of analysis for the
$q$-deformed Euclidean space.\ We use the new tools to examine $q$-deformed
eigenfunctions of the momentum or position operator within the framework of
the star product formalism. We show that these two systems of functions are
complete and orthonormal. With the $q$-deformed momentum or position
eigenfunctions, we calculate matrix elements of the momentum or position
operator. Considerations about expectation values and probability densities
conclude the studies.

\end{abstract}

\section{Introduction}

The debate whether space and time are discrete or continuous began long ago.
It was as early as in ancient times that the Greek philosopher Zenon of Elea
confused his contemporaries. He baffled them by stating that Achilles, as the
fastest runner of the Greeks, can never overtake a turtle with a one-meter
lead if the distance between Achilles and the turtle is infinitely divisible.

Of course, the idea of a continuum can form a logical basis for the ordinary
differential and integral calculus. Moreover, the physical theories formulated
with the help of this differential and integral calculus are fruitful. Certain
doubts, however, remain whether a continuum adequately describes space and
time at small distances. To overcome the problem that it is not possible to
calculate a finite self-energy of the electron with quantum field theories
formulated on a space-time continuum Werner Heisenberg, for example, used a
lattice-like space \cite{Heisenberg:1930, Heisenberg:1938}.

Even though such attempts have not yet been successful \cite{Hagar:2014}, some
considerations within the framework of a future theory of quantum gravity
suggest that space-time reveals a discrete structure at small distances
\cite{Garay:1995}. For example, the attempt to increase the accuracy of a
position measurement more and more, should disturb the background metric more
and more \cite{Mead:1966zz}. Such a fundamental uncertainty in space could
require a space-time algebra generated by non-com\-mu\-ting coordinates. Such
a non-com\-mu\-ta\-tive algebra can be obtained, for example, by
$q$-deformation. The general aim of my scientific work is to find out whether
physical theories can be formulated on $q$-de\-formed space-time algebras free
from contradictions. Furthermore, we could see if such an approach is suitable
to solve existing physical problems.

In this article, I am going to give the reader some basics for dealing with
free particles on the $q$-deformed Euclidean space. For this reason, let me
summarize some well-known facts from quantum mechanics. If we want to describe
the measurement of the position or momentum of a particle mathematically, we
must assign a linear operator to each of these measurable quantities. The
measurement corresponds to the action of the respective operator on a wave
function that represents the state of the free particle. If the particle has a
defined momentum or position after the measurement, an eigenfunction of the
momentum operator or the position operator represents the particle state. In
the case of the position measurement, these eigenfunctions are delta
functions, and in the case of the momentum measurement, they are plane waves,
i. e. exponential functions. A wave function that represents the state of a
free particle can be expanded in terms of eigenfunctions of the position
operator as well as in terms of eigenfunctions of the momentum operator.
Moreover, each expansion coefficient determines the probability for obtaining
the corresponding state of defined momentum or position in the position or
momentum measurement.

In Chap.~\ref{KapMatRep} of this article, I am going to discuss how the
concepts described above apply to $q$-deformed quantum spaces within the
framework of the so-called star product formalism. For this purpose, I am
going to summarize some algebraic basics in Chap.~\ref{KapEucQuaSpa} needed to
provide mathematical tools of analysis for $q$-de\-formed quantum spaces.
Among these tools, which we will briefly explain in the subsequent chapters,
are star products, translations, actions of partial derivatives, integrals as
well as Fourier transformations. In Chap.~\ref{KapDefImpOrtEigFkt}, we are
going to apply these new tools to introduce eigenfunctions of the
$q$-de\-formed position or momentum operator. In the subsequent chapters, we
will show that the eigenfunctions of the $q$-de\-formed position or momentum
operator are complete and orthonormal. In Chap.~\ref{KapMatDarImp}\ and
Chap.~\ref{KapMatDarOrt}, we are going to use these results to write down
transition matrix elements of the $q$-deformed position operator or the
$q$-deformed momentum operator for both the $q$-deformed momentum
eigenfunctions and the $q$-deformed position eigenfunctions. Finally, in
Chap.~\ref{KapWskDic}, we are going to determine expectation values and
probability densities for the position measurement or momentum measurement
within the new formalism.

The reflections of this article show that we can formulate many concepts on
$q$-de\-formed quantum spaces in far-reaching analogy to the undeformed case.
The non-com\-mu\-ta\-ti\-vi\-ty due to $q$-de\-formation, however, will lead
to some complexity since we have several $q$-deformed variants for the
considered mathematical objects.

\section{Elements of multidimensional $q$-analysis}

\subsection{Euclidean quantum space\label{KapEucQuaSpa}}

The $q$-de\-formed three-di\-men\-sio\-nal Euclidean quantum space
$\mathbb{R}_{q}^{3}$ is a three-di\-men\-sio\-nal representation of the Hopf
algebra $\mathcal{U}_{q}(\operatorname*{su}_{2})$ \cite{Lorek:1993tq}. The
latter is a deformation of the universal enveloping algebra of the Lie algebra
$\operatorname*{su}_{2}$ \cite{Kulish:1983md}, as the following relations of
its generators show:\footnote{If the deformation parameter $q$ tends to $1$,
we regain the commutation relations of the Lie algebra $\operatorname*{su}%
_{2}$.}%
\begin{align}
q^{-1}\hspace{0.01in}T^{+}T^{-}-q\,T^{-}T^{+}  &  =T^{3},\nonumber\\
q^{\hspace{0.01in}2}\hspace{0.01in}T^{3}T^{+}-q^{-2}\hspace{0.01in}T^{+}T^{3}
&  =(q+q^{-1})\hspace{0.01in}T^{+},\nonumber\\
q^{\hspace{0.01in}2}\hspace{0.01in}T^{-}T^{3}-q^{-2}\hspace{0.01in}T^{3}T^{-}
&  =(q+q^{-1})\hspace{0.01in}T^{-}.
\end{align}
Compared to an ordinary algebra, a Hopf algebra has not only a product and a
unit but also a co-product, a co-unit and an antipode \cite{Klimyk:1997eb}. For example, the co-product of the $\mathcal{U}%
_{q}(\operatorname*{su}_{2})$-ge\-ne\-ra\-tors takes on the following form
\cite{Lorek:1993tq}:%
\begin{align}
\Delta(T^{+})  &  =T^{+}\otimes1+\tau^{1/2}\otimes T^{+},\nonumber\\
\Delta(T^{-})  &  =T^{-}\otimes1+\tau^{1/2}\otimes T^{-},\nonumber\\
\Delta(T^{3})  &  =T^{3}\otimes1+\tau\otimes T^{3}.
\end{align}
Note that we have introduced the element $\tau=1-(q-q^{-1})\hspace
{0.01in}T^{3}$ in the above expressions. The Hopf algebra $\mathcal{U}%
_{q}(\operatorname*{su}_{2})$ has the following spin representations
\cite{Lorek:1993tq}:%
\begin{align}
T^{3}\left\vert \hspace{0.01in}j,m\right\rangle  &  =q^{-1}[[2\hspace
{0.01in}m]]_{q^{-2}}\left\vert \hspace{0.01in}j,m\right\rangle ,\nonumber\\
T^{+}\left\vert \hspace{0.01in}j,m\right\rangle  &  =q^{-1}\sqrt
{[[\hspace{0.01in}j+m+1]]_{q^{-2}}[[\hspace{0.01in}j-m]]_{q^{2}}}\left\vert
\hspace{0.01in}j,m+1\right\rangle ,\nonumber\\
T^{-}\left\vert \hspace{0.01in}j,m\right\rangle  &  =q\sqrt{[[\hspace
{0.01in}j+m]]_{q^{-2}}[[\hspace{0.01in}j-m+1]]_{q^{2}}}\left\vert
\hspace{0.01in}j,m-1\right\rangle ,\nonumber\\
\tau\left\vert \hspace{0.01in}j,m\right\rangle  &  =q^{-4m}\left\vert
\hspace{0.01in}j,m\right\rangle .
\end{align}
Note also that the above representations depend on so-called anti-symmetric
$q$-numbers:%
\begin{equation}
\lbrack\lbrack n]]_{q^{a}}=\frac{1-q^{a\hspace{0.01in}n}}{1-q^{a}}.
\end{equation}
The fundamental states of the three-di\-men\-si\-onal representation can be
identified with the generators of the three-di\-men\-sional Euclidean quantum
space:%
\begin{equation}
X^{-}=\left\vert 1,-1\right\rangle ,\qquad X^{3}=\left\vert 1,0\right\rangle
,\qquad X^{+}=\left\vert 1,1\right\rangle .
\end{equation}
The coordinates $X^{3}$, $X^{+}$, and $X^{-}$ form a light cone basis.
Therefore they are related to the usual Cartesian coordinates $x^{1}$, $x^{2}%
$, and $x^{3}$ by the following classical limit \cite{Lorek:1997eh}:%
\begin{equation}
\lim_{q\hspace{0.01in}\rightarrow\hspace{0.01in}1}X^{3}=x^{3},\qquad
\lim_{q\hspace{0.01in}\rightarrow\hspace{0.01in}1}X^{\pm}=\text{i}%
\hspace{0.01in}(x^{1}\pm\text{i\hspace{0.01in}}x^{2}).
\end{equation}

Tensor products of quantum spaces should have the same symmetry as the quantum
spaces which they are composed of. In other words, tensor products of modules
of a Hopf algebra must again be modules of the same Hopf algebra. The action
of the Hopf algebra $\mathcal{U}_{q}(\operatorname*{su}_{2})$ on a tensor
product of two Euclidean quantum spaces is calculated by using the co-product
of $\mathcal{U}_{q}(\operatorname*{su}_{2})$. Specifically, it applies to an
element $h$ of $\mathcal{U}_{q}(\operatorname*{su}_{2})$ and two elements $u$
and $v$ of the Euclidean quantum space \cite{Klimyk:1997eb}:%
\begin{equation}
h\triangleright(u\otimes v)=h_{(1)}\triangleright u\otimes h_{(2)}%
\triangleright v.
\end{equation}
Note that we have written the co-product of $h$ in the so-called Sweedler
notation, i.~e. $\Delta(h)=h_{(1)}\otimes h_{(2)}$.

Due to the non-tri\-vial co-product of $\mathcal{U}_{q}(\operatorname*{su}%
_{2})$, the multiplication on a tensor product of two quantum spaces can
generally not be performed by using the usual twist, but with the help of a
so-called braiding map. In the case of two coordinate generators this braiding
map is given by the R-matrix $\hat{R}{}{^{\hspace{0.01in}AB}}_{CD}$ for the
three-di\-men\-si\-onal $q$-deformed Euclidean space (together with a constant
$k$):%
\begin{equation}
(1\otimes Y^{A})\cdot(X^{B}\otimes1)=k\hspace{0.02in}\hat{R}{}{^{\hspace
{0.01in}AB}}_{CD}\,X^{C}\otimes Y^{D}. \label{ZopRel}%
\end{equation}
Applying the action of the $\mathcal{U}_{q}(\operatorname*{su}_{2}%
)$-ge\-ne\-ra\-tors to the above equations, a system of equations can be found
to determine the entries of the R-matrix of the Euclidean quantum space
\cite{Lorek:1993tq}.\footnote{The constant $k$ cannot be determined this way.}
The inverse matrix ${\hat{R}}^{-1}$ is obtained as a further solution of this
system:%
\begin{equation}
(1\otimes Y^{A})\cdot(X^{B}\otimes1)=k^{-1}\hspace{0.02in}({\hat{R}}%
^{-1}{)^{AB}}_{CD}\,X^{C}\otimes Y^{D}. \label{InvZopRel}%
\end{equation}
Eq.~(\ref{ZopRel}) and Eq.~(\ref{InvZopRel}) are often referred to as
\textit{braiding relations}. These braiding relations can be generalized in a
way that they apply to arbitrary elements of quantum spaces. To this end, we
introduce the so-called universal R-ma\-trix $\mathcal{R=R}_{[1]}%
\otimes\mathcal{R}_{[2]}\in\mathcal{U}_{q}(\operatorname*{su}_{2}%
)\otimes\mathcal{U}_{q}(\operatorname*{su}_{2})$ and its inverse
\cite{Majid:1996kd}, i.~e.%
\begin{equation}
(1\otimes u)\cdot(v\otimes1)=\mathcal{R}_{[2]}\triangleright v\otimes
\mathcal{R}_{[1]}\triangleright u \label{ZopRelAlg}%
\end{equation}
or%
\begin{equation}
(1\otimes u)\cdot(v\otimes1)=\mathcal{R}_{[1]}^{-1}\triangleright
v\otimes\mathcal{R}_{[2]}^{-1}\triangleright u. \label{InvZopRelAlg}%
\end{equation}

The R-ma\-trix of the Euclidean quantum space has the following projector
decomposition \cite{Lorek:1993tq}:%
\begin{equation}
\hat{R}=P_{S}+q^{-6}P_{T}-q^{-4}P_{A}.
\end{equation}
The projector $P_{A}$ is a $q$-ana\-log of the antisymmetrizer which maps on
the space of antisymmetric tensors of second rank. The projector $P_{S}$ is
the $q$-de\-formed trace-free symmetrizer and $P_{T}$ is the $q$-de\-formed trace-pro\-jec\-tor.

The antisymmetric projector $P_{A}$ defines the commutation relations for the
coordinates of the Euclidean quantum space $(A,B\in\{+,3,-\})$:%
\begin{equation}
(P_{A}){^{AB}}_{CD}\,X^{C}X^{D}=0.
\end{equation}
Explicitly, the commutation relations of the Euclidean quantum space
coordinates read as \cite{Lorek:1997eh}:%
\begin{align}
X^{3}X^{+}  &  =q^{2}X^{+}X^{3},\nonumber\\
X^{3}X^{-}  &  =q^{-2}X^{-}X^{3},\nonumber\\
X^{-}X^{+}  &  =X^{+}X^{-}+(q-q^{-1})\hspace{0.01in}X^{3}X^{3}.
\label{RelQuaEukDre}%
\end{align}

The trace projector $P_{T}$ leads us to a $q$-analog of the Euclidean metric.
Concretely, it applies to the metric $g^{AB}$ of the $q$-de\-formed Euclidean
space and its inverse $g_{CD}$ \cite{Lorek:1997eh}:%
\begin{equation}
(P_{T}){^{AB}}_{CD}=\frac{1}{g^{EF}g_{EF}}\,g^{AB}g_{CD}.
\end{equation}
This correspondence implies for the $q$-de\-formed Euclidean metric (row and
column indices have the order $+,3,-$):%
\begin{equation}
g_{AB}=g^{AB}=\left(
\begin{array}
[c]{ccc}%
0 & 0 & -\hspace{0.01in}q\\
0 & 1 & 0\\
-\hspace{0.01in}q^{-1} & 0 & 0
\end{array}
\right)  .
\end{equation}

The Euclidean quantum space $\mathbb{R}_{q}^{3}$ is also a $\ast$-al\-ge\-bra,
i.~e. it has a semilinear, involutive and anti-multiplicative mapping, which
we call \textit{quantum space conjugation}. We indicate the conjugate elements
of a quantum space by a bar. The properties of the conjugation on a quantum
space can now be written down as follows ($\alpha,\beta\in\mathbb{C}$ and
$u,v\in\mathbb{R}_{q}^{3}$):\footnote{A bar over a complex number indicates
complex conjugation.}%
\begin{equation}
\overline{\alpha\hspace{0.01in}u+\beta\hspace{0.01in}v}=\bar{\alpha}%
\hspace{0.01in}\bar{u}+\bar{\beta}\hspace{0.01in}\bar{v},\quad\overline
{\bar{u}}=u,\quad\overline{u\hspace{0.01in}v}=\bar{v}\hspace{0.01in}\bar{u}.
\end{equation}
You can show that the conjugation on the quantum space $\mathbb{R}_{q}^{3}$
respects the commutation relations in Eq.~(\ref{RelQuaEukDre}) if the
following applies \cite{Lorek:1997eh}:%
\begin{equation}
\overline{X^{A}}=X_{A}=g_{AB}\,X^{B}. \label{KovKoo}%
\end{equation}

\subsection{Star products\label{KapStePro}}

An $N$-di\-men\-sional quantum space can be seen as an algebra $V_{q}$ which
is spanned by non-com\-mu\-ta\-tive coordinates $X^{i}$ with $i=1,\ldots,N$,
i.~e. the coordinates of the quantum space satisfy certain non-tri\-vial
commutation relations. The commutation relations of the quantum space
coordinates generate a two-si\-ded ideal $\mathcal{I}$ which is invariant
under actions of the Hopf algebra describing the symmetry of $V_{q}$. From
this point of view a quantum space is a quotient algebra which is formed by
the free algebra $\mathbb{C}[X^{1},X^{2},\ldots,X^{N}]$ and the ideal
$\mathcal{I}$:%
\begin{equation}
V_{q}=\frac{\mathbb{C}[X^{1},X^{2},\ldots,X^{N}]}{\mathcal{I}}.
\end{equation}

In general, a physical theory can only be verified if it predicts certain
measurement results. The question, however, is: How can we associate the
elements of a quantum space with real numbers? One solution to this problem is
to interpret the quantum space coordinates $X^{i}$ with $i\in\{1,\ldots,N\}$
as operators acting on a ground state which is invariant under actions of the
symmetry Hopf algebra. This way, the corresponding expectation values denoted
as%
\begin{equation}
x^{i}=\langle\hspace{0.01in}X^{i}\rangle
\end{equation}
can be seen as real-valued variables with their numbers depending on the
underlying ground state. In the following, we will show how we can extend the
above identity to normal ordered monomials of quantum space coordinates.

The normal ordered monomials of the quantum space coordinates $X^{i}$ form a
basis of the quantum space $V_{q}$, i.~e. each element $F\in V_{q}$ can
uniquely be written as a finite or infinite linear combination of monomials of
a given normal ordering (\textit{Poincar\'{e}-Birkhoff-Witt property}):%
\begin{equation}
F=\sum\limits_{i_{1},\ldots,\hspace{0.01in}i_{N}}a_{\hspace{0.01in}i_{1}%
\ldots\hspace{0.01in}i_{N}}\,(X^{1})^{i_{1}}\ldots\hspace{0.01in}%
(X^{N})^{i_{N}}\quad\text{with}\quad a_{\hspace{0.01in}i_{1}\ldots
\hspace{0.01in}i_{N}}\in\mathbb{C}.
\end{equation}
Since the set of monomials $(x^{1})^{i_{1}}\ldots\hspace{0.01in}(x^{N}%
)^{i_{N}}$ with $i_{1},\ldots,i_{N}\in\mathbb{N}_{0}$ forms a basis of the
commutative algebra $V\mathcal{=\,}\mathbb{C}[\hspace{0.01in}x^{1}%
,\ldots,x^{N}]$, we can define a vector space isomorphism between $V$ and
$V_{q}$, i.~e.%
\begin{equation}
\mathcal{W}:V\rightarrow V_{q} \label{VecRauIsoInv}%
\end{equation}
with%
\begin{equation}
\mathcal{W}\big ((x^{1})^{i_{1}}\ldots\hspace{0.01in}(x^{N})^{i_{N}%
}\big )=(X^{1})^{i_{1}}\ldots\hspace{0.01in}(X^{N})^{i_{N}}. \label{StePro0}%
\end{equation}
By linear extension follows%
\begin{equation}
V\ni f\mapsto F\in V_{q},
\end{equation}
where%
\begin{align}
f  &  =\sum\limits_{i_{1},\ldots,\hspace{0.01in}i_{N}}a_{\hspace{0.01in}%
i_{1}\ldots\hspace{0.01in}i_{N}}\,(x^{1})^{i_{1}}\ldots\hspace{0.01in}%
(x^{N})^{i_{N}},\nonumber\\
F  &  =\sum\limits_{i_{1},\ldots,\hspace{0.01in}i_{N}}a_{\hspace{0.01in}%
i_{1}\ldots\hspace{0.01in}i_{N}}\,(X^{1})^{i_{1}}\ldots\hspace{0.01in}%
(X^{N})^{i_{N}}. \label{AusFfNorOrd}%
\end{align}

The vector space isomorphism $\mathcal{W}$ is nothing else but the so-called
\textit{Mo\-yal-Weyl mapping} which gives an `operator' $F\in V_{q}$ to a
complex valued function $f\in V$ \cite{Bayen:1977ha, 1997q.alg.....9040K,
Madore:2000en, Moyal:1949sk}. You can see that the inverse of the Moyal-Weyl
mapping gives each quantum space coordinate its expectation value:%
\begin{equation}
\mathcal{W}^{\hspace{0.01in}-1}(X^{i})=x^{i}=\langle X^{i}\rangle.
\end{equation}
This relation can be used for normal ordered monomials as follows:%
\begin{align}
\mathcal{W}^{\hspace{0.01in}-1}((X^{1})^{i_{1}}\ldots\hspace{0.01in}%
(X^{N})^{i_{N}})  &  =(x^{1})^{i_{1}}\ldots\hspace{0.01in}(x^{N})^{i_{N}%
}=\langle(X^{1})^{i_{1}}\rangle\ldots\langle(X^{N})^{i_{N}}\rangle\nonumber\\
&  =\langle(X^{1})^{i_{1}}\ldots\hspace{0.01in}(X^{N})^{i_{N}}\rangle.
\end{align}
We can go one step further. By linear extension, the vector space isomorphism
$\mathcal{W}^{\hspace{0.01in}-1}$ can assign an expectation value $f=\langle
F\rangle$ to any element $F$ of the quantum space $V_{q}$:%
\begin{equation}
\mathcal{W}^{\hspace{0.01in}-1}(F)=f=\langle F\rangle.
\end{equation}

As a look at Eq.~(\ref{AusFfNorOrd}) shows the expectation value $f$ is a
function of the commutative coordinates $x^{i}$. In this way, the vector space
isomorphism $\mathcal{W}^{\hspace{0.01in}-1}$ maps the non-com\-mutative
quantum space algebra $V_{q}$ onto the commutative algebra $V$ consisting of
all power series with coordinates $x^{i}$. We can even extend this vector
space isomorphism to an algebra isomorphism if a new product is introduced on
the commutative algebra $V$. This so-called \textit{star product }symbolized
by $\circledast$ satisfies the following homomorphism condition:%
\begin{equation}
\mathcal{W}^{\hspace{0.01in}-1}(F\cdot G)=\langle F\cdot G\rangle=\langle
F\rangle\circledast\langle G\hspace{0.01in}\rangle=\mathcal{W}^{\hspace
{0.01in}-1}(F)\circledast\mathcal{W}^{\hspace{0.01in}-1}(G). \label{HomMorBed}%
\end{equation}
With$\ f$ and $g$ being two formal power series of the commutative coordinates
$x^{i}$, the above condition can alternatively be written in the following
form:%
\begin{equation}
\mathcal{W}\left(  f\circledast g\right)  =\mathcal{W}\left(  f\right)
\cdot\mathcal{W}\left(  \hspace{0.01in}g\right)  . \label{HomBedWeyAbb}%
\end{equation}
Since the Mo\-yal-Weyl mapping is invertible, we can write the star product as
follows:%
\begin{equation}
f\circledast g=\mathcal{W}^{\hspace{0.01in}-1}\big (\,\mathcal{W}\left(
f\right)  \cdot\mathcal{W}\left(  \hspace{0.01in}g\right)  \big ).
\label{ForStePro}%
\end{equation}
Thus, the star product realizes the non-com\-mu\-ta\-tive product of $V_{q}$
on the commutative algebra $V$.

To get explicit formulas for calculating the star product, we must define a
suitable normal ordering for the non-commutative coordinate monomials. To
derive these formulas, we have to expand the non-com\-mu\-ta\-tive product of
two normal ordered monomials in terms of normal ordered monomials by using the
commutation relations for the quantum space coordinates:%
\begin{equation}
(X^{1})^{i_{1}}\ldots\hspace{0.01in}(X^{N})^{i_{N}}\cdot(X^{1})^{j_{1}}%
\ldots\hspace{0.01in}(X^{N})^{j_{N}}=\sum\nolimits_{\underline{k}%
}B_{\underline{k}}\,(X^{1})^{k_{1}}\ldots\hspace{0.01in}(X^{N})^{k_{N}}.
\end{equation}
In the case of the $q$-de\-formed Euclidean space, for example, we can obtain
the following formula for calculating the star-pro\-duct ($\lambda=q-q^{-1}%
$):\footnote{For the details see Ref.~\cite{Wachter:2002A}.}%
\begin{equation}
f(\mathbf{x})\circledast g(\mathbf{x})=\sum_{k\hspace{0.01in}=\hspace
{0.01in}0}^{\infty}\lambda^{k}\hspace{0.01in}\frac{(x^{3})^{2k}}{[[k]]_{q^{4}%
}!}\,q^{2(\hat{n}_{3}\hat{n}_{+}^{\prime}+\,\hat{n}_{-}\hat{n}_{3}^{\prime}%
)}D_{q^{4},\hspace{0.01in}x^{-}}^{k}f(\mathbf{x})\,D_{q^{4},\hspace
{0.01in}x^{\prime+}}^{k}g(\mathbf{x}^{\prime})\big|_{x^{\prime}\rightarrow
\hspace{0.01in}x}. \label{StaProForExp}%
\end{equation}
Note that the above expression depends on the operators%
\begin{equation}
\hat{n}_{A}=x^{A}\frac{\partial}{\partial x^{A}}\quad\text{with}\quad
q^{\hat{n}_{A}}(x^{A})^{k}=q^{k}(x^{A})^{k}%
\end{equation}
as well as the so-called Jackson derivatives \cite{Jackson:1910yd}:%
\begin{equation}
D_{q^{k},\hspace{0.01in}x}\,f=\frac{f(q^{k}x)-f(x)}{q^{k}x-x}.
\end{equation}

Eq.~(\ref{StaProForExp}) shows that the formula for calculating the star
product has the following structure:%
\begin{equation}
f\circledast g=f\hspace{0.01in}g+\sum_{k>0}\lambda^{k}\hspace{0.01in}%
W_{k}(f,g).
\end{equation}
Normally, we have%
\begin{equation}
W_{k}(f,g)\neq W_{k}(\hspace{0.01in}g,f),
\end{equation}
i.~e. the star pro\-duct modifies the ordinary product of two commutative
functions by correction terms that are responsible for the
non-com\-mu\-ta\-ti\-vity of the star pro\-duct. In the undeformed limit
$q\rightarrow1$, these correction terms disappear because they depend on
$\lambda=q-q^{-1}$.

The algebra isomorphism $\mathcal{W}^{-1}$ can also be used to carry over the
conjugation properties of the non-com\-mu\-ta\-tive quantum space algebra
$V_{q}$ to the corresponding commutative coordinate algebra $V$, i.~e. the
mapping $\mathcal{W}^{\hspace{0.01in}-1}$ is also a $\ast$-al\-ge\-bra
homomorphism. This way a conjugation is defined on the commutative coordinate
algebra $V$:%
\begin{equation}
\mathcal{W}(\hspace{0.01in}\overline{f}\hspace{0.01in})=\overline
{\mathcal{W}(f)}\qquad\Leftrightarrow\text{\qquad}\overline{f}=\mathcal{W}%
^{-1}\big (\hspace{0.01in}\overline{\mathcal{W}(f)}\hspace{0.01in}\big ).
\label{ConAlgIso}%
\end{equation}
This relation implies the following conjugation property of the star
pro\-duct:%
\begin{equation}
\overline{f\circledast g}=\overline{g}\circledast\overline{f}.
\label{KonEigSteProFkt}%
\end{equation}

Furthermore, a detailed analysis shows that a power series $f$ in the
commutative coordinates $x^{i}$ becomes under conjugation \cite{Wachter:2007A}%
:%
\begin{align}
\overline{f(x^{1},\ldots,x^{N})}  &  =\sum_{i_{1},\ldots,\hspace{0.01in}i_{N}%
}\bar{a}_{i_{1},\ldots,\hspace{0.01in}i_{N}}\,\overline{(\hspace{0.01in}%
x^{1})^{i_{1}}\ldots\hspace{0.01in}(\hspace{0.01in}x^{N})^{i_{N}}}\nonumber\\
&  =\sum_{i_{1},\ldots,\hspace{0.01in}i_{N}}\bar{a}_{i_{1},\ldots
,\hspace{0.01in}i_{N}}\,(\hspace{0.01in}x_{1})^{i_{1}}\ldots\hspace
{0.01in}(\hspace{0.01in}x_{N})^{i_{N}}\nonumber\\
&  =\sum_{i_{1},\ldots,\hspace{0.01in}i_{N}}\bar{a}_{i_{1},\ldots
,\hspace{0.01in}i_{N}}\,(\hspace{0.01in}g_{1A_{1}}\hspace{0.01in}x^{A_{1}%
})^{i_{1}}\ldots\hspace{0.01in}(\hspace{0.01in}g_{1A_{N}}\hspace
{0.01in}x^{A_{N}})^{i_{N}}\nonumber\\
&  =\bar{f}(\hspace{0.01in}x^{1},\ldots,x^{N}). \label{KonPotReiKom}%
\end{align}
In this case $\bar{a}_{i_{1},\ldots,\hspace{0.01in}i_{N}}$ designates the
complex conjugate of the coefficient $a_{i_{1},\ldots,\hspace{0.01in}i_{N}}$
and - according to Eq.~(\ref{KovKoo})\ of the previous chapter - the covariant
coordinates $x_{i}$ are given by the following expressions:%
\begin{equation}
x_{i}=g_{ij}\hspace{0.01in}X^{j}.
\end{equation}
Now one recognizes that by conjugation of a power series $f$ of the
commutative coordinates $x^{i}$ a new power series arises in which all
expansion coefficients become complex conjugate and all contravariant
coordinates are replaced by the corresponding covariant ones. In the
following, we denote this new power series with $\bar{f}$.

\subsection{Translations\label{KapTra}}

To perform translations on $q$-de\-formed quantum spaces, we replace every
coordinate generator $X^{i}$ of a $q$-de\-formed quantum space $V_{q}$ by
$X^{i}\otimes1+1\otimes Y^{i}$ \cite{Chryssomalakos:1993zm, majid-1993-34}.
Thus, we get a mapping from $V_{q}$ to the tensor product $V_{q}\otimes V_{q}%
$. Since each element of $V_{q}$ can be expanded in terms of normal ordered
monomials, you only need to know how normal ordered monomials behave under
translations. Accordingly, we apply the above substitutions to any normal
ordered monomial of quantum space coordinates. The expression obtained in this
way can again be expanded in terms of tensor products of two normal ordered
monomials:%
\begin{gather}
(X^{1}\otimes1+1\otimes Y^{1})^{i_{1}}\ldots(X^{N}\otimes1+1\otimes
Y^{N})^{i_{N}}=\nonumber\\
=\sum_{\underline{k},l}\alpha_{\underline{i};\underline{k},\underline{l}%
}\,(X^{1})^{k_{1}}\ldots(X^{N})^{k_{N}}\otimes(Y^{1})^{l_{1}}\ldots
(Y^{N})^{l_{N}}.
\end{gather}

In order to get the expansion above, you need the braiding relations between
coordinate generators of different quantum spaces [see Eq.~(\ref{ZopRel}) of
Chap.~\ref{KapEucQuaSpa}] and the commutation relations for coordinate
generators of the same quantum space. Since all non-com\-mu\-ta\-tive
monomials are normal ordered in the above expressions, we can carry over the
above formula to commutative coordinate monomials. This gives us a $q$-analog
of the multidimensional binomial formula. From this $q$-de\-formed formula we
can directly get an operator representation. With the help of this operator
representation, we can calculate $q$-de\-formed translations for all those
functions which we can write as a power series of the commutative coordinates
$x^{i}$. In the case of the $q$-de\-formed Euclidean quantum space with three
dimensions we can get the following formula for calculating $q$%
-trans\-la\-tions \cite{Wachter:2004phengl}:%
\begin{align}
f(\mathbf{x}\oplus\mathbf{y})=  &  \sum_{i_{+}=\hspace{0.01in}0}^{\infty}%
\sum_{i_{3}=\hspace{0.01in}0}^{\infty}\sum_{i_{-}=\hspace{0.01in}0}^{\infty
}\sum_{k\hspace{0.01in}=\hspace{0.01in}0}^{i_{3}}\frac{(-q^{-1}\lambda
\lambda_{+})^{k}}{[[2k]]_{q^{-2}}!!}\frac{(x^{-})^{i_{-}}(x^{3})^{i_{3}%
-\hspace{0.01in}k}(x^{+})^{i_{+}+\hspace{0.01in}k}\,(y^{-})^{k}}%
{[[i_{-}]]_{q^{-4}}!\,[[i_{3}-k]]_{q^{-2}}!\,[[i_{+}]]_{q^{-4}}!}\nonumber\\
&  \qquad\times\big (D_{q^{-4},\hspace{0.01in}y^{-}}^{i_{-}}D_{q^{-2}%
,\hspace{0.01in}y^{3}}^{i_{3}+\hspace{0.01in}k}\hspace{0.01in}D_{q^{-4}%
,\hspace{0.01in}y^{+}}^{i_{+}}f\big )(q^{2(k\hspace{0.01in}-\hspace
{0.01in}i_{3})}y^{-},q^{-2i_{+}}y^{3}).
\end{align}

As mentioned above, the derivation of the $q$-trans\-la\-tions is carried out
with the help of the braiding relations for generators of different quantum
spaces. However, there are two kinds of braiding relations [see also
Eq.~(\ref{ZopRel}) and Eq.~(\ref{InvZopRel}) of Chap.~\ref{KapEucQuaSpa}].
Accordingly, there are two versions of $q$-trans\-la\-tions on each
$q$-de\-formed quantum space. Whenever you want, the operator representations
of the two $q$-translations can be transformed into each other by simple
substitutions \cite{Wachter:2007A}.

It should be mentioned that the $q$-de\-formed quantum spaces we have
considered so far are so-called braided Hopf algebras \cite{Majid:1996kd}.
From this point of view, the two versions of $q$-trans\-lations are nothing
else but realizations of two braided co-pro\-ducts $\underline{\Delta}$ and
$\underline{\bar{\Delta}}$ on the corresponding commutative coordinate
algebras \cite{Wachter:2007A}:%
\begin{align}
f(x\oplus y)  &  =((\mathcal{W}^{\hspace{0.01in}-1}\otimes\mathcal{W}%
^{\hspace{0.01in}-1})\circ\underline{\Delta})(\mathcal{W}%
(f)),\nonumber\\[0.08in]
f(x\,\bar{\oplus}\,y)  &  =((\mathcal{W}^{\hspace{0.01in}-1}\otimes
\mathcal{W}^{-1})\circ\underline{\bar{\Delta}})(\mathcal{W}(f)).
\end{align}
In the same way, you can implement the braided antipodes $\underline{S}$ and
$\underline{\bar{S}}$ on the corresponding commutative coordinate algebras:%
\begin{align}
f(\ominus\,x)  &  =(\mathcal{W}^{\hspace{0.01in}-1}\circ\underline{S}%
\hspace{0.01in})(\mathcal{W}(f)),\nonumber\\
f(\bar{\ominus}\,x)  &  =(\mathcal{W}^{\hspace{0.01in}-1}\circ\underline
{\bar{S}}\hspace{0.01in})(\mathcal{W}(f)). \label{qInvDef}%
\end{align}

The operations in Eq.~(\ref{qInvDef})\ are referred to in the following as
$q$\textit{-in\-ver\-sions}. In the case of the $q$-de\-formed Euclidean
quantum space, we can find the following operator representation for
$q$-in\-ver\-sions \cite{Wachter:2004phengl}:%
\begin{align}
\hat{U}^{-1}f(\ominus\,\mathbf{x})=  &  \sum_{i=0}^{\infty}(-q\lambda
\lambda_{+})^{i}\,\frac{(x^{+}x^{-})^{i}}{[[2i]]_{q^{-2}}!!}\,q^{-2\hat{n}%
_{+}(\hat{n}_{+}+\hspace{0.01in}\hat{n}_{3})-2\hat{n}_{-}(\hat{n}_{-}%
+\hspace{0.01in}\hat{n}_{3})-\hat{n}_{3}\hat{n}_{3}}\nonumber\\
&  \qquad\times D_{q^{-2},\hspace{0.01in}x^{3}}^{2i}\,f(-q^{2-4i}%
x^{-},-q^{1-2i}x^{3},-q^{2-4i}x^{+}).
\end{align}
Note that the operators $\hat{U}$ and $\hat{U}^{-1}$ act on a commutative
function $f(x^{+},x^{3},x^{-})$ as follows:%
\begin{align*}
\hat{U}f  &  =\sum_{k\hspace{0.01in}=\hspace{0.01in}0}^{\infty}\left(
-\lambda\right)  ^{k}\frac{(x^{3})^{2k}}{[[k]]_{q^{-4}}!}\,q^{-2\hat{n}%
_{3}(\hat{n}_{+}+\hspace{0.01in}\hat{n}_{-}+\hspace{0.01in}k)}D_{q^{-4}%
,\hspace{0.01in}x^{+}}^{k}D_{q^{-4},\hspace{0.01in}x^{-}}^{k}f,\\
\hat{U}^{-1}f  &  =\sum_{k\hspace{0.01in}=\hspace{0.01in}0}^{\infty}%
\lambda^{k}\hspace{0.01in}\frac{(x^{3})^{2k}}{[[k]]_{q^{4}}!}\,q^{2\hat{n}%
_{3}(\hat{n}_{+}+\hspace{0.01in}\hat{n}_{-}+\hspace{0.01in}k)}D_{q^{4}%
,\hspace{0.01in}x^{+}}^{k}D_{q^{4},\hspace{0.01in}x^{-}}^{k}f.
\end{align*}
It should be mentioned that the term $q$-in\-ver\-sion arises from the fact
that the braided antipodes are subject to identities of the following form
\cite{Majid:1992sn}:%
\begin{equation}
\sum_{\underline{k},l}\alpha_{\underline{i};\underline{k},\underline{l}%
}\,(X^{1})^{k_{1}}\ldots(X^{N})^{k_{N}}\cdot\underline{S}\big ((X^{1})^{l_{1}%
}\ldots(X^{N})^{l_{N}}\big )=0.
\end{equation}
In this respect, the $q$-in\-ver\-sions are related in some way to an
operation that replaces each coordinate $x^{i}$ with $-x^{i}$.

For the sake of completeness, we will show you how $q$-trans\-la\-tions and
$q$-in\-ver\-sions behave under quantum space conjugation. Since the quantum
spaces we are looking at are so-called braided $\ast$-Hopf algebras, their
braided co-products and antipodes behave under conjugation as follows
\cite{Majid:1994ic}:\footnote{The quantum space conjugation is indicated by
$\ast$.}%
\begin{align}
\tau\circ(\ast\otimes\ast)\circ\underline{\Delta}  &  =\underline{\Delta}%
\circ\ast, & \ast\circ\underline{S}  &  =\underline{S}\circ\ast,\nonumber\\
\tau\circ(\ast\otimes\ast)\circ\underline{\bar{\Delta}}  &  =\underline
{\bar{\Delta}}\circ\ast, & \ast\circ\underline{\bar{S}}  &  =\underline
{\bar{S}}\circ\ast.
\end{align}
As we already know, $q$-trans\-lations and $q$-in\-ver\-sions are nothing else
but realizations of co-products or antipodes. For this reason, we can
immediately read off their behavior under conjugation from the identities
above:%
\begin{align}
\overline{f(x\oplus y)}  &  =\bar{f}(\,y\oplus x), & \overline{f(\ominus\,x)}
&  =\bar{f}(\ominus\,x),\nonumber\\
\overline{f(x\,\bar{\oplus}\,y)}  &  =\bar{f}(\,y\,\bar{\oplus}\,x), &
\overline{f(\bar{\ominus}\,x)}  &  =\bar{f}(\bar{\ominus}\,x).
\label{KonTraAnt}%
\end{align}

\subsection{Partial derivatives\label{ParAblKapAna}}

On $q$-de\-formed quantum spaces partial derivatives with respect to the
position coordinates can be introduced. These partial derivatives form a
$q$-de\-formed quantum space, again. They commutate among each other in the
same way as the position coordinates. Since there are two braiding mappings
for two given quantum spaces, there are also two ways of commuting
$q$-de\-formed partial derivatives with $q$-de\-formed coordinates. With the
vector representation of the R-ma\-trix, the following $q$-de\-formed Leibniz
rules can be specified \cite{CarowWatamura:1990zp, Wess:1990vh}:%
\begin{align}
\partial^{\hspace{0.01in}i}X^{j}  &  =g^{ij}+c\,(\hat{R}^{\hspace{0.01in}%
-1}){^{ij}}_{kl}\,X^{k}\partial^{\hspace{0.01in}l},\nonumber\\
\hat{\partial}^{\hspace{0.01in}i}X^{j}  &  =g^{ij}+c^{-1}\,\hat{R}%
^{\hspace{0.01in}}{^{ij}}_{kl}\,X^{k}\hat{\partial}^{\hspace{0.01in}l}.
\label{LeiRul}%
\end{align}
For the three-dimensional $q$-de\-formed Euclidean space, the constant $c$ is
$1$.

By using the Leibniz rules above, we can calculate how the partial derivatives
act on a normal ordered monomial of non-com\-mu\-ta\-tive coordinates. These
actions can be carried over to commutative coordinate monomials with the help
of the Mo\-yal-Weyl mapping:%
\begin{equation}
\partial^{i}\triangleright(x^{1})^{k_{1}}\ldots\hspace{0.01in}(x^{N})^{k_{N}%
}=\mathcal{W}^{\hspace{0.01in}-1}\big (\partial^{i}\triangleright
(X^{1})^{k_{1}}\ldots\hspace{0.01in}(X^{N})^{k_{N}}\big ).
\end{equation}
Since the Mo\-yal-Weyl mapping is linear, it is possible to extend the action
above to functions that can be expanded as a power series:%
\begin{equation}
\partial^{i}\triangleright f=\mathcal{W}^{\hspace{0.01in}-1}\big (\partial
^{i}\triangleright\mathcal{W}(f)\big ).
\end{equation}
This way we obtain operator representations for the $q$-de\-formed partial
derivatives \cite{Bauer:2003}. In the case of the covariant partial
derivatives $\partial_{A}(=g_{AB}\partial^{B})$ of the three-di\-men\-sional
Euclidean quantum space, these representations are of the following
form:\footnote{Note that the operator representations for the $q$-de\-formed
partial derivatives depend on the choice of the normal ordering of the
non-commu\-ta\-tive coordinates.}%
\begin{align}
\partial_{+}\triangleright f  &  =D_{q^{4},\hspace{0.01in}x^{+}}f,\nonumber\\
\partial_{3}\triangleright f  &  =D_{q^{2},\hspace{0.01in}x^{3}}f(q^{2}%
x^{+}),\nonumber\\
\partial_{-}\triangleright f  &  =D_{q^{4},\hspace{0.01in}x^{-}}f(q^{2}%
x^{3})+\lambda\hspace{0.01in}x^{+}D_{q^{2},\hspace{0.01in}x^{3}}^{2}f.
\end{align}

By applying the Leibniz rules of Eq.~(\ref{LeiRul}) repeatedly, we can commute
$q$-de\-formed partial derivatives from the \textit{left} side of a normal
ordered monomial of quantum space coordinates to the right side. This
procedure leads us to \textit{left}-re\-pre\-sen\-ta\-tions of partial
derivatives \cite{Bauer:2003}:%
\begin{align}
\partial^{i}X^{j}  &  =g^{ij}+c\hspace{0.01in}(\hat{R}^{\hspace{0.01in}%
-1}){^{ij}}_{kl}\,X^{k}\partial^{l} &  &  \Rightarrow &  &  \partial
^{i}\triangleright f,\nonumber\\
\hat{\partial}^{i}X^{j}  &  =g^{ij}+c^{-1}\hat{R}{}{^{\hspace{0.01in}ij}}%
_{kl}\,X^{k}\hat{\partial}^{l} &  &  \Rightarrow &  &  \hat{\partial}%
^{i}\,\bar{\triangleright}\,f. \label{FundWirkNN}%
\end{align}
We can also use the Leibniz rules to commute $q$-de\-formed partial
derivatives from the \textit{right} side of a normal ordered monomial to the
left side. This way, we get \textit{right}-re\-pre\-sen\-ta\-tions of partial
derivatives:%
\begin{align}
X^{i}\partial^{\hspace{0.01in}j}  &  =-g^{ij}+c\hspace{0.01in}(\hat
{R}^{\hspace{0.01in}-1}){^{ij}}_{kl}\,\partial^{k}X^{l} &  &  \Rightarrow &
&  f\,\bar{\triangleleft}\,\partial^{i},\nonumber\\
X^{i}\hat{\partial}^{\hspace{0.01in}j}  &  =-g^{ij}+c^{-1}\hat{R}{}%
{^{\hspace{0.01in}ij}}_{kl}\,\hat{\partial}^{k}X^{l} &  &  \Rightarrow &  &
f\triangleleft\hat{\partial}^{i}. \label{FundWirk2N}%
\end{align}

The expressions for a given representation of $q$-deformed partial derivatives
can be obtained from another representation by simple substitutions (see for
example Ref.~\cite{Bauer:2003}). Moreover, the different actions of partial
derivatives transform into each other by conjugation. In the case of the
three-dimensional $q$-de\-formed Euclidean space, for example, the following
applies \cite{Bauer:2003}:%
\begin{align}
\overline{\partial^{A}\triangleright f}  &  =-\bar{f}\,\bar{\triangleleft
}\,\partial_{A}, & \overline{f\,\bar{\triangleleft}\,\partial^{A}}  &
=-\partial_{A}\triangleright\bar{f},\nonumber\\
\overline{\hat{\partial}^{A}\,\bar{\triangleright}\,f}  &  =-\bar
{f}\triangleleft\hat{\partial}_{A}, & \overline{f\triangleleft\hat{\partial
}^{A}}  &  =-\hat{\partial}_{A}\,\bar{\triangleright}\,\bar{f}.
\label{RegConAblN}%
\end{align}

\subsection{Integration\label{KapIntegral}}

The operator representations of $q$-de\-formed partial derivatives consist of
a term $\partial_{\operatorname*{cla}}^{A}$ and a so-called correction term
$\partial_{\operatorname*{cor}}^{A}$:%
\begin{equation}
\partial^{A}\triangleright F=\left(  \partial_{\operatorname*{cla}}%
^{A}+\partial_{\operatorname*{cor}}^{A}\right)  \triangleright F.
\end{equation}
The term $\partial_{\operatorname*{cla}}^{A}$becomes an ordinary partial
derivative in the undeformed limit $q\rightarrow1$ and the term $\partial
_{\operatorname*{cor}}^{A}$ disappears in the undeformed limit. The difference
equation $\partial^{A}\triangleright F=f$ with given $f$ is solved by the
following expression:%
\begin{align}
F  &  =(\partial^{A})^{-1}\triangleright f=\left(  \partial
_{\operatorname*{cla}}^{A}+\partial_{\operatorname*{cor}}^{A}\right)
^{-1}\triangleright f\nonumber\\
&  =\sum_{k\hspace{0.01in}=\hspace{0.01in}0}^{\infty}\left[  -(\partial
_{\operatorname*{cla}}^{A})^{-1}\partial_{\operatorname*{cor}}^{A}\right]
^{k}(\partial_{\operatorname*{cla}}^{A})^{-1}\triangleright f.
\end{align}
In the case of the three-di\-men\-sional Euclidean quantum space, for example,
the above formula leads to the expressions \cite{Wachter:2004A}%
\begin{align}
(\partial_{+})^{-1}\triangleright f  &  =D_{q^{4},\hspace{0.01in}x^{+}}%
^{-1}f,\nonumber\\
(\partial_{3})^{-1}\triangleright f  &  =D_{q^{2},\hspace{0.01in}x^{3}}%
^{-1}f(q^{-2}x^{+}),
\end{align}
and%
\begin{gather}
(\partial_{-})^{-1}\triangleright f=\nonumber\\
=\sum_{k\hspace{0.01in}=\hspace{0.01in}0}^{\infty}q^{2k\left(  k\hspace
{0.01in}+1\right)  }\left(  -\lambda\,x^{+}D_{q^{4},\hspace{0.01in}x^{-}}%
^{-1}D_{q^{2},\hspace{0.01in}x^{3}}^{2}\right)  ^{k}D_{q^{4},\hspace
{0.01in}x^{-}}^{-1}f(q^{-2\left(  k\hspace{0.01in}+1\right)  }x^{3}).
\end{gather}
Note that $D_{q,\hspace{0.01in}x}^{-1}$ stands for a Jackson integral with
respect to the variable $x$ \cite{Jackson:1908}. The explicit form of this
Jackson integral depends on its integration limits and the value for the
deformation parameter $q$. If $x>0$ and $q>1$, for example, the following
applies:%
\begin{align}
\int_{0}^{\hspace{0.01in}x}\text{d}_{q}z\hspace{0.01in}f(z)  &  =(q-1)\hspace
{0.01in}x\sum_{j=1}^{\infty}q^{-j}f(q^{-j}x),\nonumber\\
\int_{x}^{\hspace{0.01in}\infty}\text{d}_{q}z\hspace{0.01in}f(z)  &
=(q-1)\hspace{0.01in}x\sum_{j=\hspace{0.01in}0}^{\infty}q^{\hspace{0.01in}%
j}f(q^{\hspace{0.01in}j}x).
\end{align}

By successively applying the $q$-in\-te\-grals for the different coordinates,
we can explain an integration over the entire coordinate space. With the
exception of a normalization factor, this integration is not dependent on the
sequence of the different integrations \cite{Wachter:2004A, Wachter:2007A}:%
\begin{equation}
\int_{-\infty}^{+\infty}\text{d}_{q}^{3}x\,f(x^{+},x^{3},x^{-})\sim
(\partial_{-})^{-1}\big |_{-\infty}^{+\infty}\,(\partial_{3})^{-1}%
\big |_{-\infty}^{+\infty}\,(\partial_{+})^{-1}\big |_{-\infty}^{+\infty
}\triangleright f.
\end{equation}
We can also show that on the right side of the above relation the integrals to
the different coordinates can be simplified to Jackson integrals
\cite{Wachter:2004A}:\footnote{This simplification results from the fact that
the integrated function must disappear at infinity.}%
\begin{equation}
\int_{-\infty}^{+\infty}\text{d}_{q}^{3}x\,f(\mathbf{x})\sim D_{q^{2}%
,\hspace{0.01in}x^{-}}^{-1}\big |_{-\infty}^{+\infty}\,D_{q,x^{3}}%
^{-1}\big |_{-\infty}^{+\infty}\,D_{q^{2},\hspace{0.01in}x^{+}}^{-1}%
\big |_{-\infty}^{+\infty}\,f(\mathbf{x}).
\end{equation}
Note that the Jackson integrals in the formula above refer to a smaller
$q$-lattice. This is not a particular restriction since we can obtain each
Jackson integral on the smaller $q$-lattice by adding two other Jackson
integrals. These two Jackson integrals refer to two wider $q$-lattices shifted
against each other. The scaling down of the $q$-lattice ensures that the
$q$-in\-te\-grals over the entire coordinate space form scalars with respect
to the symmetry of the underlying quantum space.

\begin{figure}
[ptb]
\begin{center}
\includegraphics[width=0.80\textwidth]{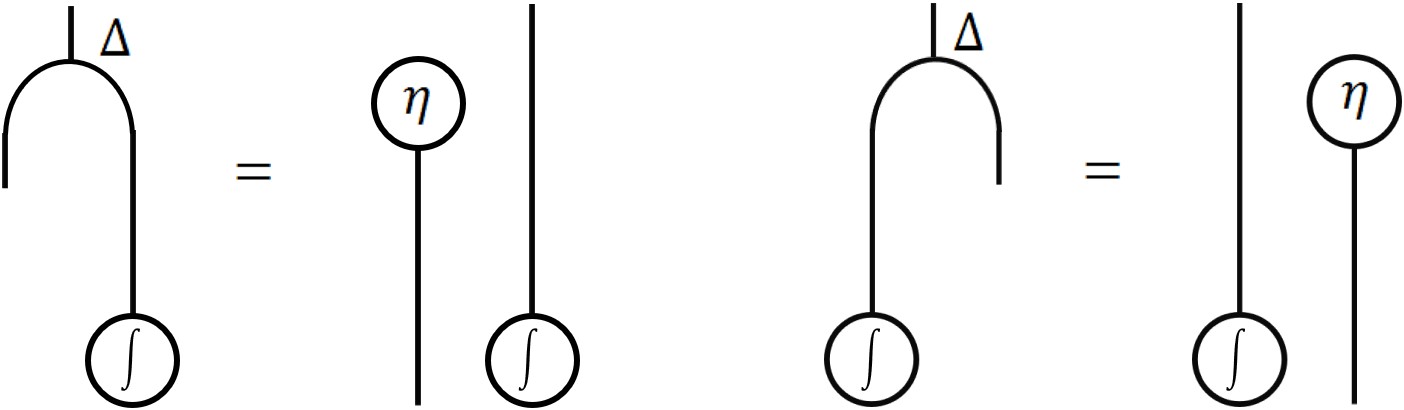}
\caption{Translational invariance of the $q$-integral over the entire space}%
\label{Fig13}%
\end{center}
\end{figure}

The $q$-de\-formed integrals over the entire coordinate space show some
important properties \cite{Wachter:2007A}. In this respect, $q$-de\-formed
versions of \textit{Stokes' theorem} apply to the $q$-in\-te\-gra\-tion over
the entire $q$-de\-formed Euclidean coordinate space:%
\begin{align}
\int_{-\infty}^{+\infty}\text{d}_{q}^{3}x\,\partial^{A}\triangleright f  &
=\int_{-\infty}^{+\infty}\text{d}_{q}^{3}x\,f\,\bar{\triangleleft}%
\,\partial^{A}=0,\nonumber\\
\int_{-\infty}^{+\infty}\text{d}_{q}^{3}x\,\hat{\partial}^{A}\,\bar
{\triangleright}\,f  &  =\int_{-\infty}^{+\infty}\text{d}_{q}^{3}%
x\,f\triangleleft\hat{\partial}^{A}=0.
\end{align}
This implies that $q$-in\-te\-grals over the entire coordinate space are
invariant with respect to $q$-de\-formed translations:%
\begin{align}
\int_{-\infty}^{+\infty}\text{d}_{q}^{3}x\,f(x)  &  =\int_{-\infty}^{+\infty
}\text{d}_{q}^{3}x\,f(\hspace{0.01in}y\oplus x)=\int_{-\infty}^{+\infty
}\text{d}_{q}^{3}x\,f(\hspace{0.01in}y\,\bar{\oplus}\,x)\nonumber\\
&  =\int_{-\infty}^{+\infty}\text{d}_{q}^{3}x\,f(x\oplus y)=\int_{-\infty
}^{+\infty}\text{d}_{q}^{3}x\,f(x\,\bar{\oplus}\,y). \label{TraInvQuaVolInt}%
\end{align}
For later purposes, we provide a graphical representation\footnote{In
App.~\ref{AppA} we have collected some information on the graphical calculus
we use.} of these identities\ in Fig.~\ref{Fig13}. The $q$-de\-formed Stokes'
theorem also implies the following rules for integration by parts:%
\begin{align}
\int_{-\infty}^{+\infty}\text{d}_{q}^{3}x\,f\circledast(\partial
^{A}\triangleright g)  &  =\int_{-\infty}^{+\infty}\text{d}_{q}^{3}%
x\,(f\triangleleft\partial^{A})\circledast g,\nonumber\\
\int_{-\infty}^{+\infty}\text{d}_{q}^{3}x\,f\circledast(\hat{\partial}%
^{A}\,\bar{\triangleright}\,g)  &  =\int_{-\infty}^{+\infty}\text{d}_{q}%
^{3}x\,(f\,\bar{\triangleleft}\,\hat{\partial}^{A})\circledast g.
\label{PatIntUneRaumInt}%
\end{align}

Finally, it should be mentioned that the invariant integral for the
$q$-de\-formed Euclidean quantum space behaves as follows under quantum space
conjugation:%
\begin{equation}
\overline{\int_{-\infty}^{+\infty}\text{d}_{q}^{3}x\,f}=\int_{-\infty
}^{+\infty}\text{d}_{q}^{3}x\,\bar{f}. \label{KonEigVolInt}%
\end{equation}%

\subsection{Exponentials\label{KapExp}}

\begin{figure}
[ptb]
\begin{center}
\includegraphics[width=0.85\textwidth]{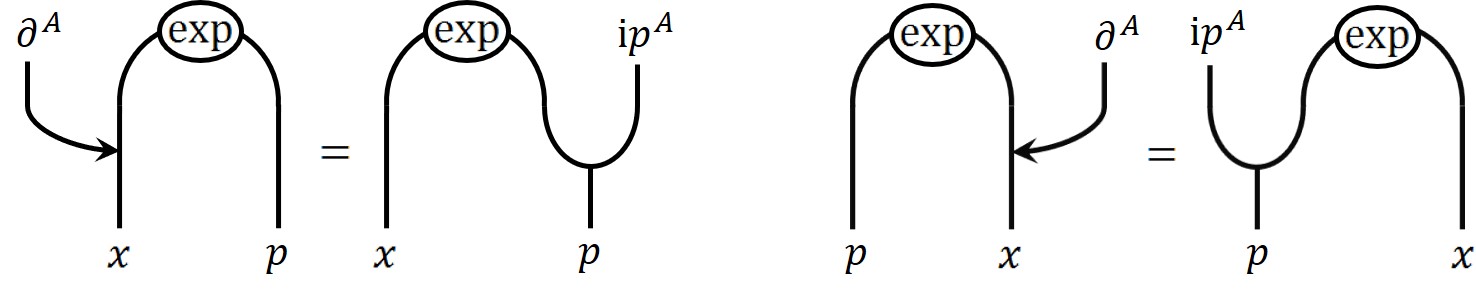}%
\caption{Eigenvalue equations of $q$-exponentials}%
\label{Fig1}%
\end{center}
\end{figure}

The $q$-de\-formed exponentials are eigenfunctions of the $q$-de\-formed
partial derivatives of a given quantum space \cite{Majid:1993ud}. In the
following, we consider those $q$-de\-formed exponentials that are
eigenfunctions of unconjugated left derivatives or conjugated right
derivatives:%
\begin{align}
\text{i}^{-1}\partial^{A}\triangleright\exp_{q}(x|\text{i}p)  &  =\exp
_{q}(x|\text{i}p)\circledast p^{A},\nonumber\\
\exp_{q}(\text{i}^{-1}p|x)\,\bar{\triangleleft}\,\partial^{A}\text{i}^{-1}  &
=p^{A}\circledast\exp_{q}(\text{i}^{-1}p|x). \label{EigGl1N}%
\end{align}
For a better understanding, the above eigenvalue equations are shown
graphically in Fig.~\ref{Fig1}. The $q$-ex\-po\-nen\-tials are uniquely
defined by their eigenvalue equations in connection with the following
normalization conditions:%
\begin{align}
\exp_{q}(x|\text{i}p)|_{x\hspace{0.01in}=\hspace{0.01in}0}  &  =\exp
_{q}(x|\text{i}p)|_{p\hspace{0.01in}=\hspace{0.01in}0}=1,\nonumber\\
\exp_{q}(\text{i}^{-1}p|x)|_{x\hspace{0.01in}=\hspace{0.01in}0}  &  =\exp
_{q}(\text{i}^{-1}p|x)|_{p\hspace{0.01in}=\hspace{0.01in}0}=1.
\end{align}%

In order to get explicit formulas for the $q$-ex\-ponentials, we best consider
the dual pairings between the coordinate algebra of the $q$-de\-formed
position space and that of the corresponding $q$-de\-formed momentum space
\cite{Majid:1993ud}:%
\begin{align}
\big\langle f(\hspace{0.01in}p),g(x)\big\rangle_{p,\bar{x}}  &  \equiv
\lbrack\hspace{0.01in}f(\text{i}^{-1}\partial)\triangleright g(x)]_{x\hspace
{0.01in}=\hspace{0.01in}0},\nonumber\\
\big\langle f(x),g(\hspace{0.01in}p)\big\rangle_{x,\bar{p}}  &  \equiv
\lbrack\hspace{0.01in}f(x)\,\bar{\triangleleft}\,g(\partial\hspace
{0.01in}\text{i}^{-1})]_{x\hspace{0.01in}=\hspace{0.01in}0}.
\end{align}
Let $\{e^{a}\}$ be a basis of the $q$-de\-formed position space algebra and
let $\{f_{b}\}$ be a basis of the corresponding $q$-de\-formed momentum
algebra. Furthermore, the elements of the two bases shall be dually paired in
the following sense:\footnote{$\mathcal{W}_{x}$ and $\mathcal{W}_{p}$ denote
the Moyal-Weyl mapping for the $q$-de\-formed position space algebra and that
for the $q$-de\-formed momentum space algebra, respectively.}%
\begin{equation}
\big\langle\mathcal{W}_{p}(f_{b}),\mathcal{W}_{x}(e^{a})\big\rangle_{p,\bar
{x}}=\delta_{b}^{a},\qquad\big\langle\mathcal{W}_{x}(e^{a}),\mathcal{W}%
_{p}(f_{b})\big\rangle_{x,\bar{p}}=\delta_{b}^{a},
\end{equation}
Now, we are in a position to write the $q$-ex\-po\-nen\-tials as canonical
elements:%
\begin{align}
\exp_{q}(x|\text{i}p)  &  \equiv\sum\nolimits_{a}\mathcal{W}_{x}(e^{a}%
)\otimes\mathcal{W}_{p}(f_{a}),\nonumber\\
\exp_{q}(\text{i}p|x)  &  \equiv\sum\nolimits_{a}\mathcal{W}_{p}(f_{a}%
)\otimes\mathcal{W}_{x}(e^{a}).
\end{align}
The normal ordered monomials of non-commutative coordinates establish a basis
of the quantum space under consideration. The elements of the dual basis can
be obtained by the action of the partial derivatives on these normal ordered
monomials. This way, we have found the following expressions for the
$q$-ex\-ponen\-tials of the three-di\-mensional Euclidean quantum space
\cite{Wachter:2004ExpA}:%
\begin{align}
\exp_{q}(x|\text{i}p)  &  =\sum_{\underline{n}\,=\,0}^{\infty}\frac
{(q\hspace{0.01in}x^{+})^{n_{+}}(x^{3})^{n_{3}}(q^{-1}x^{-})^{n_{-}}%
(\text{i}^{-1}p^{+})^{n_{-}}(\text{i}p^{3})^{n_{3}}(\text{i}^{-1}p^{-}%
)^{n_{+}}}{[[n_{+}]]_{q^{4}}!\,[[n_{3}]]_{q^{2}}!\,[[n_{-}]]_{q^{4}}%
!},\nonumber\\
\exp_{q}(\text{i}^{-1}p|x)  &  =\sum_{\underline{n}\,=\,0}^{\infty}%
\frac{(\text{i}p^{+})^{n_{+}}(\text{i}^{-1}p^{3})^{n_{3}}(\text{i}%
p^{-})^{n_{-}}(q^{-1}x^{+})^{n_{-}}(x^{3})^{n_{3}}(q\hspace{0.01in}%
x^{-})^{n_{+}}}{[[n_{+}]]_{q^{4}}!\,[[n_{3}]]_{q^{2}}!\,[[n_{-}]]_{q^{4}}}.
\label{ExpEukExp}%
\end{align}%
\begin{figure}
[ptb]
\begin{center}
\includegraphics[width=0.40\textwidth]{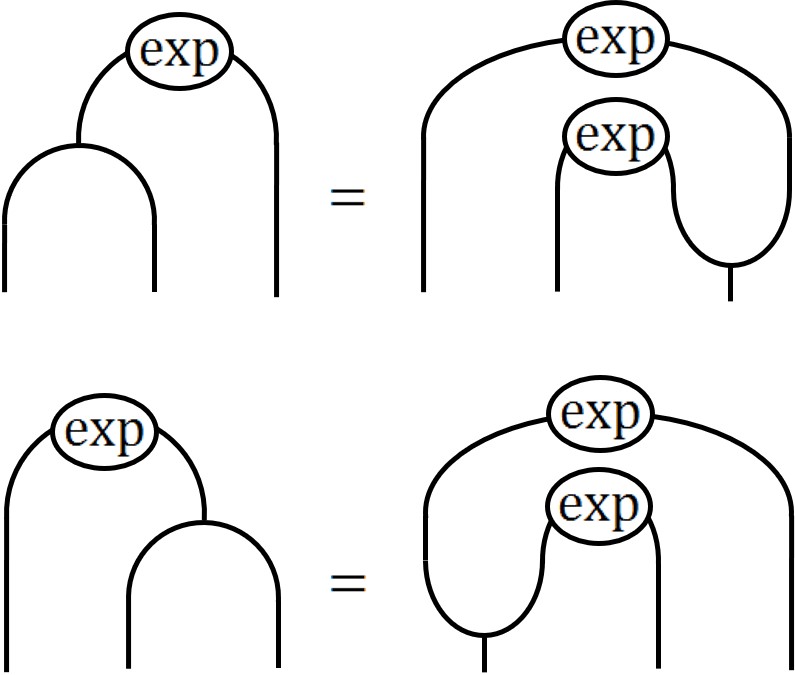}
\caption{Laws of addition for $q$-exponentials}%
\label{Fig2}%
\end{center}
\end{figure}

In analogy to the undeformed case, the $q$-ex\-ponen\-tials satisfy addition
theorems of the following form \cite{Majid:1993ud, Wachter:2007A}:%
\begin{align}
\exp_{q}(x\,\bar{\oplus}\,y|\text{i}p)  &  =\exp_{q}(x|\exp_{q}(\hspace
{0.01in}y|\text{i}p)\circledast\text{i}p),\nonumber\\
\exp_{q}(\text{i}x|p\,\bar{\oplus}\,\tilde{p})  &  =\exp_{q}(x\circledast
\exp_{q}(x|\text{i}p)|\text{i}\tilde{p}). \label{AddTheExp}%
\end{align}
Note that we can obtain further addition theorems from the above identities by
substituting the position coordinates by momentum coordinates and vice versa.
For a better understanding of the meaning of the two addition theorems in
Eq.~(\ref{AddTheExp}), we have depicted them graphically in Fig.~\ref{Fig2}.%
\begin{figure}
[ptb]
\begin{center}
\includegraphics[width=0.55\textwidth]{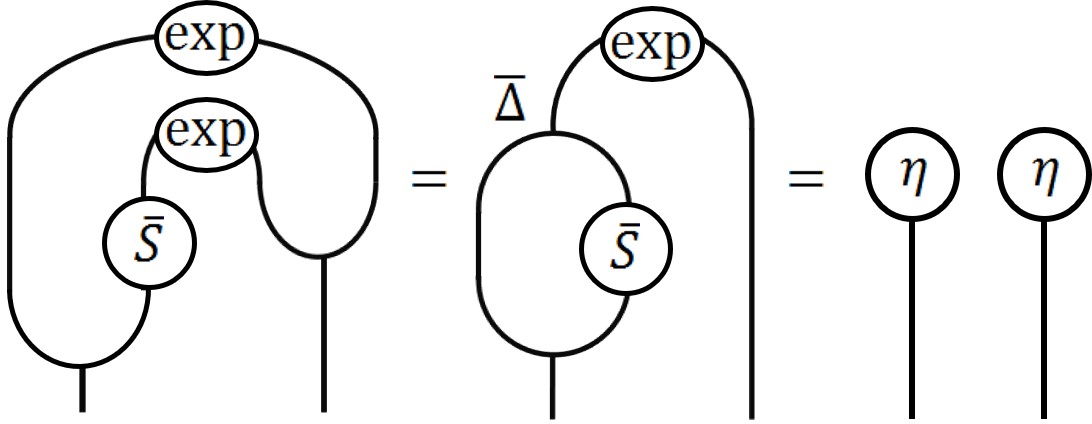}
\caption{Invertibility of $q$-exponentials}%
\label{Fig3}%
\end{center}
\end{figure}
\begin{figure}
[ptbptb]
\begin{center}
\includegraphics[width=0.38\textwidth]{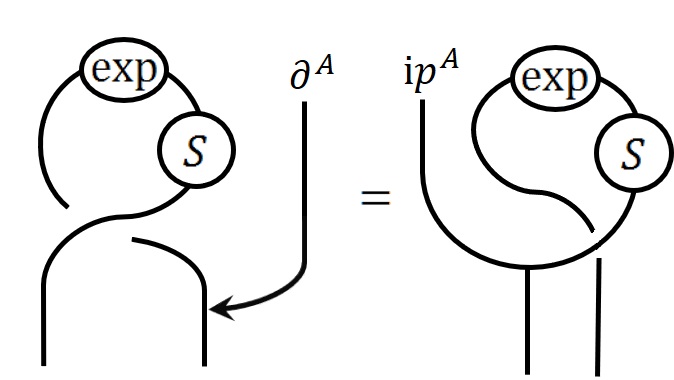}
\caption{Eigenvalue equation of twisted $q$-exponential}%
\label{Fig4}%
\end{center}
\end{figure}

We can also introduce inverse $q$-ex\-ponen\-tials by using the $q$%
-in\-ver\-sions from Chap.~\ref{KapTra}:%
\begin{equation}
\exp_{q}(\bar{\ominus}\,x|\text{i}p)=\exp_{q}(\text{i}x|\text{{}}\bar{\ominus
}\,p). \label{InvExpAlgDefKom}%
\end{equation}
Due to the addition theorems and the normalization conditions of the
$q$-ex\-ponen\-tials, the following applies:%
\begin{equation}
\exp_{q}(\text{i}x\circledast\exp_{q}(\bar{\ominus}\,x|\text{i}p)\circledast
p)=\exp_{q}(x\,\bar{\oplus}\,(\bar{\ominus}\,x)|\text{i}p)=\exp_{q}%
(x|\text{i}p)|_{x=0}=1.
\end{equation}
For a better understanding of these identities, we have given their graphic
representation in Fig.~\ref{Fig3}.

Next, we will describe another way of obtaining $q$-ex\-ponentials. For this
purpose, we exchange the two tensor factors of a $q$-ex\-ponential using the
inverse universal R-matrix [see also Eq.~(\ref{InvZopRelAlg}) of
Chap.~\ref{KapEucQuaSpa}\textbf{\ }and the graphic representation in
Fig.~\ref{Fig4}]:%
\begin{align}
\exp_{q}^{\ast}(\text{i}p|x)  &  \equiv\tau\circ\lbrack(\mathcal{R}_{[2]}%
^{-1}\otimes\mathcal{R}_{[1]}^{-1})\triangleright\exp_{q}(\text{i}%
x|\hspace{-0.03in}\ominus\hspace{-0.01in}p)]\nonumber\\
&  =\sum\nolimits_{a}\mathcal{W}_{p}\big (\mathcal{R}_{[1]}^{-1}%
\triangleright\underline{S}\hspace{0.01in}f^{a}\big )\otimes\mathcal{W}%
_{x}\big (\mathcal{R}_{[2]}^{-1}\triangleright e_{a}\big ), \label{DuaExp1}%
\\[0.05in]
\exp_{q}^{\ast}(x|\text{i}p)  &  \equiv\tau\circ\lbrack(\mathcal{R}_{[2]}%
^{-1}\otimes\mathcal{R}_{[1]}^{-1})\triangleright\exp_{q}(\ominus
\hspace{0.02in}p\hspace{0.01in}|\text{i}x)]\nonumber\\
&  =\sum\nolimits_{a}\mathcal{W}_{x}\big (\mathcal{R}_{[1]}^{-1}\triangleright
e_{a}\big )\otimes\mathcal{W}_{p}\big (\mathcal{R}_{[2]}^{-1}\triangleright
\underline{S}\hspace{0.01in}f^{a}\big ). \label{DuaExp2}%
\end{align}
Note that in the expressions above $\tau$ denotes the ordinary twist operator.
You can show that the new $q$-ex\-ponentials satisfy the following eigenvalue
equations (also cf. Fig.~\ref{Fig4}):%
\begin{align}
\exp_{q}^{\ast}(\text{i}p|x)\triangleleft\partial^{A}  &  =\text{i}%
p^{A}\circledast\exp_{q}^{\ast}(\text{i}p|x),\nonumber\\
\partial^{A}\,\bar{\triangleright}\,\exp_{q}^{\ast}(x|\text{i}^{-1}p)  &
=\exp_{q}^{\ast}(x|\text{i}^{-1}p)\circledast\text{i}p^{A}.
\label{EigGleExpQueAbl}%
\end{align}

Finally, we write down how the $q$-ex\-ponentials of the three-di\-men\-sional
quantum space behave under conjugation:%
\begin{align}
\overline{\exp_{q}(x|\text{i}p)}  &  =\exp_{q}(\text{i}^{-1}p|x),\nonumber\\
\overline{\exp_{q}^{\ast}(\text{i}p|x)}  &  =\exp_{q}^{\ast}(x|\text{i}%
^{-1}p). \label{KonEigExpQua}%
\end{align}
The first identity can be derived directly from the expression in
Eq.~(\ref{ExpEukExp}). The second identity then follows by taking the
conjugation properties of the universal R-ma\-trix into account.

\section{Fourier transformations on quantum spaces\label{KapFourTra}}

\subsection{Definition\label{KapFTDef}}

In Ref.~\cite{Kempf:1994yd}, it was shown how to define Fourier
transformations between two $q$-de\-formed quantum space algebras. Recall that
the star product formalism allows us to identify $q$-de\-formed quantum space
algebras with commutative coordinate algebras. Thus the different operations
introduced for $q$-de\-formed quantum space algebras can be carried over to
the corresponding commutative coordinate algebras. For this reason, we are
able to implement the Fourier transformations for $q$-de\-formed quantum
spaces on the corresponding commutative coordinate algebras. By using the star
products from Chap.~\ref{KapStePro}, the improper $q$-in\-te\-grals from
Chap.~\ref{KapIntegral}\ and the $q$-ex\-po\-nenti\-als from
Chap.~\ref{KapExp} we get:%
\begin{align}
\mathcal{F}_{L}(f)(\hspace{0.01in}p)  &  =\int\nolimits_{-\infty}^{+\infty
}\text{d}_{q}^{N}\hspace{-0.02in}x\,f(x)\circledast\exp_{q}(x|\text{i}%
p),\nonumber\\
\mathcal{F}_{R}(f)(\hspace{0.01in}p)  &  =\int\nolimits_{-\infty}^{+\infty
}\text{d}_{q}^{N}\hspace{-0.02in}x\,\exp_{q}(\text{i}^{-1}p|x)\circledast
f(x). \label{DefFouKomLR}%
\end{align}

In analogy to the undeformed case, the $q$-de\-formed Fourier transformations
of the function $f(x)=1$ result in $q$-de\-formed delta functions:%
\begin{align}
\delta_{L}^{N}(\hspace{0.01in}p)  &  =\mathcal{F}_{L}(1)(\hspace
{0.01in}p)=\int\nolimits_{-\infty}^{+\infty}\text{d}_{q}^{N}\hspace
{-0.02in}x\,\exp_{q}(x|\text{i}p),\nonumber\\
\delta_{R}^{N}(\hspace{0.01in}p)  &  =\mathcal{F}_{R}(1)(\hspace
{0.01in}p)=\int\nolimits_{-\infty}^{+\infty}\text{d}_{q}^{N}\hspace
{-0.02in}x\,\exp_{q}(\text{i}^{-1}p|x). \label{DefDelLR}%
\end{align}
Integration of the $q$-de\-formed delta functions, in turn, leads to the
so-called $q$-de\-formed volume elements:%
\begin{align}
\operatorname*{vol}\nolimits_{L}  &  =\int\nolimits_{-\infty}^{+\infty
}\text{d}_{q}^{N}\hspace{-0.02in}p\,\delta_{L}^{N}(\hspace{0.01in}%
p)=\int\nolimits_{-\infty}^{+\infty}\text{d}_{q}^{N}\hspace{-0.02in}%
x\int\nolimits_{-\infty}^{+\infty}\text{d}_{q}^{N}\hspace{-0.02in}p\,\exp
_{q}(x|\text{i}p),\nonumber\\
\operatorname*{vol}\nolimits_{R}  &  =\int\nolimits_{-\infty}^{+\infty
}\text{d}_{q}^{N}\hspace{-0.02in}p\,\delta_{R}^{N}(\hspace{0.01in}%
p)=\int\nolimits_{-\infty}^{+\infty}\text{d}_{q}^{N}\hspace{-0.02in}%
p\int\nolimits_{-\infty}^{+\infty}\text{d}_{q}^{N}\hspace{-0.02in}x\,\exp
_{q}(\text{i}^{-1}p|x). \label{DelVolLR}%
\end{align}

As we can see from Eq.~(\ref{ExpEukExp}), the $q$-ex\-ponentials $\exp_{q}%
(x|$i$p)$ and $\exp_{q}($i$^{-1}p|x)$ transform into each other by the
substitutions $x^{A}\rightarrow p^{A}$ and $p^{A}\rightarrow-x^{A}$, i.~e. the
two $q$-ex\-ponentials differ from each other by an exchange of position
coordinates and momentum coordinates. Therefore, the integration of both
$q$-ex\-ponentials over the entire phase space must lead to the same result.
For this reason, the two volume elements in Eq.~(\ref{DelVolLR}) are
identical:%
\begin{equation}
\operatorname*{vol}\equiv\operatorname*{vol}\nolimits_{L}=\operatorname*{vol}%
\nolimits_{R}. \label{ZusVolLRqLqR}%
\end{equation}

Up to now, we have considered Fourier transformations which map from the
$q$-de\-formed \textit{position} algebra into the $q$-de\-formed
\textit{momentum} algebra. If we want to formulate Fourier transformations
which map from the $q$-de\-formed \textit{momentum} algebra into the
$q$-de\-formed \textit{position} algebra, we have to use the $q$%
-ex\-po\-nenti\-als given in Eq.~(\ref{DuaExp1}) and Eq.~(\ref{DuaExp2})\ of
Chap.~\ref{KapExp}:%
\begin{align}
\mathcal{F}_{R}^{\hspace{0.01in}\ast}(f)(x)  &  =\int\nolimits_{-\infty
}^{+\infty}\text{d}_{q}^{N}\hspace{-0.02in}p\hspace{0.01in}\exp_{q}^{\ast
}(x|\text{i}^{-1}p)\circledast f(\hspace{0.01in}p),\nonumber\\
\mathcal{F}_{L}^{\hspace{0.01in}\ast}(f)(x)  &  =\int\nolimits_{-\infty
}^{+\infty}\text{d}_{q}^{N}\hspace{-0.02in}p\,f(\hspace{0.01in}p)\circledast
\exp_{q}^{\ast}(\text{i}p|x). \label{FTtype1}%
\end{align}

From their defining expressions and the conjugation properties of
$q$-in\-te\-grals and $q$-ex\-po\-nentials, we can directly deduce that the
$q$-de\-formed Fourier transformations of three-di\-men\-sional Euclidean
quantum space behave as follows under conjugation:%
\begin{equation}
\overline{\mathcal{F}_{L}(f)}=\mathcal{F}_{R}(\bar{f}),\qquad\overline
{\mathcal{F}_{L}^{\ast}(f)}=\mathcal{F}_{R}^{\ast}(\bar{f}).
\end{equation}
Correspondingly, we find for the volume element:%
\begin{equation}
\overline{\operatorname*{vol}}=\overline{\operatorname*{vol}\nolimits_{L}%
}=\operatorname*{vol}\nolimits_{R}=\operatorname*{vol}.
\label{KonEigVolEleHer}%
\end{equation}
In addition to this, the $q$-de\-formed delta functions are subject to the
following identities:%
\begin{equation}
\overline{\delta_{L}^{N}(\hspace{0.01in}p)}=\overline{\mathcal{F}%
_{L}(1)(\hspace{0.01in}p)}=\mathcal{F}_{R}(1)(\hspace{0.01in}p)=\delta_{R}%
^{N}(\hspace{0.01in}p). \label{KonEigDelFkt}%
\end{equation}

\subsection{Invertibility\label{KapFourInv}}

The Fourier transformations on coordinate algebras of $q$-de\-formed quantum
spaces can be inverted \cite{Kempf:1994yd}. Specifically, we have the
identities%
\begin{align}
(\mathcal{F}_{L}\circ\mathcal{F}_{L}^{\text{\hspace{0.01in}}\ast}%
)(f(\hspace{0.01in}p))  &  =\operatorname*{vol}\nolimits_{L}f(\hspace
{0.01in}p),\nonumber\\
(\mathcal{F}_{R}\circ\mathcal{F}_{R}^{\text{\hspace{0.01in}}\ast}%
)(f(\hspace{0.01in}p))  &  =\operatorname*{vol}\nolimits_{R}f(\hspace
{0.01in}p), \label{InvFourAnf2a}%
\end{align}
and%
\begin{align}
(\mathcal{F}_{L}^{\text{\hspace{0.01in}}\ast}\circ\mathcal{F}_{L})(f(x))  &
=\operatorname*{vol}\nolimits_{L}f(x),\nonumber\\
(\mathcal{F}_{R}^{\text{\hspace{0.01in}}\ast}\circ\mathcal{F}_{R})(f(x))  &
=\operatorname*{vol}\nolimits_{R}f(x). \label{InvFourAnf1a}%
\end{align}
In App.~\ref{AnhB} we explain how to prove the above identities.

\subsection{Transformation of actions and products\label{KapFouActPro}}

The Fourier transformations we have introduced in Eq.~(\ref{DefFouKomLR}) show
the characteristic property that they transform the action of a derivative
operator into a product with a momentum coordinate \cite{Kempf:1994yd}:%
\begin{align}
\mathcal{F}_{L}(f(x)\triangleleft\partial_{x}^{k})  &  =\text{i\hspace
{0.01in}}\mathcal{F}_{L}(f(x))\circledast p^{k},\nonumber\\
\mathcal{F}_{R}(\partial_{x}^{k}\hspace{0.03in}\bar{\triangleright}\,f(x))  &
=\text{i\hspace{0.01in}}p^{k}\circledast\mathcal{F}_{R}(f(x)).
\label{FunProp1}%
\end{align}
We can derive these identities in a direct way, as the following calculation
helps to demonstrate:%
\begin{align}
\mathcal{F}_{L}(f(x)\triangleleft\partial_{x}^{k})  &  =\int\nolimits_{-\infty
}^{+\infty}\text{d}_{q}^{N}\hspace{-0.02in}x\,(f(x)\triangleleft\partial
_{x}^{k})\circledast\exp_{q}(x|\text{i}p)\nonumber\\
&  =\int\nolimits_{-\infty}^{+\infty}\text{d}_{q}^{N}\hspace{-0.02in}%
x\,f(x)\circledast\big (\partial_{x}^{k}\triangleright\exp_{q}(x|\text{i}%
p)\big )\nonumber\\
&  =\text{i}\int\nolimits_{-\infty}^{+\infty}\text{d}_{q}^{N}\hspace
{-0.02in}x\,f(x)\circledast\exp_{q}(x|\text{i}p)\circledast p^{k}%
\nonumber\\[0.03in]
&  =\text{i\hspace{0.01in}}\mathcal{F}_{L}(f(x))\circledast p^{k}.
\label{HerFourId}%
\end{align}
The second identity follows from integration by parts [cf.
Eq.~(\ref{PatIntUneRaumInt}) of\ Chap.~\ref{KapIntegral}] and the third
identity is a consequence of the eigenvalue equations of the $q$%
-ex\-po\-nen\-tial [cf. Eq.~(\ref{EigGl1N}) of Chap.~\ref{KapExp}].

In the other direction, the multiplication by coordinates is transformed into
the action of partial derivatives:%
\begin{align}
\mathcal{F}_{L}(f(x)\circledast x^{k})  &  =\mathcal{F}_{L}(f(x))\,\bar
{\triangleleft}\,\partial_{p}^{k}\text{i},\nonumber\\
\mathcal{F}_{R}(x^{k}\circledast f(x))  &  =\text{i}\partial_{p}%
^{k}\triangleright\mathcal{F}_{R}(f(x)). \label{FunProp2a}%
\end{align}
These identities can be derived by calculations similar to those of
Eq.~(\ref{HerFourId}):%
\begin{align}
\mathcal{F}_{L}(f(x)\circledast x^{k})  &  =\int\nolimits_{-\infty}^{+\infty
}\text{d}_{q}^{N}\hspace{-0.02in}x\,f(x)\circledast x^{k}\circledast\exp
_{q}(x|\text{i}p)\nonumber\\
&  =\text{i}\int\nolimits_{-\infty}^{+\infty}\text{d}_{q}^{N}\hspace
{-0.02in}x\,f(x)\circledast\exp_{q}(x|\text{i}p)\,\bar{\triangleleft
}\,\partial_{p}^{k}\nonumber\\[0.03in]
&  =\text{i\hspace{0.01in}}\mathcal{F}_{L}(f(x))\,\bar{\triangleleft
}\,\partial_{p}^{k}. \label{HerFLMulAbl}%
\end{align}

The $q$-de\-formed Fourier transformations given in Eq.~(\ref{FTtype1}) also
transform actions of derivatives into products with coordinates and vice
versa, i.~e.%
\begin{align}
\mathcal{F}_{R}^{\text{\hspace{0.01in}}\ast}(\hspace{0.01in}p^{k}\circledast
f(\hspace{0.01in}p))  &  =\text{i}^{-1}\partial_{x}^{k}\,\bar{\triangleright
}\,\mathcal{F}_{R}^{\text{\hspace{0.01in}}\ast}(f(\hspace{0.01in}%
p)),\nonumber\\
\mathcal{F}_{L}^{\text{\hspace{0.01in}}\ast}(f(\hspace{0.01in}p)\circledast
p^{k})  &  =\mathcal{F}_{L}^{\text{\hspace{0.01in}}\ast}(f(\hspace
{0.01in}p))\triangleleft\partial_{x}^{k}\text{i}^{-1}, \label{InfTransFour}%
\end{align}
and%
\begin{align}
\mathcal{F}_{R}^{\text{\hspace{0.01in}}\ast}(\text{i}\partial_{p}%
^{k}\text{\hspace{0.01in}}\triangleright f(\hspace{0.01in}p))  &
=x^{k}\circledast\mathcal{F}_{R}^{\text{\hspace{0.01in}}\ast}(f(\hspace
{0.01in}p)),\nonumber\\
\mathcal{F}_{L}^{\text{\hspace{0.01in}}\ast}(f(\hspace{0.01in}p)\,\bar
{\triangleleft}\,\partial_{p}^{k}\text{i})  &  =\mathcal{F}_{L}^{\text{\hspace
{0.01in}}\ast}(f(\hspace{0.01in}p))\circledast x^{k}. \label{FunPropInv1}%
\end{align}
We show that the identities of Eq.~(\ref{InfTransFour}) can be derived from
those of Eq.~(\ref{FunProp1}):%
\begin{align}
&  & \mathcal{F}_{L}(f(x)\triangleleft\partial_{x}^{k})  &  =\text{i\hspace
{0.01in}}\mathcal{F}_{L}(f(x))\circledast p^{k}\nonumber\\
&  \Rightarrow & f(x)\triangleleft\partial_{x}^{k}  &  =\operatorname*{vol}%
\nolimits_{L}^{-1}\mathcal{F}_{L}^{\text{\hspace{0.01in}}\ast}(\text{i\hspace
{0.01in}}\mathcal{F}_{L}(f(x))\circledast p^{k})(x)\nonumber\\
&  \Rightarrow & \mathcal{F}_{L}^{\text{\hspace{0.01in}}\ast}(f(\hspace
{0.01in}p))\triangleleft\partial_{x}^{k}  &  =\mathcal{F}_{L}^{\text{\hspace
{0.01in}}\ast}(\text{i}f(\hspace{0.01in}p)\circledast p^{k})(x).
\label{HerFouSteAbl}%
\end{align}
In the first step, the Fourier transformation $\mathcal{F}_{L}^{\text{\hspace
{0.01in}}\ast}$ was applied to the first identity of Eq.~(\ref{FunProp1}) and
then we made us of Eq.~(\ref{InvFourAnf1a}). In the second step, $f$ was
replaced by $\mathcal{F}_{L}^{\text{\hspace{0.01in}}\ast}(f(\hspace
{0.01in}p))$ and then Eq.~(\ref{InvFourAnf2a}) was used. Note that the
identities in Eq.~(\ref{FunPropInv1}) can be derived in a similar way, as the
following calculation shows:%
\begin{align}
&  & \mathcal{F}_{L}(f\circledast x^{k})  &  =\text{i\hspace{0.01in}%
}\mathcal{F}_{L}(f)\,\bar{\triangleleft}\,\partial_{p}^{k}\nonumber\\
&  \Rightarrow & f\circledast x^{k}  &  =\operatorname*{vol}\nolimits_{L}%
^{-1}\mathcal{F}_{L}^{\text{\hspace{0.01in}}\ast}(\text{i\hspace{0.01in}%
}\mathcal{F}_{L}(f)\,\bar{\triangleleft}\,\partial_{p}^{k})(x)\nonumber\\
&  \Rightarrow & \mathcal{F}_{L}^{\text{\hspace{0.01in}}\ast}(f(\hspace
{0.01in}p))\circledast x^{k}  &  =\mathcal{F}_{L}^{\text{\hspace{0.01in}}\ast
}(\text{i}f(\hspace{0.01in}p)\,\bar{\triangleleft}\,\partial_{p}^{k})(x).
\label{HerFouSteMulKoo}%
\end{align}

\subsection{Fourier transforms of exponentials and delta
functions\label{KapDelExpFou}}

For some physical applications, the Fourier transforms of exponentials or
delta functions are required. In the following we show that $q$%
-ex\-po\-nen\-tials and $q$-del\-ta functions are transformed into each other
by $q$-de\-formed Fourier transformations.

\begin{figure}
[ptb]
\begin{center}
\includegraphics[width=0.90\textwidth]{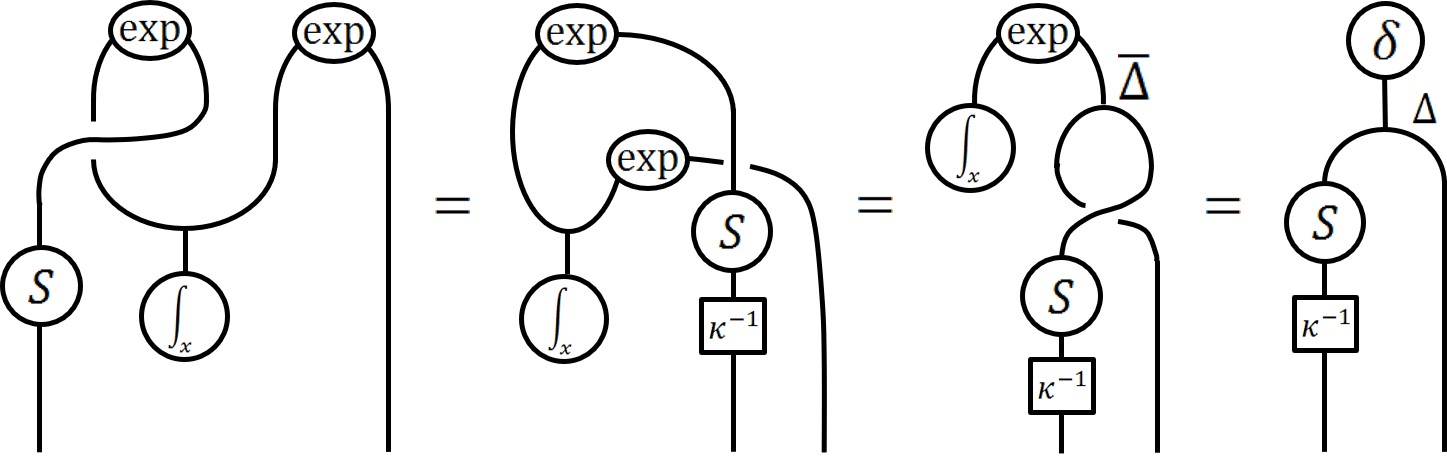}
\caption{Graphical proof of Eq.~(\ref{FTExp1})}%
\label{Fig15}%
\end{center}
\end{figure}

We first calculate $q$-de\-formed Fourier transforms of `dual' $q$%
-ex\-po\-nen\-tials. With the help of graphical methods we get:%
\begin{align}
\mathcal{F}_{L}(\exp_{q}^{\ast}(\text{i}p^{\prime}|x))(\hspace{0.01in}p)  &
=\int_{-\infty}^{+\infty}\text{d}_{q}^{N}\hspace{-0.02in}x\,\exp_{q}^{\ast
}(\text{i}p^{\prime}|\text{\hspace{0.01in}}x)\circledast\exp_{q}%
(x|\text{i}p)\nonumber\\[0.02in]
&  =\delta_{L}^{N}((\ominus\text{\hspace{0.01in}}\kappa^{-1}p^{\prime})\oplus
p),\label{FTExp1}\\[0.16in]
\mathcal{F}_{R}(\exp_{q}^{\ast}(x|\text{i}^{-1}p^{\prime}))(\hspace{0.01in}p)
&  =\int_{-\infty}^{+\infty}\text{d}_{q}^{N}\hspace{-0.02in}x\,\exp
_{q}(\text{i}^{-1}p|x)\circledast\exp_{q}^{\ast}(x|\text{i}^{-1}p^{\prime
})\nonumber\\[0.02in]
&  =\delta_{R}^{N}(\hspace{0.01in}p\oplus(\ominus\text{\hspace{0.01in}}%
\kappa^{-1}p^{\prime})). \label{FTExp1a}%
\end{align}
The graphical proof for Eq.~(\ref{FTExp1}) is shown in Fig.~\ref{Fig15}. We
briefly explain the different steps of this graphical calculation. First, we
pull the left strand past the integral as well as the $q$-exponential on the
right side. By doing so, we take account of the identities shown in
Fig.~\ref{Fig8} of App.~\ref{AppA}\ as well as the trivial braiding properties
of the $q$-ex\-po\-nen\-tials. Now, we can apply the addition theorem for
$q$-ex\-po\-nen\-tials (cf. Fig.~\ref{Fig2} of Chap.~\ref{KapExp}). We then
identify the expression for the $q$-de\-formed delta function and rewrite the
`conjugated' co-pro\-duct into its `unconjugated' counterpart by taking into
account the following identities \cite{Majid:1992sn, Majid:1995hr}%
:\footnote{In our convention the mappings $\Psi$ and $\Psi^{-1}$ are exchanged
in comparison with Ref.~\cite{Majid:1992sn} or Ref.~\cite{Majid:1995hr}.}%
\begin{equation}
\underline{\Delta}=\Psi^{-1}\circ\hspace{0.01in}\underline{\bar{\Delta}%
},\qquad\underline{\bar{\Delta}}=\Psi\circ\underline{\Delta}.
\end{equation}
Similar considerations lead to the following identities:%
\begin{align}
\mathcal{F}_{L}^{\text{\hspace{0.01in}}\ast}(\exp_{q}(\hspace{0.01in}%
y|\text{i}p))(x)  &  =\int_{-\infty}^{+\infty}\text{d}_{q}^{N}\hspace
{-0.02in}p\,\exp_{q}(\hspace{0.01in}y|\text{i}p)\circledast\exp_{q}^{\ast
}(\text{i}p|x)\nonumber\\[0.02in]
&  =\delta_{R}^{N}(\hspace{0.01in}y\oplus(\ominus\text{\hspace{0.01in}}%
\kappa^{-1}x)),\label{FTexpSte2}\\[0.16in]
\mathcal{F}_{R}^{\text{\hspace{0.01in}}\ast}(\exp_{q}(\text{i}^{-1}%
p\hspace{0.01in}|\hspace{0.01in}y))(x)  &  =\int_{-\infty}^{+\infty}%
\text{d}_{q}^{N}\hspace{-0.02in}p\,\exp_{q}^{\ast}(x|\text{i}^{-1}%
p)\circledast\exp_{q}(\text{i}^{-1}p\hspace{0.01in}|\hspace{0.01in}%
y)\nonumber\\[0.02in]
&  =\delta_{L}^{N}((\ominus\text{\hspace{0.01in}}\kappa^{-1}x)\oplus y).
\label{FTexp2}%
\end{align}

From the identities of Eq.~(\ref{FTExp1}) and Eq.~(\ref{FTExp1a}) one can read
off \textit{orthonormality relations} for the $q$-ex\-po\-nen\-tials:%
\begin{align}
\delta_{L}^{N}((\ominus\text{\hspace{0.01in}}\kappa^{-1}p^{\prime})\oplus p)
&  =\int_{-\infty}^{+\infty}\text{d}_{q}^{N}\hspace{-0.02in}x\,\exp_{q}^{\ast
}(\text{i}p^{\prime}|x)\circledast\exp_{q}(x|\text{i}p),\nonumber\\[0.04in]
\delta_{R}^{N}(\hspace{0.01in}p\oplus(\ominus\text{\hspace{0.01in}}\kappa
^{-1}p^{\prime}))  &  =\int_{-\infty}^{+\infty}\text{d}_{q}^{N}\hspace
{-0.02in}x\,\exp_{q}(\text{i}^{-1}p|x)\circledast\exp_{q}^{\ast}%
(x|\text{i}^{-1}p^{\prime}). \label{VolRelImp1V}%
\end{align}
The identities of Eq.~(\ref{FTexpSte2}) and Eq.~(\ref{FTexp2}) can be
interpreted as \textit{completeness relations} for the $q$-ex\-po\-nen\-tials:%
\begin{align}
\delta_{R}^{N}(\hspace{0.01in}y\oplus(\ominus\text{\hspace{0.01in}}\kappa
^{-1}x))  &  =\int_{-\infty}^{+\infty}\text{d}_{q}^{N}\hspace{-0.02in}%
p\,\exp_{q}(\hspace{0.01in}y|\text{i}p)\circledast\exp_{q}^{\ast}%
(\text{i}p|x),\nonumber\\[0.04in]
\delta_{L}^{N}((\ominus\text{\hspace{0.01in}}\kappa^{-1}x)\oplus y)  &
=\int_{-\infty}^{+\infty}\text{d}_{q}^{N}\hspace{-0.02in}p\,\exp_{q}^{\ast
}(x|\text{i}^{-1}p)\circledast\exp_{q}(\text{i}^{-1}p\hspace{0.01in}%
|\hspace{0.01in}y). \label{VolRelExp}%
\end{align}

Next, we calculate Fourier transforms of $q$-del\-ta functions. To this end,
we apply the Fourier transformation $\mathcal{F}_{L}$ or $\mathcal{F}_{R}$ to
the identity of Eq.~(\ref{FTexpSte2}) or Eq.~(\ref{FTexp2}) and take
into\textbf{ }account Eq.~(\ref{InvFourAnf2a}). Thus, we obtain:%
\begin{align}
\mathcal{F}_{L}(\delta_{R}^{N}(\hspace{0.01in}y\oplus(\ominus\text{\hspace
{0.01in}}\kappa^{-1}x)))(\hspace{0.01in}p)  &  =\int_{-\infty}^{+\infty
}\text{d}_{q}^{N}\hspace{-0.02in}x\,\delta_{R}^{N}(\hspace{0.01in}%
y\oplus(\ominus\text{\hspace{0.01in}}\kappa^{-1}x))\circledast\exp
_{q}(x|\text{i}p)\nonumber\\[0.02in]
&  =\operatorname*{vol}\nolimits_{L}\exp_{q}(\hspace{0.01in}y|\text{i}%
p),\label{FouDelt1}\\[0.16in]
\mathcal{F}_{R}(\delta_{L}^{N}((\ominus\text{\hspace{0.01in}}\kappa
^{-1}x)\oplus y))(\hspace{0.01in}p)  &  =\int_{-\infty}^{+\infty}\text{d}%
_{q}^{N}\hspace{-0.02in}x\,\exp_{q}(\text{i}^{-1}p|x)\circledast\delta_{L}%
^{N}((\ominus\text{\hspace{0.01in}}\kappa^{-1}x)\oplus y)\nonumber\\[0.02in]
&  =\operatorname*{vol}\nolimits_{R}\exp_{q}(\text{i}^{-1}p\hspace
{0.01in}|\hspace{0.01in}y).
\end{align}
Similarly, we can apply the Fourier transformation $\mathcal{F}_{L}%
^{\text{\hspace{0.01in}}\ast}$ or $\mathcal{F}_{R}^{\text{\hspace{0.01in}}%
\ast}$ to Eq.~(\ref{FTExp1})\ or Eq.~(\ref{FTExp1a})\ while taking into
account Eq.~(\ref{InvFourAnf1a}):%
\begin{align}
\mathcal{F}_{L}^{\text{\hspace{0.01in}}\ast}(\delta_{L}^{N}((\ominus
\text{\hspace{0.01in}}\kappa^{-1}p^{\prime})\oplus p))(x)  &  =\int_{-\infty
}^{+\infty}\text{d}_{q}^{N}\hspace{-0.02in}p\,\delta_{L}^{N}((\ominus
\text{\hspace{0.01in}}\kappa^{-1}p^{\prime})\oplus p)\circledast\exp_{q}%
^{\ast}(\text{i}p|x)\nonumber\\[0.02in]
&  =\operatorname*{vol}\nolimits_{L}\exp_{q}^{\ast}(\text{i}p^{\prime
}|x),\\[0.16in]
\mathcal{F}_{R}^{\text{\hspace{0.01in}}\ast}(\delta_{R}^{N}(\hspace
{0.01in}p\oplus(\ominus\text{\hspace{0.01in}}\kappa^{-1}p^{\prime})))(x)  &
=\int_{-\infty}^{+\infty}\text{d}_{q}^{N}\hspace{-0.02in}p\,\exp_{q}^{\ast
}(x|\text{i}^{-1}p)\circledast\delta_{L}^{N}(\hspace{0.01in}p\oplus
(\ominus\text{\hspace{0.01in}}\kappa^{-1}p^{\prime}))\nonumber\\[0.02in]
&  =\operatorname*{vol}\nolimits_{R}\exp_{q}^{\ast}(x|\text{i}^{-1}p^{\prime
}).
\end{align}%

\subsection{Characteristic identities of delta functions\label{KapChaIde}}

The $q$-delta functions satisfy some useful identities. In this respect, we
have ($G\in\{L,R\}$):%
\begin{align}
f(x)  &  =\operatorname*{vol}\nolimits_{\hspace{0.01in}G}^{-1}\int
\nolimits_{-\infty}^{+\infty}\text{d}_{q}^{N}\hspace{-0.02in}y\,\delta_{G}%
^{N}((\ominus\hspace{0.01in}\kappa^{-1}x)\oplus y)\circledast f(\hspace
{0.01in}y)\nonumber\\
&  =\operatorname*{vol}\nolimits_{\hspace{0.01in}G}^{-1}\int\nolimits_{-\infty
}^{+\infty}\text{d}_{q}^{N}\hspace{-0.02in}y\,f(\hspace{0.01in}y)\circledast
\delta_{G}^{N}(\hspace{0.01in}y\oplus(\ominus\hspace{0.01in}\kappa^{-1}x)).
\label{AlgChaIdeqDelFkt}%
\end{align}
Similarly, it holds:%
\begin{align}
f(x)  &  =\operatorname*{vol}\nolimits_{\hspace{0.01in}G}^{-1}\int
\nolimits_{-\infty}^{+\infty}\text{d}_{q}^{N}\hspace{-0.02in}y\,f(\hspace
{0.01in}y)\circledast\delta_{G}^{N}((\ominus\hspace{0.01in}\kappa^{-1}y)\oplus
x)\nonumber\\
&  =\operatorname*{vol}\nolimits_{\hspace{0.01in}G}^{-1}\int\nolimits_{-\infty
}^{+\infty}\text{d}_{q}^{N}\hspace{-0.02in}y\,\delta_{G}^{N}(x\oplus
(\ominus\hspace{0.01in}\kappa^{-1}y))\circledast f(\hspace{0.01in}y).
\label{AlgChaIdeqDelFkt1}%
\end{align}
The following calculation serves as an example to show how the above
identities can be proven. For this calculation, we use the completeness
relations in Eq.~(\ref{VolRelExp}) and the invertibility of the $q$-de\-formed
Fourier transformations:%
\begin{align}
&  \operatorname*{vol}\nolimits_{\hspace{0.01in}R}^{-1}\int\nolimits_{-\infty
}^{+\infty}\text{d}_{q}^{N}\hspace{-0.02in}y\,\delta_{G}^{N}((\ominus
\hspace{0.01in}\kappa^{-1}x)\oplus y)\circledast f(\hspace{0.01in}%
y)=\nonumber\\
&  \quad=\operatorname*{vol}\nolimits_{\hspace{0.01in}R}^{-1}\int
\nolimits_{-\infty}^{+\infty}\text{d}_{q}^{N}\hspace{-0.02in}y\,\int_{-\infty
}^{+\infty}\text{d}_{q}^{N}\hspace{-0.02in}p\,\exp_{q}^{\ast}(x|\text{i}%
^{-1}p)\circledast\exp_{q}(\text{i}^{-1}p|y)\circledast f(\hspace
{0.01in}y)\nonumber\\
&  \quad=\operatorname*{vol}\nolimits_{\hspace{0.01in}R}^{-1}\int_{-\infty
}^{+\infty}\text{d}_{q}^{N}\hspace{-0.02in}p\,\exp_{q}^{\ast}(x|\text{i}%
^{-1}p)\circledast\mathcal{F}_{R}(f)(\hspace{0.01in}p)\nonumber\\
&  \quad=\operatorname*{vol}\nolimits_{\hspace{0.01in}R}^{-1}\mathcal{F}%
_{R}^{\hspace{0.01in}\ast}(\mathcal{F}_{R}(f))(x)=f.
\end{align}

We show now that some versions of the $q$-de\-formed delta function are
identical. To this end, we replace the function $f$ in the identities of
Eq.~(\ref{AlgChaIdeqDelFkt}) and Eq.~(\ref{AlgChaIdeqDelFkt1}) by a delta
function ($G\in\{L,R\}$):%
\begin{align}
&  \int\nolimits_{-\infty}^{+\infty}\text{d}_{q}^{N}\hspace{-0.02in}%
z\,\delta_{G}^{N}((\ominus\hspace{0.01in}\kappa^{-1}x)\oplus z)\circledast
\delta_{G}^{N}(z\oplus(\ominus\text{\hspace{0.01in}}\kappa^{-1}y))=\nonumber\\
&  =\operatorname*{vol}\nolimits_{\hspace{0.01in}G}\delta_{G}^{N}%
((\ominus\hspace{0.01in}\kappa^{-1}x)\oplus y)=\operatorname*{vol}%
\nolimits_{G}\delta_{G}^{N}(x\oplus(\ominus\text{\hspace{0.01in}}\kappa
^{-1}y)). \label{HerZusDelFkt}%
\end{align}
This results in the following relation for $q$-de\-formed delta functions:%
\begin{equation}
\delta_{G}^{N}((\ominus\hspace{0.01in}\kappa^{-1}x)\oplus y)=\delta_{G}%
^{N}(x\oplus(\ominus\text{\hspace{0.01in}}\kappa^{-1}y)). \label{ZusDelFkt1}%
\end{equation}
Similar reasonings using Eq.~(\ref{ZusVolLRqLqR}) lead to:%
\begin{equation}
\delta_{R}^{N}((\ominus\hspace{0.01in}\kappa^{-1}x)\oplus y)=\delta_{L}%
^{N}(x\oplus(\ominus\text{\hspace{0.01in}}\kappa^{-1}y)). \label{ZusQDelFktN}%
\end{equation}
Using the above result in combination with Eq.~(\ref{ZusDelFkt1}) leads to:%
\begin{equation}
\delta_{R}^{N}(x\oplus y)=\delta_{L}^{N}(x\oplus y). \label{IdeDelRLTra}%
\end{equation}
If the $y$-co\-or\-di\-na\-tes in the last two identities are set equal to
zero, we can see the following identities:%
\begin{equation}
\delta_{q}^{N}(x)\equiv\delta_{R}^{N}(x)=\delta_{L}^{N}(x).
\label{ZusDelRqLRLq}%
\end{equation}
Let us mention that the identities in Eqns.~(\ref{HerZusDelFkt}%
)-(\ref{IdeDelRLTra}) remain valid if we carry out the following
substitutions:%
\begin{equation}
(\oplus,\ominus)\leftrightarrow(\bar{\oplus},\bar{\ominus}),\text{\qquad
}\kappa\leftrightarrow\kappa^{-1}. \label{ErsKonZPrQTraKap}%
\end{equation}

Finally, we give the conjugation properties of the $q$-de\-formed delta
functions. Eq.~(\ref{KonTraAnt}) of Chap.~\ref{KapTra} and
Eq.~(\ref{KonEigDelFkt}) lead to the following result:%
\begin{equation}
\overline{\delta_{L}^{N}((\ominus\text{\hspace{0.01in}}\kappa^{-1}x)\oplus
y)}=\delta_{R}^{N}(\hspace{0.01in}y\oplus(\ominus\text{\hspace{0.01in}}%
\kappa^{-1}x)). \label{KonEigDelFktTra}%
\end{equation}
\ 

\subsection{Fourier-Plancherel identities\label{KapFourPlaIde}}

Certain $q$-analogs of the so-called Fourier-Plancherel identities apply to
the $q$-de\-formed Fourier transformations. These identities read as%
\begin{equation}
\int\text{d}_{q}^{N}\hspace{-0.02in}x\,f(x)\circledast
g(x)=\operatorname*{vol}\nolimits^{-1}\int\text{d}_{q}^{N}\hspace
{-0.02in}p\,\mathcal{F}_{L}^{\hspace{0.01in}\ast}(f)(p)\circledast
\mathcal{F}_{R}(g)(p) \label{FouPla1}%
\end{equation}
and%
\begin{equation}
\int\text{d}_{q}^{N}\hspace{-0.02in}x\,f(x)\circledast
g(x)=\operatorname*{vol}\nolimits^{-1}\int\text{d}_{q}^{N}\hspace
{-0.02in}p\,\mathcal{F}_{L}(f)(p)\circledast\mathcal{F}_{R}^{\hspace
{0.01in}\ast}(g)(p). \label{FouPla2}%
\end{equation}
In the following, we show how to prove the first identity (similar
considerations hold for the second identity):%
\begin{align}
&  \int\text{d}_{q}^{N}\hspace{-0.02in}p\,\mathcal{F}_{L}^{\hspace{0.01in}%
\ast}(f)(p)\circledast\mathcal{F}_{R}(g)(p)=\nonumber\\
&  \qquad=\int\text{d}_{q}^{N}\hspace{-0.02in}x\int\text{d}_{q}^{N}%
\hspace{-0.02in}y\,f(x)\circledast\int\text{d}_{q}^{N}\hspace{-0.02in}%
p\,\exp_{q}^{\ast}(x|\text{i}^{-1}p)\circledast\exp_{q}(\text{i}%
^{-1}p|y)\circledast g(\hspace{0.01in}y)\nonumber\\
&  \qquad=\int\text{d}_{q}^{N}\hspace{-0.02in}x\,f(x)\circledast\int
\text{d}_{q}^{N}\hspace{-0.02in}y\,\delta_{q}^{N}((\ominus\text{\hspace
{0.01in}}\kappa^{-1}x)\oplus y)\circledast g(\hspace{0.01in}y)\nonumber\\
&  \qquad=\operatorname*{vol}\int\text{d}_{q}^{N}\hspace{-0.02in}%
x\,f(x)\circledast g(x).
\end{align}
In the first step of the above calculation, we used the expressions for the
Fourier transformations $\mathcal{F}_{L}^{\hspace{0.01in}\ast}$ and
$\mathcal{F}_{R}$ [see Eq.(\ref{DefFouKomLR}) and Eq.~(\ref{FTtype1}%
)].\footnote{The expression for $\mathcal{F}_{L}^{\hspace{0.01in}\ast}$
differs slightly from the expression of Eq.~(\ref{FTtype1}). See the
explanations in the paragraph after Eq. (\ref{DelVolLR}).} In the second step,
we have written the momentum integral as a $q$-de\-formed delta function [see
Eq.~(\ref{VolRelExp})]. Using the characteristic identities of the
$q$-de\-formed delta functions [cf. Eq.~(\ref{AlgChaIdeqDelFkt})] is our final step.

\section{Matrix representations on position and momentum
space\label{KapMatRep}}

\subsection{Eigenfunctions of the momentum or position
operator\label{KapDefImpOrtEigFkt}}

The partial derivatives of a quantum space determine a momentum operator in
analogy to the undeformed case. However, there are several ways for partial
derivatives to act on a quantum space [cf. Eq.~(\ref{FundWirkNN}) and
Eq.~(\ref{FundWirk2N}) of Chap.~\ref{ParAblKapAna}], and for this reason, we
also have different $q$-de\-formed \textit{momentum operators}:%
\begin{align}
P^{k}\triangleright f(x)  &  =\text{i}^{-1}\partial^{k}\triangleright f(x), &
f(x)\,\bar{\triangleleft}\,P^{k}  &  =f(x)\,\bar{\triangleleft}\,\partial
^{k}\hspace{0.01in}\text{i}^{-1},\nonumber\\
P^{k}\,\bar{\triangleright}\,f(x)  &  =\text{i}^{-1}\hat{\partial}^{k}%
\,\bar{\triangleright}\,f(x), & f(x)\triangleleft P^{k}  &  =f(x)\triangleleft
\hat{\partial}^{k}\hspace{0.01in}\text{i}^{-1}.
\end{align}

The momentum eigenfunctions describe states with defined momentum, i.~e. plane
waves. So they must be eigenfunctions of the partial derivatives of the
quantum space in question. According to the different actions of partial
derivatives, you have the following ways to formulate \textit{eigenvalue
equations for momentum eigenfunctions}:\footnote{Do not confuse the function
$\bar{u}_{\hspace{0.01in}p}(x)$ with the conjugate of $u_{\hspace{0.01in}%
p}(x)$ [see also Eq.~(\ref{KonEigImp})].}%
\begin{align}
\text{i}^{-1}\partial^{k}\triangleright u_{\hspace{0.01in}p}(x)  &
=u_{\hspace{0.01in}p}(x)\circledast p^{k}, & u^{p}(x)\,\bar{\triangleleft
}\,\partial^{k}\hspace{0.01in}\text{i}^{-1}  &  =p^{k}\circledast
u^{p}(x),\nonumber\\
\text{i}^{-1}\hat{\partial}^{k}\,\bar{\triangleright}\,\bar{u}_{\hspace
{0.01in}p}(x)  &  =\bar{u}_{\hspace{0.01in}p}(x)\circledast p^{k}, & \bar
{u}^{p}(x)\triangleleft\hat{\partial}^{k}\hspace{0.01in}\text{i}^{-1}  &
=p^{k}\circledast\bar{u}^{p}(x). \label{EigGleImpOpeImpEigFkt0}%
\end{align}
We get the results for the eigenfunctions in the second line of
Eq.~(\ref{EigGleImpOpeImpEigFkt0}) from those of the eigenfunctions in the
first line by simple substitutions. For this reason, we do not consider the
eigenfunctions of the second line in the following.

Remember that the quantum space exponentials of Chap.~\ref{KapExp} are
eigenfunctions of partial derivatives of a given quantum space. Consequently,
the $q$-de\-formed \textit{momentum eigenfunctions} take on the following
form:%
\begin{equation}
u_{\hspace{0.01in}p}(x)=\operatorname*{vol}\nolimits^{-1/2}\exp_{q}%
(x|\text{i}p),\qquad u^{p}(x)=\operatorname*{vol}\nolimits^{-1/2}\exp
_{q}(\text{i}^{-1}p\hspace{0.01in}|x). \label{ImpEigFktqDef}%
\end{equation}
We can also introduce `dual'\ momentum eigenfunctions [cf. Eq.~(\ref{DuaExp1})
and Eq.~(\ref{DuaExp2}) of Chap.~\ref{KapExp}]:%
\begin{equation}
(u^{\ast})_{p}(x)=\operatorname*{vol}\nolimits^{-1/2}\exp_{q}^{\ast}%
(\text{i}p|x),\qquad(u^{\ast})^{p}(x)=\operatorname*{vol}\nolimits^{-1/2}%
\exp_{q}^{\ast}(x|\text{i}^{-1}p). \label{DefDuaImpEigFktWdh}%
\end{equation}
These `dual'\ momentum eigenfunctions satisfy the following eigenvalue
equations [cf. also Eq.~(\ref{EigGleExpQueAbl}) of Chap.~\ref{KapExp}]:%
\begin{align}
(u^{\ast})_{p}(x)\triangleleft\partial^{k}\hspace{0.01in}\text{i}^{-1}%
\hspace{-0.01in}  &  =p^{k}\circledast(u^{\ast})_{p}(x),\nonumber\\
\text{i}^{-1}\partial^{k}\,\bar{\triangleright}\,(u^{\ast})^{p}(x)  &
=(u^{\ast})^{p}(x)\circledast p^{k}. \label{ImpEigFktqDef2}%
\end{align}

The components of the \textit{position operator} act on the functions of the
position space as multiplication operators:%
\begin{equation}
X^{k}\triangleright f(x)=x^{k}\circledast f(x),\qquad f(x)\triangleleft
X^{k}=f(x)\circledast x^{k}. \label{DefWirOrtOpe1}%
\end{equation}
The corresponding \textit{position eigenfunctions}\ are determined again by
eigenvalue equations:%
\begin{align}
X^{k}\triangleright u_{\hspace{0.01in}y}(x)  &  =x^{k}\circledast
u_{\hspace{0.01in}y}(x)=u_{\hspace{0.01in}y}(x)\circledast y^{k},\nonumber\\
(u^{\ast})_{y}(x)\triangleleft X^{k}  &  =(u^{\ast})_{y}(x)\circledast
x^{k}=y^{k}\circledast(u^{\ast})_{y}(x). \label{DefGleOrtEigFkt}%
\end{align}
The $q$-de\-formed delta functions satisfy the eigenvalue equations above:%
\begin{align}
u_{\hspace{0.01in}y}(x)  &  =\operatorname*{vol}\nolimits^{-1}\hspace
{-0.01in}\delta_{q}^{N}(x\oplus(\ominus\hspace{0.01in}\kappa^{-1}%
y))=\operatorname*{vol}\nolimits^{-1}\hspace{-0.01in}\delta_{q}^{N}%
((\ominus\hspace{0.01in}\kappa^{-1}x)\oplus y)=(u^{\ast})^{y}(x),\nonumber\\
(u^{\ast})_{y}(x)  &  =\operatorname*{vol}\nolimits^{-1}\hspace{-0.01in}%
\delta_{q}^{N}(\hspace{0.01in}y\oplus(\ominus\hspace{0.01in}\kappa
^{-1}x))=\operatorname*{vol}\nolimits^{-1}\hspace{-0.01in}\delta_{q}%
^{N}((\ominus\hspace{0.01in}\kappa^{-1}y)\oplus x)=u^{y}(x).
\label{DefEigPos1}%
\end{align}
To prove this, we calculate e.~g.%
\begin{align}
\int\text{d}_{q}^{N}\hspace{-0.02in}x\,(u^{\ast})_{y}(x)\circledast
x^{k}\circledast f(x)  &  =\int\text{d}_{q}^{N}\hspace{-0.02in}x\hspace
{0.01in}\operatorname*{vol}\nolimits^{-1}\hspace{-0.01in}\delta_{q}%
^{N}((\ominus\hspace{0.01in}\kappa^{-1}y)\oplus x)\circledast x^{k}\circledast
f(x)\nonumber\\
&  =y^{k}\circledast f(\hspace{0.01in}y) \label{HerOrtGle1}%
\end{align}
and%
\begin{align}
y^{k}\circledast f(\hspace{0.01in}y)  &  =\operatorname*{vol}\nolimits^{-1}%
\hspace{-0.01in}y^{k}\circledast\int\text{d}_{q}^{N}\hspace{-0.02in}%
x\,\delta_{q}^{N}((\ominus\hspace{0.01in}\kappa^{-1}y)\oplus x)\circledast
f(x)\nonumber\\
&  =\int\text{d}_{q}^{N}\hspace{-0.02in}x\,y^{k}\circledast(u^{\ast}%
)_{y}(x)\circledast f(x). \label{HerOrtGle2}%
\end{align}
Note that in both calculations, we have made use of the characteristic
identities of the delta functions [cf. Eq.~(\ref{AlgChaIdeqDelFkt}) and
Eq.~(\ref{AlgChaIdeqDelFkt1}) of Chap.~\ref{KapChaIde}]. Comparing the results
of Eq.~(\ref{HerOrtGle1}) and Eq.~(\ref{HerOrtGle2}) shows:%
\begin{equation}
\int\text{d}_{q}^{N}\hspace{-0.02in}x\hspace{0.01in}\big ((u^{\ast}%
)_{y}(x)\circledast x^{k}-y^{k}\circledast(u^{\ast})_{y}(x)\big )\circledast
f(x)=0.
\end{equation}
If we substitute $f(x)$ by a $q$-de\-formed delta function and take the
characteristic identities of the delta functions into account again, we
finally get the eigenvalue equation for $(u^{\ast})_{y}$:%
\begin{align}
0  &  =\int\text{d}_{q}^{N}\hspace{-0.02in}x^{\prime}\hspace{0.01in}%
\big ((u^{\ast})_{y}(x^{\prime})\circledast x^{\prime k}-\hspace{0.01in}%
y^{k}\circledast(u^{\ast})_{y}(x^{\prime})\big )\circledast\delta_{q}%
^{N}(x^{\prime}\oplus(\ominus\hspace{0.01in}\kappa^{-1}x))\operatorname*{vol}%
\nolimits^{-1}\nonumber\\
&  =(u^{\ast})_{y}(x)\circledast x^{k}-\hspace{0.01in}y^{k}\circledast
(u^{\ast})_{y}(x).
\end{align}

For the sake of completeness, we must not forget that the position
eigenfunctions satisfy the following identities:%
\begin{align}
\partial_{x}^{k}\triangleright u_{\hspace{0.01in}y}(x)  &  =u_{\hspace
{0.01in}y}(x)\triangleleft\partial_{y}^{k},\nonumber\\
(u^{\ast})_{y}(x)\,\bar{\triangleleft}\,\partial_{x}^{k}  &  =\partial_{y}%
^{k}\,\bar{\triangleright}\,(u^{\ast})_{y}(x). \label{OrtEigFktAblIde}%
\end{align}
The following calculation serves as an example for proving the above
identities:%
\begin{align}
\partial_{x}^{k}\triangleright\delta_{q}^{N}(x\oplus(\ominus\hspace
{0.01in}\kappa^{-1}y))  &  =\int_{-\infty}^{+\infty}\text{d}_{q}^{N}%
\hspace{-0.02in}p\,\partial_{x}^{k}\triangleright\exp_{q}(x|\text{i}%
p)\circledast\exp_{q}^{\ast}(\text{i}p|y)=\nonumber\\
&  =\text{i}\int_{-\infty}^{+\infty}\text{d}_{q}^{N}\hspace{-0.02in}%
p\,\exp_{q}(x|\text{i}p)\circledast p^{k}\circledast\exp_{q}^{\ast}%
(\text{i}p|y)=\nonumber\\
&  =\int_{-\infty}^{+\infty}\text{d}_{q}^{N}\hspace{-0.02in}p\,\exp
_{q}(x|\text{i}p)\circledast\exp_{q}^{\ast}(\text{i}p|y)\triangleleft
\partial_{y}^{k}=\nonumber\\
&  =\delta_{q}^{N}(x\oplus(\ominus\hspace{0.01in}\kappa^{-1}y))\triangleleft
\partial_{y}^{k}.
\end{align}
The first step uses the completeness relations for $q$-de\-formed exponentials
[cf. Eq.~(\ref{VolRelExp}) of Chap.~\ref{KapDelExpFou}]. The second step
follows from the eigenvalue equations of the $q$-exponentials [cf.
Eq.~(\ref{EigGl1N}) and Eq.~(\ref{EigGleExpQueAbl}) of Chap.~\ref{KapExp}].
The last step again is a consequence of the completeness relations for
$q$-de\-formed exponentials.

The momentum eigenfunctions in Eq.~(\ref{ImpEigFktqDef}) refer to a
representation of the \textit{momentum operator in position space}.
Correspondingly, the position eigenfunctions in Eq.~(\ref{DefEigPos1}) refer
to a representation of the \textit{position operator in position space}. To a
large extent, position space and momentum space can be treated in the same
way. For this reason, we can replace the momentum variables by position
variables and vice versa in Eq.~(\ref{ImpEigFktqDef}) and
Eq.~(\ref{DefEigPos1}). This way, we obtain position or momentum
eigenfunctions referring to a representation of the \textit{position or
momentum operator in momentum space}.

We know that $q$-ex\-ponential functions and $q$-del\-ta functions transform
into each other by $q$-de\-formed Fourier transformations. This correspondence
is carried over to the momentum and position eigenfunctions. Thus, we have
[cf. Eq.~(\ref{FTExp1a}) and Eq.~(\ref{FTexpSte2}) of Chap.~\ref{KapDelExpFou}%
]%
\begin{equation}
\operatorname*{vol}\nolimits^{-1/2}\mathcal{F}_{R}((u^{\ast})_{p}%
(\hspace{0.01in}y))(x)=u_{\hspace{0.01in}y}(x)
\end{equation}
and%
\begin{equation}
\operatorname*{vol}\nolimits^{-1/2}\mathcal{F}_{L}^{\hspace{0.01in}\ast
}(u_{\hspace{0.01in}p}(\hspace{0.01in}y))(x)=(u^{\ast})_{y}(x).
\end{equation}

Finally, we write down how momentum and position eigenfunctions behave under
conjugation. Using Eq.~(\ref{KonEigExpQua}) from Chap.~\ref{KapExp}\ and
Eq.~(\ref{KonEigVolEleHer}) from Chap.~\ref{KapFTDef}\ follows for the
momentum eigenfunctions:%
\begin{equation}
\overline{u_{\hspace{0.01in}p}(x)}=u^{p}(x),\qquad\overline{(u^{\ast})_{p}%
(x)}=(u^{\ast})^{p}(x). \label{KonEigImp}%
\end{equation}
Due to Eq.~(\ref{KonEigDelFktTra}) of Chap.~\ref{KapChaIde} the position
eigenfunctions are subject to:%
\begin{equation}
\overline{u_{\hspace{0.01in}y}(x)}=u^{y}(x),\qquad\overline{(u^{\ast})_{y}%
(x)}=(u^{\ast})^{y}(x).
\end{equation}

\subsection{Completeness of momentum eigenfunctions \label{KapVolRelImpOrt}}

In this section, we show that the momentum eigenfunctions constitute a
complete system of functions. To this end, we remember that for a given wave
function different expansions in terms of $q$-de\-formed momentum
eigenfunctions exist. These expansions follow from the definitions of the
$q$-de\-formed Fourier transformations [cf. Eq.~(\ref{DefFouKomLR}) and
Eq.~(\ref{FTtype1}) of Chap.~\ref{KapFTDef}] if we write the $q$%
-ex\-po\-nentials as momentum eigenfunctions, i.~e.%
\begin{align}
\psi_{R}(x)  &  =\int\text{d}_{q}^{N}\hspace{-0.02in}p\,u_{\hspace{0.01in}%
p}(x)\circledast c_{\hspace{0.01in}p},\nonumber\\
\psi_{L}(x)  &  =\int\text{d}_{q}^{N}\hspace{-0.02in}p\,c^{\hspace{0.01in}%
p}\circledast u^{p}(x) \label{EntIm2}%
\end{align}
and%
\begin{align}
\psi_{R}^{\ast}(x)  &  =\int\text{d}_{q}^{N}\hspace{-0.02in}p\,(u^{\ast}%
)^{p}(x)\circledast(c^{\ast})^{p},\nonumber\\
\psi_{L}^{\ast}(x)  &  =\int\text{d}_{q}^{N}\hspace{-0.02in}p\,(c^{\ast}%
)_{p}\circledast(u^{\ast})_{p}(x). \label{EntImp1}%
\end{align}

It remains to determine the expansion coefficients. We start with the
coefficients $c_{p}$ for the wave function $\psi_{R}$. Using the last identity
in Eq.~(\ref{InvFourAnf2a}) of Chap.~\ref{KapFourInv} we can write $\psi
_{R}(x)$ as follows:%
\begin{align}
\psi_{R}(x)  &  =\operatorname*{vol}\nolimits^{-1}\hspace{-0.01in}%
\mathcal{F}_{R}(\mathcal{F}_{R}^{\hspace{0.01in}\ast}(\psi_{R}))(x)\nonumber\\
&  =\operatorname*{vol}\nolimits^{-1/2}\hspace{-0.01in}\int\text{d}_{q}%
^{N}\hspace{-0.02in}p\,u_{\hspace{0.01in}p}(x)\circledast\mathcal{F}%
_{R}^{\hspace{0.01in}\ast}(\psi_{R})(\hspace{0.01in}p). \label{BerFouKoeWel}%
\end{align}
If we compare the last expression in Eq.~(\ref{BerFouKoeWel}) with the
corresponding expansion in Eq.~(\ref{EntIm2}), we read off:%
\begin{equation}
c_{\hspace{0.01in}p}=\operatorname*{vol}\nolimits^{-1/2}\mathcal{F}%
_{R}^{\hspace{0.01in}\ast}(\psi_{R})(\hspace{0.01in}p)\hspace{0.01in}%
=\int\text{d}_{q}^{N}\hspace{-0.02in}x\,(u^{\ast})_{p}(x)\circledast\psi
_{R}(x). \label{CPUnt}%
\end{equation}
In the same way, we can show:%
\begin{equation}
c^{\hspace{0.01in}p}=\operatorname*{vol}\nolimits^{-1/2}\mathcal{F}%
_{L}^{\hspace{0.01in}\ast}(\psi_{L})(\hspace{0.01in}p)\hspace{0.01in}%
=\int\text{d}_{q}^{N}\hspace{-0.02in}x\,\psi_{L}(x)\circledast(u^{\ast}%
)^{p}(x). \label{CPUnt2}%
\end{equation}

Similar reasonings enable us to determine the coefficients $(c^{\ast})_{p}$
for the wave function $\psi_{L}^{\ast}$. Using the first identity in
Eq.~(\ref{InvFourAnf1a}) of Chap.~\ref{KapFourInv} follows:%
\begin{align}
\psi_{L}^{\ast}(x)  &  =\operatorname*{vol}\nolimits^{-1}\hspace
{-0.01in}\mathcal{F}_{L}^{\hspace{0.01in}\ast}(\mathcal{F}_{L}(\psi_{L}^{\ast
}))(x)\nonumber\\
&  =\operatorname*{vol}\nolimits^{-1/2}\hspace{-0.01in}\int\text{d}_{q}%
^{N}\hspace{-0.02in}p\,\mathcal{F}_{L}(\psi_{L}^{\ast})(\hspace{0.01in}%
p)\circledast(u^{\ast})_{p}(x).
\end{align}
From this result, we read off:%
\begin{equation}
(c^{\ast})_{p}=\operatorname*{vol}\nolimits^{-1/2}\mathcal{F}_{L}(\psi
_{L}^{\ast})(\hspace{0.01in}p)\hspace{0.01in}=\int\text{d}_{q}^{N}%
\hspace{-0.02in}x\,\psi_{L}^{\ast}(x)\circledast u_{\hspace{0.01in}p}(x).
\label{DPUnt}%
\end{equation}
Similarly, we get:%
\begin{equation}
(c^{\ast})^{p}=\operatorname*{vol}\nolimits^{-1/2}\mathcal{F}_{R}(\psi
_{R}^{\ast})(\hspace{0.01in}p)\hspace{0.01in}=\int\text{d}_{q}^{N}%
\hspace{-0.02in}x\,u^{p}(x)\circledast\psi_{R}^{\ast}(x). \label{DPUnt2}%
\end{equation}

Now, we can write down \textit{completeness relations} for $q$-de\-formed
momentum eigenfunctions. To this end, we perform the following calculation
[also see Eq.~(\ref{BerFouKoeWel})]:%
\begin{align}
\psi_{R}(x)  &  =\operatorname*{vol}\nolimits^{-1}\hspace{-0.01in}%
\mathcal{F}_{R}(\mathcal{F}_{R}^{\hspace{0.01in}\ast}(\psi_{R}))(x)\nonumber\\
&  =\int\text{d}_{q}^{N}\hspace{-0.02in}p\,u_{\hspace{0.01in}p}(x)\circledast
\int\text{d}_{q}^{N}\hspace{-0.02in}y\,(u^{\ast})_{p}(\hspace{0.01in}%
y)\circledast\psi_{R}(\hspace{0.01in}y)\nonumber\\
&  =\int\text{d}_{q}^{N}\hspace{-0.02in}y\left(  \int\text{d}_{q}^{N}%
\hspace{-0.02in}p\,u_{\hspace{0.01in}p}(x)\circledast(u^{\ast})_{p}%
(\hspace{0.01in}y)\right)  \circledast\psi_{R}(\hspace{0.01in}y).
\label{UmfWelFktVolRel}%
\end{align}
If you compare the result of the above calculation with the characteristic
identities of the $q$-del\-ta functions [cf. Eq.~(\ref{AlgChaIdeqDelFkt1}) of
Chap.~\ref{KapChaIde}], you can see:%
\begin{equation}
\int\text{d}_{q}^{N}\hspace{-0.02in}p\,u_{\hspace{0.01in}p}(x)\circledast
(u^{\ast})_{p}(\hspace{0.01in}y)=\operatorname*{vol}\nolimits^{-1}%
\hspace{-0.01in}\delta_{q}^{N}(x\oplus(\ominus\hspace{0.01in}\kappa^{-1}y)).
\label{VolRelImpEigFkt1}%
\end{equation}
Similar considerations lead to:%
\begin{equation}
\int\text{d}_{q}^{N}\hspace{-0.02in}p\,(u^{\ast})^{p}(\hspace{0.01in}%
y)\circledast u^{p}(x)=\operatorname*{vol}\nolimits^{-1}\hspace{-0.01in}%
\delta_{q}^{N}((\ominus\hspace{0.01in}\kappa^{-1}y)\oplus x).
\label{VolRelImpEigFkt2n}%
\end{equation}

\subsection{Completeness of position eigenfunctions\label{KapVolRelOrtEigFkt}}

We can develop each wave function in position space in terms of the position
eigenfunctions. We have introduced them in Eq.~(\ref{DefEigPos1}) of
Chap.~\ref{KapDefImpOrtEigFkt}. In this respect the following expansions
apply:%
\begin{align}
\psi_{R}(x)  &  =\int\text{d}_{q}^{N}\hspace{-0.02in}y\,u_{\hspace{0.01in}%
y}(x)\circledast c_{\hspace{0.01in}y}=\int\text{d}_{q}^{N}\hspace
{-0.02in}y\,(u^{\ast})^{y}(x)\circledast(c^{\ast})^{y}=\psi_{R}^{\ast
}(x),\nonumber\\
\psi_{L}(x)  &  =\int\text{d}_{q}^{N}\hspace{-0.02in}y\,c^{\hspace{0.01in}%
y}\circledast u^{y}(x)=\int\text{d}_{q}^{N}\hspace{-0.02in}y\,(c^{\ast}%
)_{y}\circledast(u^{\ast})_{y}(x)=\psi_{L}^{\ast}(x). \label{EntOrtEigFkt}%
\end{align}
The above expansions are a direct consequence of the characteristic identities
of the $q$-del\-ta functions [cf. Eq.~(\ref{AlgChaIdeqDelFkt}) and
Eq.~(\ref{AlgChaIdeqDelFkt1}) of Chap.~\ref{KapChaIde}]. They also result from
the fact that the position eigenfunctions are nothing else but $q$-del\-ta functions.

The expansion coefficients in Eq.~(\ref{EntOrtEigFkt}) are identical to the
corresponding wave functions. To see this, let us take another look at the
characteristic identities of the $q$-del\-ta functions:%
\begin{align}
\psi_{R}(x)  &  =\operatorname*{vol}\nolimits^{-1}\hspace{-0.02in}\int
\text{d}_{q}^{N}\hspace{-0.02in}y\,\delta_{q}^{N}((\ominus\hspace
{0.01in}\kappa^{-1}x)\oplus y)\circledast\psi_{R}(\hspace{0.01in}%
y),\nonumber\\
\psi_{L}(x)  &  =\operatorname*{vol}\nolimits^{-1}\hspace{-0.02in}\int
\text{d}_{q}^{N}\hspace{-0.02in}y\,\psi_{L}(\hspace{0.01in}y)\circledast
\delta_{q}^{N}(\hspace{0.01in}y\oplus(\ominus\hspace{0.01in}\kappa^{-1}x)).
\label{BesEntKoeOrtEig}%
\end{align}
If you compare these identities with the expansions in Eq.~(\ref{EntOrtEigFkt}%
) and take into account the expressions for the position eigenfunctions in
Eq.~(\ref{DefEigPos1}) then you will recognize:%
\begin{align}
c_{\hspace{0.01in}y}  &  =\psi_{R}(\hspace{0.01in}y)=\psi_{R}^{\ast}%
(\hspace{0.01in}y)=(c^{\ast})^{y},\nonumber\\
c^{\hspace{0.01in}y}  &  =\psi_{L}(\hspace{0.01in}y)=\psi_{L}^{\ast}%
(\hspace{0.01in}y)=(c^{\ast})_{y}. \label{EntKoeOrtEigFkt}%
\end{align}
Note that due to the above results, we can write the expansion coefficients as
follows:%
\begin{align}
c_{\hspace{0.01in}y}  &  =\int\text{d}_{q}^{N}\hspace{-0.02in}x\,(u^{\ast
})_{y}(x)\circledast\psi_{R}(x)\hspace{0.01in}=\int\text{d}_{q}^{N}%
\hspace{-0.02in}x\,u^{y}(x)\circledast\psi_{R}^{\ast}(x)=(c^{\ast}%
)^{y},\nonumber\\
c^{\hspace{0.01in}y}  &  =\int\text{d}_{q}^{N}\hspace{-0.02in}x\,\psi
_{L}(x)\circledast(u^{\ast})^{y}(x)\hspace{0.01in}=\int\text{d}_{q}^{N}%
\hspace{-0.02in}x\,\psi_{L}^{\ast}(x)\circledast u_{\hspace{0.01in}%
y}(x)=(c^{\ast})_{y}. \label{EntWicKoeIntDar}%
\end{align}

The position eigenfunctions fulfill certain \textit{completeness relations},
which we obtain by the following calculation:%
\begin{align}
\psi_{L}(x)  &  =\int\text{d}_{q}^{N}\hspace{-0.02in}y\,c^{\hspace{0.01in}%
y}\circledast u^{y}(x)=\int\text{d}_{q}^{N}\hspace{-0.02in}y\int\text{d}%
_{q}^{N}\hspace{-0.02in}x^{\prime}\,\psi_{L}(x^{\prime})\circledast(u^{\ast
})^{y}(x^{\prime})\circledast u^{y}(x)\nonumber\\
&  =\operatorname*{vol}\nolimits^{-2}\hspace{-0.02in}\int\text{d}_{q}%
^{N}\hspace{-0.02in}y\int\text{d}_{q}^{N}\hspace{-0.02in}x^{\prime}\,\psi
_{L}(x^{\prime})\circledast\delta_{q}^{N}(x^{\prime}\oplus(\ominus
\hspace{0.01in}\kappa^{-1}y))\circledast\delta_{q}^{N}(\hspace{0.01in}%
y\oplus(\ominus\hspace{0.01in}\kappa^{-1}x))\nonumber\\
&  =\operatorname*{vol}\nolimits^{-1}\hspace{-0.02in}\int\text{d}_{q}%
^{N}\hspace{-0.02in}x^{\prime}\,\psi_{L}(x^{\prime})\circledast\delta_{q}%
^{N}(x^{\prime}\oplus(\ominus\hspace{0.01in}\kappa^{-1}x)).
\label{BerVolRelOrtEigFkt}%
\end{align}
If in the above calculation, we compare the third with the last expression, we
find the wanted relations:%
\begin{align}
\operatorname*{vol}\nolimits^{-1}\hspace{-0.01in}\delta_{q}^{N}(x^{\prime
}\oplus(\ominus\hspace{0.01in}\kappa^{-1}x))  &  =\int\text{d}_{q}^{N}%
\hspace{-0.02in}y\,(u^{\ast})^{y}(x^{\prime})\circledast u^{y}(x)\nonumber\\
&  =\int\text{d}_{q}^{N}\hspace{-0.02in}y\,u_{\hspace{0.01in}y}(x^{\prime
})\circledast(u^{\ast})_{y}(x). \label{VolRelOrtEigFkt}%
\end{align}

\subsection{Orthogonality of momentum or position
eigenfunctions\label{KapOrtImOrtEigFkgt}}

We know from the last section that we can expand a wave function in terms of
$q$-deformed momentum eigenfunctions. For this reason, the $q$-de\-formed
momentum eigenfunctions fulfil certain completeness relations [cf.
Eq.~(\ref{VolRelImpEigFkt1}) and Eq.~(\ref{VolRelImpEigFkt2n}) of
Chap.~\ref{KapVolRelImpOrt}]. From an algebraic point of view, the
$q$-de\-formed position and momentum coordinates behave in the same way.
Therefore, we can replace the momentum coordinates in the completeness
relations of the $q$-de\-formed momentum eigenfunctions by position
coordinates and vice versa. In doing so, we can obtain orthonormality
relations for the $q$-de\-formed momentum eigenfunctions from their
completeness relations. We demonstrate this procedure with an example. We
begin our considerations with the following completeness relation [cf.
Eq.~(\ref{VolRelImpEigFkt1}) of Chap.~\ref{KapVolRelImpOrt}]:%
\begin{align}
\int\text{d}_{q}^{N}\hspace{-0.02in}p\,u_{\hspace{0.01in}p}(x)\circledast
(u^{\ast})_{p}(\hspace{0.01in}y)  &  =\operatorname*{vol}\nolimits^{-1}%
\hspace{-0.02in}\int\text{d}_{q}^{N}\hspace{-0.02in}p\,\exp_{q}(x|\text{i}%
p)\circledast\exp_{q}^{\ast}(\text{i}p|y)\nonumber\\
&  =\operatorname*{vol}\nolimits^{-1}\hspace{-0.01in}\delta_{q}^{N}%
(x\oplus(\ominus\hspace{0.01in}\kappa^{-1}y)). \label{VolRelEbeWelDarStel1}%
\end{align}
The exchange of position coordinates and momentum coordinates leads to the
wanted orthonormality relation:%
\begin{align}
\int\text{d}_{q}^{N}\hspace{-0.02in}x\,u^{p}(x)\circledast(u^{\ast
})^{p^{\prime}}(x)  &  =\operatorname*{vol}\nolimits^{-1}\hspace{-0.02in}%
\int\text{d}_{q}^{N}\hspace{-0.02in}x\,\exp_{q}(\text{i}^{-1}p|x)\circledast
\exp_{q}^{\ast}(x|\text{i}^{-1}p^{\prime})\nonumber\\
&  =\operatorname*{vol}\nolimits^{-1}\hspace{-0.01in}\delta_{q}^{N}%
(\hspace{0.01in}p\oplus(\ominus\hspace{0.01in}\kappa^{-1}p^{\prime})).
\label{OrtRelImpEigDarKap}%
\end{align}
In the same way, we obtain the following orthonormality relation:%
\begin{align}
\int\text{d}_{q}^{N}\hspace{-0.02in}x\,(u^{\ast})_{p}(x)\circledast
u_{\hspace{0.01in}p^{\prime}}(x)  &  =\operatorname*{vol}\nolimits^{-1}%
\hspace{-0.02in}\int\text{d}_{q}^{N}\hspace{-0.02in}x\,\exp_{q}^{\ast
}(\text{i}p|x)\circledast\exp_{q}(x|\text{i}p^{\prime})\nonumber\\
&  =\operatorname*{vol}\nolimits^{-1}\hspace{-0.01in}\delta_{q}^{N}%
((\ominus\hspace{0.01in}\kappa^{-1}p)\oplus p^{\prime}).
\label{OrtRelImpEigDarKap2}%
\end{align}

Using a similar procedure, we find orthonormality relations for the
$q$-de\-formed position eigenfunctions. For this purpose, we consider the
completeness relations for the position eigenfunctions [cf.
Eq.~(\ref{VolRelOrtEigFkt}) from the last chapter]:%
\begin{align}
\int\text{d}_{q}^{N}\hspace{-0.02in}y\,u_{\hspace{0.01in}y}(x)\circledast
(u^{\ast})_{y}(x^{\prime})  &  =\int\text{d}_{q}^{N}\hspace{-0.02in}%
y\,(u^{\ast})^{y}(x)\circledast u^{y}(x^{\prime})\nonumber\\
&  =\operatorname*{vol}\nolimits^{-2}\hspace{-0.02in}\int\text{d}_{p}%
^{N}\hspace{-0.02in}y\,\delta_{q}^{N}(x\oplus(\ominus\hspace{0.01in}%
\kappa^{-1}y))\circledast\delta_{q}^{N}((\ominus\hspace{0.01in}\kappa
^{-1}y)\oplus x^{\prime})\nonumber\\
&  =\operatorname*{vol}\nolimits^{-1}\hspace{-0.01in}\delta_{q}^{N}%
(x\oplus(\ominus\hspace{0.01in}\kappa^{-1}x^{\prime})).
\end{align}
If we exchange the $x$- and $y$-co\-or\-di\-nates, we get the following
result:%
\begin{align}
\int\text{d}_{q}^{N}\hspace{-0.02in}x\,u^{y}(x)\circledast(u^{\ast
})^{y^{\prime}}(x)  &  =\int\text{d}_{q}^{N}\hspace{-0.02in}x\,(u^{\ast}%
)_{y}(x)\circledast u_{\hspace{0.01in}y^{\prime}}(x)\nonumber\\
&  =\operatorname*{vol}\nolimits^{-2}\hspace{-0.02in}\int\text{d}_{q}%
^{N}\hspace{-0.02in}x\,\delta_{q}^{N}(\hspace{0.01in}y\oplus(\ominus
\hspace{0.01in}\kappa^{-1}x))\circledast\delta_{q}^{N}((\ominus\hspace
{0.01in}\kappa^{-1}x)\oplus y^{\prime})\nonumber\\
&  =\operatorname*{vol}\nolimits^{-1}\hspace{-0.01in}\delta_{q}^{N}%
(\hspace{0.01in}y\oplus(\ominus\hspace{0.01in}\kappa^{-1}y^{\prime})).
\label{OrtNorBedOrtEig2}%
\end{align}

\subsection{Matrix representations on momentum space\label{KapMatDarImp}}

The expansion coefficients in Eq.~(\ref{CPUnt}) or Eq.~(\ref{CPUnt2}) of
Chap.~\ref{KapVolRelImpOrt} form a vector of infinite dimension, with the
momentum variable playing the role of an index to distinguish the components
of the vector. From this point of view, the expansion coefficients under
consideration can describe a physical state with respect to a basis of
$q$-de\-formed momentum eigenfunctions. The same applies to the expansion
coefficients in Eq.~(\ref{DPUnt}) or Eq.~(\ref{DPUnt2}) of
Chap.~\ref{KapVolRelImpOrt}.

We now continue these considerations by determining matrix representations of
the momentum or position operator with respect to a basis of $q$-de\-formed
momentum eigenfunctions. There are different versions of these matrix
representations since there are also different expansion coefficients for the
same wave function. One way to represent the momentum operator in a momentum
basis is [also see Eq.~(\ref{FunPropInv1}) of Chap.~\ref{KapFouActPro}]:%
\begin{align}
(P^{A})_{p^{\prime}p}  &  =\operatorname*{vol}\nolimits^{-1/2}\mathcal{F}%
_{R}^{\hspace{0.01in}\ast}(\text{i}^{-1}\partial_{x}^{A}\triangleright
u_{\hspace{0.01in}p})(\hspace{0.01in}p^{\prime})\nonumber\\[0.02in]
&  =\int\text{d}_{q}^{3}x\,(u^{\ast})_{p^{\prime}}(x)\circledast\lbrack
\hspace{0.01in}\text{i}^{-1}\partial_{x}^{A}\triangleright u_{p}%
(x)]\nonumber\\
&  =\int\text{d}_{q}^{3}x\,(u^{\ast})_{p^{\prime}}(x)\circledast
u_{\hspace{0.01in}p}(x)\circledast p^{A}\nonumber\\[0.02in]
&  =\operatorname*{vol}\nolimits^{-1}\hspace{-0.01in}\delta_{q}^{3}%
((\ominus\hspace{0.01in}\kappa^{-1}p^{\prime})\oplus p)\circledast p^{A}.
\label{MatEleImpOpe}%
\end{align}
In the penultimate step of the above calculation, we have used the eigenvalue
equation of the momentum operator. The last identity follows from the
orthonormality relation in Eq.~(\ref{OrtRelImpEigDarKap2}) of the previous
chapter. A similar proceeding provides us with another way of calculating
matrix elements of the $q$-de\-formed momentum operator:%
\begin{align}
(P^{A})^{p\hspace{0.01in}p^{\prime}}  &  =\operatorname*{vol}\nolimits^{-1/2}%
\mathcal{F}_{L}^{\hspace{0.01in}\ast}(u^{p}\,\bar{\triangleleft}\,\partial
_{x}^{A}\text{i}^{-1})(\hspace{0.01in}p^{\prime})\nonumber\\[0.02in]
&  =\int\text{d}_{q}^{3}x\,[\hspace{0.01in}u^{p}(x)\,\bar{\triangleleft
}\,\partial_{x}^{A}\text{i}^{-1}]\circledast(u^{\ast})^{p^{\prime}%
}(x)\nonumber\\
&  =\int\text{d}_{q}^{3}x\,p^{A}\circledast u^{p}(x)\circledast(u^{\ast
})^{p^{\prime}}(x)\nonumber\\[0.02in]
&  =\operatorname*{vol}\nolimits^{-1}p^{A}\circledast\delta_{q}^{3}%
(\hspace{0.01in}p\oplus(\ominus\hspace{0.01in}\kappa^{-1}p^{\prime})).
\label{MatEleImpOpe2}%
\end{align}

Next, we show that the action of the $q$-de\-formed momentum operator in
momentum space is nothing else but multiplication by a momentum variable. To
this end, we replace the $q$-de\-formed plane wave $u_{\hspace{0.01in}p}(x)$
in the second expression of Eq.~(\ref{MatEleImpOpe}) by a general wave
function. In doing so, we obtain in analogy to the result from
Eq.~(\ref{FunPropInv1}) of Chap.~\ref{KapFouActPro}:%
\begin{align}
(P^{A}\triangleright\psi_{R})_{p}  &  =\operatorname*{vol}\nolimits^{-1/2}%
\hspace{-0.01in}\mathcal{F}_{R}^{\ast}(\text{i}^{-1}\partial_{x}%
^{A}\triangleright\psi_{R})(\hspace{0.01in}p)\nonumber\\
&  =\operatorname*{vol}\nolimits^{-1/2}p^{A}\circledast\mathcal{F}_{R}^{\ast
}(\psi)(\hspace{0.01in}p)=p^{A}\circledast c_{\hspace{0.01in}p}.
\label{MatMulImpOpe1}%
\end{align}
In the last step of the above calculation, we could identify the expansion
coefficient $c_{\hspace{0.01in}p}$ by its expression in Eq.~(\ref{CPUnt}) of
Chap.~\ref{KapVolRelImpOrt}. Due to the characteristic identities of the
$q$-de\-formed delta functions, we can also write the above result as a kind
of matrix multiplication [cf. Eq.~(\ref{AlgChaIdeqDelFkt}) and
Eq.~(\ref{AlgChaIdeqDelFkt1}) of Chap.~\ref{KapChaIde}]:%
\begin{align}
p^{A}\circledast c_{\hspace{0.01in}p}  &  =\operatorname*{vol}\nolimits^{-1}%
\hspace{-0.02in}\int\text{d}_{q}^{3}p^{\prime}\,\delta_{q}^{3}((\ominus
\hspace{0.01in}\kappa^{-1}p)\oplus p^{\prime})\circledast p^{\prime
A}\circledast c_{\hspace{0.01in}p^{\prime}}\nonumber\\
&  =\int\text{d}_{q}^{3}p^{\prime}\,(P^{A})_{p\hspace{0.01in}p^{\prime}%
}\circledast c_{\hspace{0.01in}p^{\prime}}. \label{MatMulImpOpe2}%
\end{align}
Note that in the last step, we were able to identify the expression for the
momentum matrix element given in Eq.~(\ref{MatEleImpOpe}). Finally, we
summarize the results in Eq.~(\ref{MatMulImpOpe1}) and
Eq.~(\ref{MatMulImpOpe2}) as follows:%
\begin{equation}
(P^{A}\triangleright\psi_{R})_{p}=\int\text{d}_{q}^{3}p^{\prime}%
\,(P^{A})_{p\hspace{0.01in}p^{\prime}}\circledast c_{\hspace{0.01in}p^{\prime
}}\hspace{-0.01in}=p^{A}\circledast c_{\hspace{0.01in}p}.
\label{VekImpOpeWelFkt1}%
\end{equation}
In the same way, one can show:%
\begin{align}
(\psi_{L}\,\bar{\triangleleft}\,P^{A})^{p}  &  =\operatorname*{vol}%
\nolimits^{-1}\hspace{-0.01in}\mathcal{F}_{L}^{\ast}(\psi_{L}\,\bar
{\triangleleft}\,\hspace{0.01in}\partial_{x}^{A}\text{i}^{-1})(\hspace
{0.01in}p)\nonumber\\
&  =\int\text{d}_{q}^{3}p^{\prime}\,c^{\hspace{0.01in}p^{\prime}}%
\hspace{-0.01in}\circledast(P^{A})^{p^{\prime}p}=c^{\hspace{0.01in}%
p}\circledast p^{A}. \label{VekImpOpeWelFkt2}%
\end{align}

If two momentum operators act successively on a wave function, their
representation matrices in the momentum space are multiplied with each other.
We can read it off from repeated application of the first identity of
Eq.~(\ref{VekImpOpeWelFkt1}):%
\begin{align}
(P^{A}P^{B}\triangleright\psi_{R})_{p}  &  =(P^{A}\triangleright
(P^{B}\triangleright\psi_{R}))_{p}\hspace{0.01in}=\int\text{d}_{q}%
^{3}p^{\prime}\,(P^{A})_{p\hspace{0.01in}p^{\prime}}\circledast(P^{B}%
\triangleright\psi_{R})_{p^{\prime}}\nonumber\\
&  =\int\text{d}_{q}^{3}p^{\prime}\hspace{-0.01in}\int\text{d}_{q}%
^{3}p^{\prime\prime}\,(P^{A})_{p\hspace{0.01in}p^{\prime}}\circledast
(P^{B})_{p^{\prime}p^{\prime\prime}}\circledast c_{\hspace{0.01in}%
p^{\prime\prime}}. \label{MatMulImpEle}%
\end{align}
Similarly, we have:%
\begin{equation}
(\psi_{L}\,\bar{\triangleleft}\,P^{A}P^{B})^{p}=\int\text{d}_{q}^{3}p^{\prime
}\hspace{-0.01in}\int\text{d}_{q}^{3}p^{\prime\prime}\,c^{\hspace
{0.01in}p^{\prime\prime}}\hspace{-0.02in}\circledast(P^{A})^{p^{\prime\prime
}p^{\prime}}\hspace{-0.01in}\circledast(P^{B})^{p^{\prime}p}.
\label{MatMulImpEle2}%
\end{equation}

Next, we calculate matrix representations of the position operator with
respect to a basis of $q$-de\-formed momentum eigenfunctions. We do this in a
similar way as in Eq.~(\ref{MatEleImpOpe}):%
\begin{align}
(X^{A})_{p^{\prime}p}  &  =\operatorname*{vol}\nolimits^{-1/2}\mathcal{F}%
_{R}^{\ast}(x^{A}\circledast u_{\hspace{0.01in}p})(\hspace{0.01in}p^{\prime
})\nonumber\\[0.02in]
&  =\int\text{d}_{q}^{3}x\,(u^{\ast})_{p^{\prime}}(x)\circledast\lbrack
\hspace{0.01in}x^{A}\circledast u_{\hspace{0.01in}p}(x)]\nonumber\\
&  =\int\text{d}_{q}^{3}x\,(u^{\ast})_{p^{\prime}}(x)\circledast
u_{\hspace{0.01in}p}(x)\,\bar{\triangleleft}\,\partial_{p}^{A}\text{i}%
\nonumber\\[0.02in]
&  =\operatorname*{vol}\nolimits^{-1}\hspace{-0.01in}\delta_{q}^{3}%
((\ominus\hspace{0.01in}\kappa^{-1}p^{\prime})\oplus p)\,\bar{\triangleleft
}\,\partial_{p}^{A}\text{i}. \label{OrtMatEleImpRau1}%
\end{align}
Alternatively, we can calculate matrix elements of the position operator as
follows:%
\begin{align}
(X^{A})^{p\hspace{0.01in}p^{\prime}}  &  =\operatorname*{vol}\nolimits^{-1/2}%
\mathcal{F}_{L}^{\ast}(u^{p}\circledast x^{A})(\hspace{0.01in}p^{\prime
})\nonumber\\[0.02in]
&  =\int\text{d}_{q}^{3}x\,[\hspace{0.01in}u^{p}(x)\circledast x^{A}%
]\circledast(u^{\ast})^{p^{\prime}}\hspace{-0.01in}(x)\nonumber\\
&  =\int\text{d}_{q}^{3}x\,\text{i}\partial_{p}^{A}\triangleright
u^{p}(x)\circledast(u^{\ast})^{p^{\prime}}\hspace{-0.01in}%
(x)\nonumber\\[0.02in]
&  =\operatorname*{vol}\nolimits^{-1}\text{i}\partial_{p}^{A}\triangleright
\delta_{q}^{3}(\hspace{0.01in}p\oplus(\ominus\hspace{0.01in}\kappa
^{-1}p^{\prime})).
\end{align}

The action of the $q$-de\-formed position operator on a wave function becomes
the action of a derivative operator on the corresponding expansion
coefficients in the momentum space since we have in analogy to
Eq.~(\ref{MatMulImpOpe1}):%
\begin{align}
(X^{A}\,\bar{\triangleright}\,\psi_{R})_{p}  &  =\operatorname*{vol}%
\nolimits^{-1/2}\mathcal{F}_{R}^{\ast}(x^{A}\circledast\psi_{R})(\hspace
{0.01in}p)\nonumber\\
&  =\operatorname*{vol}\nolimits^{-1/2}\text{i}\partial_{p}^{A}\,\bar
{\triangleright}\,\mathcal{F}_{R}^{\ast}(\psi_{R})(\hspace{0.01in}%
p)=\text{i}\partial_{p}^{A}\,\bar{\triangleright}\,c_{p}.
\label{BerMatOrtMul1}%
\end{align}
In the second step of the above calculation, we have made use of the
characteristic identities of the $q$-Fourier transformations [cf.
Eq.~(\ref{InfTransFour}) of Chap.~\ref{KapFouActPro}]. In the last step, we
have identified the expansion coefficients of $\psi_{R}$ [cf. Eq.~(\ref{CPUnt}%
) of Chap.~\ref{KapVolRelImpOrt}]. In addition to this, the following applies:%
\begin{align}
\text{i}\partial_{p}^{A}\,\bar{\triangleright}\,c_{\hspace{0.01in}p}  &
=\operatorname*{vol}\nolimits^{-1}\hspace{-0.02in}\int\text{d}_{q}%
^{3}p^{\prime}\hspace{0.01in}\delta_{q}^{3}((\ominus\hspace{0.01in}\kappa
^{-1}p)\oplus p^{\prime})\circledast\text{i}\partial_{p^{\prime}}^{A}%
\,\bar{\triangleright}\,c_{\hspace{0.01in}p^{\prime}}\nonumber\\
&  =\operatorname*{vol}\nolimits^{-1}\hspace{-0.02in}\int\text{d}_{q}%
^{3}p^{\prime}\hspace{0.01in}\delta_{q}^{3}((\ominus\hspace{0.01in}\kappa
^{-1}p)\oplus p^{\prime})\,\bar{\triangleleft}\,\partial_{p^{\prime}}%
^{A}\text{i}\circledast c_{\hspace{0.01in}p^{\prime}}\nonumber\\
&  =\int\text{d}_{q}^{3}p^{\prime}\,(X^{A})_{p\hspace{0.01in}p^{\prime}%
}\circledast c_{\hspace{0.01in}p^{\prime}}. \label{BerMatOrtMul2}%
\end{align}
Note that we have first used the characteristic identities of the $q$-del\-ta
functions and the rules for integration by parts. The last step then follows
from Eq.~(\ref{OrtMatEleImpRau1}).\ If we summarize the results of
Eq.~(\ref{BerMatOrtMul1}) and Eq.~(\ref{BerMatOrtMul2}), we get:%
\begin{equation}
(X^{A}\,\bar{\triangleright}\,\psi_{R})_{p}=\int\text{d}_{q}^{3}p^{\prime
}\,(X^{A})_{p\hspace{0.01in}p^{\prime}}\circledast c_{\hspace{0.01in}%
p^{\prime}}=\text{i}\partial_{p}^{A}\,\bar{\triangleright}\,c_{\hspace
{0.01in}p}. \label{VekOrtOpeWelFkt1}%
\end{equation}
In the same way, we can show:%
\begin{equation}
(\psi_{L}\triangleleft X^{A})^{p}=\int\text{d}_{q}^{3}p^{\prime}%
\,c^{\hspace{0.01in}p^{\prime}}\hspace{-0.01in}\circledast(X^{A})^{p^{\prime
}p}=c^{\hspace{0.01in}p}\triangleleft\partial_{p}^{A}\text{i}.
\label{VekOrtOpeWelFkt2}%
\end{equation}

For the sake of completeness, we write down how the given matrix elements of
the $q$-de\-formed momentum operator and that of the $q$-de\-formed position
operator behave under conjugation. We can show that the different versions of
matrix elements transform into each other by conjugation in the following way:%
\[
\overline{(P^{A})^{p\hspace{0.01in}p^{\prime}}}=(P_{A})_{p^{\prime}p}%
,\qquad\overline{(X^{A})^{p\hspace{0.01in}p^{\prime}}}=(X_{A})_{p^{\prime}p}.
\]

So far, we have only considered matrix representations referring to the
expansion coefficients $c_{\hspace{0.01in}p}$ or $c^{\hspace{0.01in}p}$ [cf.
Eq.~(\ref{CPUnt}) and Eq.~(\ref{CPUnt2}) of Chap.~\ref{KapVolRelImpOrt}].
However, we can also give matrix representations corresponding to the
expansion coefficients $(c^{\ast})_{p}$ or $(c^{\ast})^{p}$ [cf.
Eq.~(\ref{DPUnt}) and Eq.~(\ref{DPUnt2}) of Chap.~\ref{KapVolRelImpOrt}]:%
\begin{align}
(P^{A})_{p\hspace{0.01in}p^{\prime}}  &  =\operatorname*{vol}\nolimits^{-1/2}%
\mathcal{F}_{L}((u^{\ast})_{p}\triangleleft\partial_{x}^{A}\text{i}%
^{-1})(\hspace{0.01in}p^{\prime})\nonumber\\[0.02in]
&  =\int\text{d}_{q}^{3}x\,[(u^{\ast})_{p}(x)\triangleleft\partial^{A}%
\text{i}^{-1}]\circledast u_{\hspace{0.01in}p^{\prime}}(x)\nonumber\\
&  =\int\text{d}_{q}^{3}x\,p^{A}\circledast(u^{\ast})_{p}(x)\circledast
u_{\hspace{0.01in}p^{\prime}}(x)\nonumber\\[0.02in]
&  =\operatorname*{vol}\nolimits^{-1}p^{A}\circledast\delta_{q}^{3}%
((\ominus\hspace{0.01in}\kappa^{-1}p)\oplus p^{\prime}).
\label{ImpMatEle2Bas1}%
\end{align}
In the second step of the above calculation, we have written out the
$q$-Fourier transformation. The penultimate step is a consequence of the
eigenvalue equations for the $q$-de\-formed momentum eigenfunctions [cf.
Eq.~(\ref{ImpEigFktqDef2}) of Chap.~\ref{KapDefImpOrtEigFkt}]. The last step
follows from the orthonormality relations of the $q$-de\-formed momentum
eigenfunctions [also see Eq.~(\ref{OrtRelImpEigDarKap2}) of the previous
chapter]. Similar considerations lead to:%
\begin{align}
(P^{A})^{p^{\prime}\hspace{-0.01in}p}  &  =\operatorname*{vol}\nolimits^{-1/2}%
\mathcal{F}_{R}(\text{i}^{-1}\partial_{x}^{A}\,\bar{\triangleright}\,(u^{\ast
})^{p})(\hspace{0.01in}p^{\prime})\nonumber\\
&  =\int\text{d}_{q}^{3}x\,u^{p^{\prime}}\hspace{-0.01in}(x)\circledast
\text{i}^{-1}\partial_{x}^{A}\,\bar{\triangleright}\,(u^{\ast})^{p}%
(x)\nonumber\\
&  =\operatorname*{vol}\nolimits^{-1}\hspace{-0.01in}\delta_{q}^{3}%
(\hspace{0.01in}p^{\prime}\oplus(\ominus\hspace{0.01in}\kappa^{-1}%
p))\circledast p^{A}. \label{ImpMatEle2Bas2}%
\end{align}

Next, we consider the action of the $q$-de\-formed momentum operator on a wave
function represented in the momentum space by the coefficients $(c^{\ast}%
)^{p}$ or $(c^{\ast})_{p}$, i.~e. [also see Eq.~(\ref{FunProp1}) of
Chap.~\ref{KapFouActPro}\ and Eq.~(\ref{DPUnt}) of Chap.~\ref{KapVolRelImpOrt}%
]%
\begin{align}
(\psi_{L}^{\ast}\triangleleft P^{A})_{p}  &  =\operatorname*{vol}%
\nolimits^{-1/2}\mathcal{F}_{L}(\psi_{L}^{\ast}\triangleleft\partial_{x}%
^{A}\text{i}^{-1})(\hspace{0.01in}p)\nonumber\\
&  =\operatorname*{vol}\nolimits^{-1/2}\mathcal{F}_{L}(\psi_{L}^{\ast
})(\hspace{0.01in}p)\circledast p^{A}=(c^{\ast})_{p}\circledast p^{A}
\label{MatMulImpOpeR0}%
\end{align}
and%
\begin{align}
(P^{A}\,\bar{\triangleright}\,\psi_{R}^{\ast})^{p}  &  =\operatorname*{vol}%
\nolimits^{-1/2}\mathcal{F}_{R}(\text{i}^{-1}\partial_{x}^{A}\,\bar
{\triangleright}\,\psi_{R}^{\ast})(\hspace{0.01in}p)\nonumber\\
&  =\operatorname*{vol}\nolimits^{-1/2}p^{A}\circledast\mathcal{F}_{R}%
(\psi_{R}^{\ast})(\hspace{0.01in}p)=p^{A}\circledast(c^{\ast})^{p}.
\label{MatMulImpOpeR}%
\end{align}
On the other hand, the characteristic identities of the $q$-del\-ta function
together with Eq.~(\ref{ImpMatEle2Bas1}) and Eq.~(\ref{ImpMatEle2Bas2}) imply:%
\begin{align}
(c^{\ast})_{p}\circledast p^{A}  &  =\operatorname*{vol}\nolimits^{-1}%
\hspace{-0.01in}\int\text{d}_{q}^{3}p^{\prime}\,(c^{\ast})_{p^{\prime}%
}\circledast p^{\prime A}\circledast\delta_{q}^{3}((\ominus\hspace
{0.01in}\kappa^{-1}p^{\prime})\oplus p)\nonumber\\
&  =\int\text{d}_{q}^{3}p^{\prime}\,(c^{\ast})_{p^{\prime}}\circledast
(P^{A})_{p^{\prime}p},\\[0.1in]
p^{A}\circledast(c^{\ast})^{p}  &  =\operatorname*{vol}\nolimits^{-1}%
\hspace{-0.01in}\int\text{d}_{q}^{3}p^{\prime}\,\delta_{q}^{3}(\hspace
{0.01in}p\oplus(\ominus\hspace{0.01in}\kappa^{-1}p^{\prime}))\circledast
p^{\prime A}\circledast(c^{\ast})^{p^{\prime}}\nonumber\\
&  =\int\text{d}_{q}^{3}p^{\prime}\,(P^{A})^{p\hspace{0.01in}p^{\prime}%
}\hspace{-0.01in}\circledast(c^{\ast})^{p^{\prime}}.
\end{align}
Finally, from the above considerations follows:%
\begin{align}
(\psi_{L}^{\ast}\triangleleft P^{A})_{p}  &  =(c^{\ast})_{p}\circledast
p^{A}=\int\text{d}_{q}^{3}p^{\prime}\,(c^{\ast})_{p^{\prime}}\circledast
(P^{A})_{p^{\prime}p},\nonumber\\
(P^{A}\,\bar{\triangleright}\,\psi_{R}^{\ast})^{p}  &  =p^{A}\circledast
(c^{\ast})^{p}=\int\text{d}_{q}^{3}p^{\prime}\,(P^{A})^{p\hspace
{0.01in}p^{\prime}}\circledast(c^{\ast})^{p^{\prime}}.
\label{VekImpOpeWelFkt3}%
\end{align}
Accordingly, we can write the successive application of two $q$-de\-formed
momentum operators as a kind of matrix multiplication:%
\begin{align}
(\psi_{L}^{\ast}\triangleleft P^{A}P^{B})_{p}  &  =\int\text{d}_{q}%
^{3}p^{\prime}\int\text{d}_{q}^{3}p^{\prime\prime}\,(c^{\ast})_{p^{\prime
\prime}}\circledast(P^{A})_{p^{\prime\prime}p^{\prime}}\circledast
(P^{B})_{p^{\prime}p},\nonumber\\
(P^{A}P^{B}\,\bar{\triangleright}\,\psi_{R}^{\ast})^{p}  &  =\int\text{d}%
_{q}^{3}p^{\prime}\int\text{d}_{q}^{3}p^{\prime\prime}\,(P^{A})^{p\hspace
{0.01in}p^{\prime}}\hspace{-0.01in}\circledast(P^{B})^{p^{\prime}%
p^{\prime\prime}}\circledast(c^{\ast})^{p^{\prime\prime}}.
\label{MatMulImpEle3}%
\end{align}

Again, we turn to the matrix representations of the $q$-de\-formed position
operator. We can obtain these matrix representations by a calculation
analogous to that of Eq.~(\ref{ImpMatEle2Bas1}):%
\begin{align}
(X^{A})_{p\hspace{0.01in}p^{\prime}}  &  =\operatorname*{vol}\nolimits^{-1/2}%
\mathcal{F}_{L}((u^{\ast})_{p}\circledast x^{A})(\hspace{0.01in}p^{\prime
})\nonumber\\[0.02in]
&  =\int\text{d}_{q}^{3}x\,[(u^{\ast})_{p}\circledast x^{A}]\circledast
u_{\hspace{0.01in}p^{\prime}}(x)\nonumber\\
&  =\int\text{d}_{q}^{3}x\,[\hspace{0.01in}\text{i}\partial_{p}^{A}%
\,\bar{\triangleright}\,(u^{\ast})_{p}]\circledast u_{\hspace{0.01in}%
p^{\prime}}(x)\nonumber\\[0.02in]
&  =\operatorname*{vol}\nolimits^{-1}\text{i}\partial_{p}^{A}\,\bar
{\triangleright}\,\delta_{q}^{3}((\ominus\hspace{0.01in}\kappa^{-1}p)\oplus
p^{\prime}).
\end{align}
In the same way, we have:%
\begin{align}
(X^{A})^{p^{\prime}\hspace{-0.01in}p}  &  =\operatorname*{vol}\nolimits^{-1/2}%
\mathcal{F}_{R}(x^{A}\circledast(u^{\ast})^{p})(\hspace{0.01in}p^{\prime
})\nonumber\\[0.02in]
&  =\int\text{d}_{q}^{3}x\,u^{p^{\prime}}\hspace{-0.01in}(x)\circledast
\lbrack\hspace{0.01in}x^{A}\circledast(u^{\ast})^{p}(x)]\nonumber\\
&  =\int\text{d}_{q}^{3}x\,u^{p^{\prime}}\hspace{-0.01in}(x)\circledast
\lbrack(u^{\ast})^{p}(x)\triangleleft\partial_{p}^{A}\text{i}%
]\nonumber\\[0.02in]
&  =\operatorname*{vol}\nolimits^{-1}\hspace{-0.01in}\delta_{q}^{3}%
(\hspace{0.01in}p^{\prime}\oplus(\ominus\hspace{0.01in}\kappa^{-1}%
p))\triangleleft\partial_{p}^{A}\text{i}.
\end{align}

Next, we consider the action of the $q$-de\-formed position operator on a wave
function. With the help of Eq.~(\ref{FunProp2a}) from Chap.~\ref{KapFouActPro}%
\ we find:%
\begin{align}
(\psi_{L}^{\ast}\,\bar{\triangleleft}\,X^{A})_{p}  &  =\operatorname*{vol}%
\nolimits^{-1/2}\mathcal{F}_{L}(\psi_{L}^{\ast}\circledast x^{A}%
)(\hspace{0.01in}p)\nonumber\\
&  =\operatorname*{vol}\nolimits^{-1/2}\mathcal{F}_{L}(\psi_{L}^{\ast
})(\hspace{0.01in}p)\,\bar{\triangleleft}\,\partial_{p}^{A}\text{i}=(c^{\ast
})_{p}\,\bar{\triangleleft}\,\partial_{p}^{A}\text{i}.
\label{MatOrtOpeKonRecWir}%
\end{align}
Moreover, we have:%
\begin{align}
(c^{\ast})_{p}\,\bar{\triangleleft}\,\partial_{p}^{A}\text{i}  &
=\operatorname*{vol}\nolimits^{-1}\hspace{-0.01in}\int\text{d}_{q}%
^{3}p^{\prime}\,[(c^{\ast})_{p^{\prime}}\,\bar{\triangleleft}\,\hspace
{0.01in}\partial_{p^{\prime}}^{A}\text{i}]\circledast\delta_{q}^{3}%
((\ominus\hspace{0.01in}\kappa^{-1}p^{\prime})\oplus p)\nonumber\\
&  =\operatorname*{vol}\nolimits^{-1}\hspace{-0.01in}\int\text{d}_{q}%
^{3}p^{\prime}\,(c^{\ast})_{p^{\prime}}\circledast\lbrack\hspace
{0.01in}\text{i}\partial_{p^{\prime}}^{A}\,\bar{\triangleright}\,\delta
_{q}^{3}((\ominus\hspace{0.01in}\kappa^{-1}p^{\prime})\oplus p)]\nonumber\\
&  =\int\text{d}_{q}^{3}p^{\prime}\,(c^{\ast})_{p^{\prime}}\circledast
(X^{A})_{p^{\prime}p}.
\end{align}
Finally, it follows:%
\begin{equation}
(\psi_{L}^{\ast}\,\bar{\triangleleft}\,X^{A})_{p}=\int\text{d}_{q}%
^{3}p^{\prime}\,(c^{\ast})_{p^{\prime}}\circledast(X^{A})_{p^{\prime}%
p}=(c^{\ast})_{p}\,\bar{\triangleleft}\,\partial_{p}^{A}\text{i}.
\label{VekOrtOpeWelFkt3}%
\end{equation}
In the same way, you can show:%
\begin{equation}
(X^{A}\triangleright\psi_{R}^{\ast})^{p}=\int\text{d}_{q}^{3}p^{\prime
}\,(X^{A})^{p\hspace{0.01in}p^{\prime}}\hspace{-0.01in}\circledast(c^{\ast
})^{p^{\prime}}\hspace{-0.01in}=\text{i}\partial_{p}^{A}\triangleright
(c^{\ast})^{p}. \label{VekOrtOpeWelFkt4}%
\end{equation}

\subsection{Matrix representations on position space\label{KapMatDarOrt}}

In this chapter, we determine matrix representations of the position or
momentum operator for a basis of $q$-de\-formed position eigenfunctions. We
start with the matrix elements of the position operator. Using the
characteristic identities for the $q$-de\-formed delta functions, we obtain
the following expressions for the matrix elements of the position operator
[cf. Eq.~(\ref{AlgChaIdeqDelFkt}) and Eq.~(\ref{AlgChaIdeqDelFkt1}) of
Chap.~\ref{KapChaIde}]:%
\begin{align}
(X^{A})_{y^{\prime}y}  &  =\int\text{d}_{q}^{3}x\,(u^{\ast})_{y^{\prime}%
}(x)\circledast x^{A}\circledast u_{\hspace{0.01in}y}(x)\nonumber\\[0.03in]
&  =\operatorname*{vol}\nolimits^{-2}\hspace{-0.01in}\int\text{d}_{q}%
^{3}x\,\delta_{q}^{3}((\ominus\hspace{0.01in}\kappa^{-1}y^{\prime})\oplus
x)\circledast x^{A}\circledast\delta_{q}^{3}((\ominus\hspace{0.01in}%
\kappa^{-1}x)\oplus y)\nonumber\\
&  =\operatorname*{vol}\nolimits^{-1}\hspace{-0.01in}\delta_{q}^{3}%
((\ominus\hspace{0.01in}\kappa^{-1}y^{\prime})\oplus y)\circledast y^{A}.
\label{MatEleXASub}%
\end{align}
In the second step of the above calculation, we have written the position
eigenfunctions in the form of $q$-del\-ta functions [cf. Eq.~(\ref{DefEigPos1}%
) of Chap.~\ref{KapDefImpOrtEigFkt}]. In the same manner follows:%
\begin{align}
(X^{A})^{y^{\prime}\hspace{-0.01in}y}  &  =\int\text{d}_{q}^{3}x\,u^{y^{\prime
}}\hspace{-0.01in}(x)\circledast x^{A}\circledast(u^{\ast})^{y}(x)\nonumber\\
&  =\operatorname*{vol}\nolimits^{-1}y^{\prime A}\circledast\delta_{q}%
^{3}(\hspace{0.01in}y^{\prime}\oplus(\ominus\hspace{0.01in}\kappa^{-1}y)).
\label{MatEleXASub2}%
\end{align}
Note that the matrix element of Eq.~(\ref{MatEleXASub}) is equal to that of
Eq.(\ref{MatEleXASub2}) due to the following distribution equation:%
\begin{equation}
\delta_{q}^{3}((\ominus\hspace{0.01in}\kappa^{-1}y^{\prime})\oplus
y)\circledast y^{A}=y^{\prime A}\circledast\delta_{q}^{3}(\hspace
{0.01in}y^{\prime}\oplus(\ominus\hspace{0.01in}\kappa^{-1}y)).
\end{equation}

If we conjugate the above expressions for the matrix elements of the position
operator and take into account the conjugation properties of volume elements,
delta functions, and position coordinates, we will get the following result:%
\begin{equation}
\overline{(X^{A})_{y^{\prime}y}}=\operatorname*{vol}\nolimits^{-1}%
y_{A}\circledast\delta_{q}^{3}(\hspace{0.01in}y\oplus(\ominus\hspace
{0.01in}\kappa^{-1}y^{\prime}))=(X_{A})^{y\hspace{0.01in}y^{\prime}}.
\label{MatEleXASubKon}%
\end{equation}

Next, we consider the action of the $q$-de\-formed position operator on a wave
function. If the $q$-de\-formed position operator acts on a wave function of
position space from the left, the given wave function will be multiplied by
position coordinates from the left:%
\begin{align}
(X^{A}\triangleright\psi_{R})_{y}  &  =\int\text{d}_{q}^{3}x\,(u^{\ast}%
)_{y}(x)\circledast x^{A}\circledast\psi_{R}(x)\nonumber\\
&  =\operatorname*{vol}\nolimits^{-1}\hspace{-0.01in}\int\text{d}_{q}%
^{3}x\,\delta_{q}^{3}(\hspace{0.01in}y\oplus(\ominus\hspace{0.01in}\kappa
^{-1}x))\circledast x^{A}\circledast\psi_{R}(x)\nonumber\\[0.02in]
&  =y^{A}\circledast\psi_{R}(\hspace{0.01in}y)=y^{A}\circledast c_{\hspace
{0.01in}y}. \label{MatEleOrtOpeOrtRau}%
\end{align}
In the final step of the above calculation, we took into account that the wave
function can be identified with its expansion coefficients\ referring to a
basis of $q$-de\-formed position eigenfunctions [see
Eq.~(\ref{EntKoeOrtEigFkt}) of Chap.~\ref{KapVolRelOrtEigFkt}]. On the other
hand, we can also express $(X^{A}\triangleright\psi_{R})_{y}$ by using the
matrix elements from Eq.~(\ref{MatEleXASub}):%
\begin{align}
(X^{A}\triangleright\psi_{R})_{y}  &  =\operatorname*{vol}\nolimits^{-1}%
\hspace{-0.01in}\int\text{d}_{q}^{3}y^{\prime}\,\delta_{q}^{3}(\hspace
{0.01in}y\oplus(\ominus\hspace{0.01in}\kappa^{-1}y^{\prime}))\circledast
y^{\prime A}\circledast\psi_{R}(\hspace{0.01in}y^{\prime})\nonumber\\
&  =\int\text{d}_{q}^{3}y^{\prime}\,(X^{A})_{y\hspace{0.01in}y^{\prime}%
}\circledast\psi_{R}(\hspace{0.01in}y^{\prime})=\int\text{d}_{q}^{3}y^{\prime
}\,(X^{A})_{y\hspace{0.01in}y^{\prime}}\circledast c_{\hspace{0.01in}%
y^{\prime}}. \label{MatEleOrtOpeOrtRau2}%
\end{align}
We can summarize the results of Eq.~(\ref{MatEleOrtOpeOrtRau}) and
Eq.~(\ref{MatEleOrtOpeOrtRau2}) as follows:%
\begin{equation}
(X^{A}\triangleright\psi_{R})_{y}=\int\text{d}_{q}^{3}y^{\prime}%
\,(X^{A})_{y\hspace{0.01in}y^{\prime}}\circledast c_{\hspace{0.01in}y^{\prime
}}\hspace{-0.01in}=y^{A}\circledast c_{\hspace{0.01in}y}.
\label{ZusOrtEleOrtBas}%
\end{equation}

If the $q$-de\-formed position operator acts on the wave function from the
right, the wave function must be multiplied by the position coordinates from
the right:%
\begin{align}
(\psi_{L}^{\ast}\triangleleft X^{A})_{y}  &  =\int\text{d}_{q}^{3}x\,\psi
_{L}^{\ast}(x)\circledast x^{A}\circledast u_{\hspace{0.01in}y}(x)\nonumber\\
&  =\operatorname*{vol}\nolimits^{-1}\hspace{-0.01in}\int\text{d}_{q}%
^{3}x\,\psi_{L}^{\ast}(x)\circledast x^{A}\circledast\delta_{q}^{3}%
((\ominus\hspace{0.01in}\kappa^{-1}x)\oplus y)\nonumber\\[0.02in]
&  =\psi_{L}^{\ast}(\hspace{0.01in}y)\circledast y^{A}=(c^{\ast}%
)_{y}\circledast y^{A}.
\end{align}
In analogy to Eq.~(\ref{ZusOrtEleOrtBas}), we can also write this result as
follows:%
\begin{align}
(\psi_{L}^{\ast}\triangleleft X^{A})_{y}  &  =\int\text{d}_{q}^{3}y^{\prime
}\,(c^{\ast})_{y^{\prime}}\circledast\hspace{0.01in}(X^{A})_{y^{\prime}%
y}\nonumber\\
&  =(c^{\ast})_{y^{\prime}}\circledast\hspace{0.01in}y^{A}.
\label{ZusOrtEleOrtBasRec}%
\end{align}

We now turn to the momentum operator and calculate its matrix elements for a
basis of $q$-de\-formed position eigenfunctions. We proceed similarly to the
calculation of the matrix elements of the position operator, i.~e. we write
the position eigenfunctions as $q$-de\-formed delta functions and apply the
characteristic identities of the $q$-de\-formed delta functions. This way, we
obtain the following expression for the matrix elements of the momentum
operator:%
\begin{align}
(P^{A})_{y^{\prime}y}  &  =\int\text{d}_{q}^{3}x\,(u^{\ast})_{y^{\prime}%
}(x)\circledast\lbrack\hspace{0.01in}\text{i}^{-1}\partial_{x}^{A}%
\,\bar{\triangleright}\,u_{\hspace{0.01in}y}(x)]\nonumber\\
&  =\int\text{d}_{q}^{3}x\,[(u^{\ast})_{y^{\prime}}(x)\,\bar{\triangleleft
}\,\partial_{x}^{A}\text{i}^{-1}]\circledast u_{\hspace{0.01in}y}%
(x)\nonumber\\
&  =\operatorname*{vol}\nolimits^{-2}\hspace{-0.01in}\int\text{d}_{q}%
^{3}x\,[\delta_{q}^{3}((\ominus\hspace{0.01in}\kappa^{-1}y^{\prime})\oplus
x)\,\bar{\triangleleft}\,\partial_{x}^{A}\text{i}^{-1}]\circledast\delta
_{q}^{3}((\ominus\hspace{0.01in}\kappa^{-1}x)\oplus y)\nonumber\\[0.03in]
&  =\operatorname*{vol}\nolimits^{-1}\delta_{q}^{3}((\ominus\hspace
{0.01in}\kappa^{-1}y^{\prime})\oplus y)\,\bar{\triangleleft}\,\partial_{y}%
^{A}\text{i}^{-1}. \label{MatEleImpOpeOrtDar1}%
\end{align}
Note that in the second step we made use of the $q$-de\-formed Stokes' theorem
[cf. Eq.~(\ref{PatIntUneRaumInt}) of Chap.~\ref{KapIntegral}]. It follows from
similar arguments:%
\begin{align}
(P^{A})^{y\hspace{0.01in}y^{\prime}}  &  =\int\text{d}_{q}^{3}x\,[\hspace
{0.01in}u^{y}(x)\triangleleft\partial_{x}^{A}\text{i}^{-1}]\circledast
(u^{\ast})^{y^{\prime}}\hspace{-0.01in}(x)\nonumber\\[0.03in]
&  =\operatorname*{vol}\nolimits^{-1}\text{i}^{-1}\partial_{y}^{A}%
\triangleright\delta_{q}^{3}(\hspace{0.01in}y\oplus(\ominus\hspace
{0.01in}\kappa^{-1}y^{\prime})). \label{MatEleImpOpeOrtDar2}%
\end{align}

Taking into account the conjugation properties of volume elements, delta
functions, and partial derivatives, the results of
Eq.~(\ref{MatEleImpOpeOrtDar1}) and Eq.~(\ref{MatEleImpOpeOrtDar2}) imply that
the matrix elements of the $q$-de\-formed momentum operator behave as follows
under conjugation:%
\[
\overline{(P^{A})_{y^{\prime}y}}=(P_{A})^{y\hspace{0.01in}y^{\prime}}.
\]

Finally, we consider the action of the momentum operator on a general wave
function. To this end, we proceed in analogy to the calculation of
Eq.~(\ref{MatEleOrtOpeOrtRau2}):%
\begin{align}
(P^{A}\,\bar{\triangleright}\,\psi_{R})_{y}  &  =\int\text{d}_{q}%
^{3}x\,(u^{\ast})_{y}(x)\circledast\lbrack\hspace{0.01in}\text{i}^{-1}%
\partial_{x}^{A}\,\bar{\triangleright}\,\psi_{R}(x)]\nonumber\\
&  =\operatorname*{vol}\nolimits^{-1}\hspace{-0.02in}\int\text{d}_{q}%
^{3}x\,\delta_{q}^{3}((\ominus\hspace{0.01in}\kappa^{-1}y)\oplus
x)\circledast\lbrack\hspace{0.01in}\text{i}^{-1}\partial_{x}^{A}%
\,\bar{\triangleright}\,\psi_{R}(x)]\nonumber\\
&  =\operatorname*{vol}\nolimits^{-1}\hspace{-0.02in}\int\text{d}_{q}%
^{3}y^{\prime}\,[\delta_{q}^{3}((\ominus\hspace{0.01in}\kappa^{-1}y)\oplus
y^{\prime})\,\bar{\triangleleft}\,\partial_{y^{\prime}}^{A}\text{i}%
^{-1}]\circledast\psi_{R}(\hspace{0.01in}y^{\prime})\nonumber\\
&  =\int\text{d}_{q}^{3}y\,(P^{A})_{yy^{\prime}}\circledast c_{\hspace
{0.01in}y^{\prime}}.
\end{align}
On the other hand, we have:%
\begin{align}
(P^{A}\,\bar{\triangleright}\,\psi_{R})_{y}  &  =\operatorname*{vol}%
\nolimits^{-1}\hspace{-0.02in}\int\text{d}_{q}^{3}x\,\delta_{q}^{3}%
(\hspace{0.01in}y\oplus(\ominus\hspace{0.01in}\kappa^{-1}x))\circledast
\lbrack\hspace{0.01in}\text{i}^{-1}\partial_{x}^{A}\,\bar{\triangleright
}\,\psi_{R}(x)]\nonumber\\
&  =\text{i}^{-1}\partial_{y}^{A}\,\bar{\triangleright}\,\psi_{R}%
(\hspace{0.01in}y)=\text{i}^{-1}\partial_{y}^{A}\,\bar{\triangleright
}\,c_{\hspace{0.01in}y}.
\end{align}
For this reason, it holds:%
\begin{equation}
(P^{A}\,\bar{\triangleright}\,\psi_{R})_{y}=\int\text{d}_{q}^{3}y^{\prime
}\,(P^{A})_{yy^{\prime}}\circledast c_{\hspace{0.01in}y^{\prime}}%
=\text{i}^{-1}\partial_{y}^{A}\,\bar{\triangleright}\,c_{\hspace{0.01in}y}.
\label{WirImpOpeWelFkt1}%
\end{equation}
Similar arguments lead to:%
\begin{equation}
(\psi_{L}\triangleleft P^{A})^{y}=\int\text{d}_{q}^{3}y^{\prime}%
\,c^{\hspace{0.01in}y^{\prime}}\hspace{-0.01in}\circledast(P^{A})^{y^{\prime
}\hspace{-0.01in}y}=c^{\hspace{0.01in}y}\triangleleft\partial_{y}^{A}%
\text{i}^{-1}. \label{WirImpOpeWelFkt2}%
\end{equation}

Let us mention that in the Eqns.~(\ref{MatEleImpOpeOrtDar1}%
)-(\ref{WirImpOpeWelFkt2}) we may carry out the following replacements:%
\begin{equation}
\bar{\triangleright}\,\leftrightarrow\triangleright,\qquad\triangleleft
\,\leftrightarrow\,\bar{\triangleleft}.
\end{equation}
It should also be noted that all formulas in this chapter remain valid if we
make the following replacements:%
\begin{align}
u_{\hspace{0.01in}y}  &  \leftrightarrow(u^{\ast})^{y}, & c_{\hspace
{0.01in}y}  &  \leftrightarrow(c^{\ast})^{y}, & \psi_{R}  &  \leftrightarrow
\psi_{R}^{\ast},\nonumber\\
(u^{\ast})_{y}  &  \leftrightarrow u^{y}, & (c^{\ast})_{y}  &  \leftrightarrow
c^{\hspace{0.01in}y}, & \psi_{L}^{\ast}  &  \leftrightarrow\psi_{L}.
\end{align}

\subsection{Expectation values and probability densities\label{KapWskDic}}

To begin with, we describe a physical state on a $q$-de\-formed quantum space
by a wave function $\psi_{R}(x)$ and a corresponding `dual' wave function
$\psi_{L}^{\ast}(x)$. Due to their probabilistic interpretation, these wave
functions should fulfill a \textit{normalization condition} of the following
form:%
\begin{equation}
\int\text{d}_{q}^{3}x\,\psi_{L}^{\ast}(x)\circledast\psi_{R}(x)=1.
\label{NorBed12}%
\end{equation}
As the indices $L$ and $R$ indicate, the wave function $\psi_{L}^{\ast}$ is
the left factor in the star product of the above expression and the wave
function $\psi_{R}$ is the right factor.

In quantum mechanics, an operator $\hat{O}$ is assigned to a measurable
quantity. If the $q$-de\-formed wave functions $\psi_{R}$ and $\psi_{L}^{\ast
}$ describe the state of a physical system, the measured values of the
physical quantity $\hat{O}$ scatter around an \textit{expectation value} given
by the following expression:%
\begin{equation}
\langle\hat{O}\rangle_{\psi}=\int\text{d}_{q}^{3}x\,\psi_{L}^{\ast
}(x)\circledast\hat{O}\triangleright\psi_{R}(x)=\int\text{d}_{q}^{3}%
x\,\psi_{L}^{\ast}(x)\triangleleft\hat{O}\circledast\psi_{R}(x).
\label{DefErwObsEin}%
\end{equation}

We require that the expectation value behaves as follows under conjugation:%
\begin{equation}
\overline{\langle\hat{O}\rangle_{\psi}}=\langle\,\overline{\hat{O}}%
\,\rangle_{\psi}.
\end{equation}
Using the conjugation properties of star product and $q$-de\-formed integral,
we can show that the above condition is satisfied if $\psi_{R}(x)$ and
$\psi_{L}^{\ast}(x)$ are transformed into each other by quantum space
conjugation, i.~e.%
\begin{equation}
\overline{\psi_{R}(x)}=\psi_{L}^{\ast}(x). \label{KonBedWell}%
\end{equation}
If the operator $\hat{O}$ is self-adjoint, i.~e.%
\[
\overline{\hat{O}}=\hat{O},
\]
its expectation value is real in the following way:%
\begin{equation}
\overline{\langle\hat{O}\rangle_{\psi}}=\langle\hat{O}\rangle_{\psi}.
\label{ReeErwOpe}%
\end{equation}

We illustrate the above considerations using the example of the
three-di\-men\-sional $q$-de\-formed position operator made up of the
following components [also see Eq.~(\ref{RelQuaEukDre}) of
Chap.~\ref{KapEucQuaSpa}]:%
\begin{align}
X^{1}  &  =\frac{\text{i}}{2}(-\hspace{0.01in}q^{-1/2}X^{+}-\hspace
{0.01in}q^{1/2}X^{-}),\nonumber\\
X^{2}  &  =\frac{1}{2}(-\hspace{0.01in}q^{-1/2}X^{+}+\hspace{0.01in}%
q^{1/2}X^{-}),\nonumber\\
X^{3}  &  =X^{3}. \label{RealKoor3dimOrt}%
\end{align}
Note that due to the conjugation properties of the quantum space generators
$X^{+}$, $X^{3}$, and $X^{-}$ [cf. Eq.~(\ref{KovKoo}) of
Chap.~\ref{KapEucQuaSpa}],\ the components in Eq.~(\ref{RealKoor3dimOrt}) are
self-adjoint in the following way ($i\in\{1,2,3\}$):%
\begin{equation}
\overline{X^{i}}=X^{i}.
\end{equation}
Next, we consider the expectation value of $X^{i}$. Using the conjugation
properties of the invariant integral and those of the star product, we find:%
\begin{align}
\overline{\langle X^{i}\rangle_{\psi}}  &  =\overline{\int\text{d}_{q}%
^{3}x\,\psi_{L}^{\ast}\circledast(X^{i}\triangleright\psi_{R})}=\overline
{\int\text{d}_{q}^{3}x\,\psi_{L}^{\ast}\circledast(x^{i}\circledast\psi_{R}%
)}\nonumber\\
&  =\int\text{d}_{q}^{3}x\,\overline{(x^{i}\circledast\psi_{R})}%
\circledast\overline{\psi_{L}^{\ast}}=\int\text{d}_{q}^{3}x\,\overline
{\psi_{R}}\circledast x^{i}\circledast\overline{\psi_{L}^{\ast}}\nonumber\\
&  =\int\text{d}_{q}^{3}x\,\psi_{L}^{\ast}\circledast(X^{i}\triangleright
\psi_{R})=\langle X^{i}\rangle_{\psi}. \label{BerReeOrtErw}%
\end{align}

According to Eq.~(\ref{EntOrtEigFkt}) of Chap.~\ref{KapVolRelOrtEigFkt}, the
wave functions $\psi_{R}$ and $\psi_{L}^{\ast}$ can be expanded in terms of
position eigenfunctions as follows:%
\begin{align}
\psi_{R}(x)  &  =\int\text{d}_{q}^{3}y\,u_{\hspace{0.01in}y}(x)\circledast
c_{\hspace{0.01in}y},\nonumber\\
\psi_{L}^{\ast}(x)  &  =\int\text{d}_{q}^{3}y\,(c^{\ast})_{y}\circledast
(u^{\ast})_{y}(x). \label{EntWelKonWelOrt}%
\end{align}
Due to the formulas in Eq.~(\ref{EntWicKoeIntDar}) of
Chap.~\ref{KapVolRelOrtEigFkt}, we also have the identities%
\begin{align}
\psi_{R}(x)  &  =\int\text{d}_{q}^{3}y\,u_{\hspace{0.01in}y}(x)\circledast
c_{\hspace{0.01in}y}=\int\text{d}_{q}^{3}y\,u_{\hspace{0.01in}y}%
(x)\circledast\hspace{-0.01in}\int\text{d}_{q}^{3}x^{\prime}\hspace
{0.01in}(u^{\ast})_{y}(x^{\prime})\circledast\psi_{R}(x^{\prime})\nonumber\\
&  =\int\text{d}_{q}^{3}y\,P_{y}\triangleright\psi_{R}(x)
\label{DefProOrtEig1}%
\end{align}
and%
\begin{align}
\psi_{L}^{\ast}(x)  &  =\int\text{d}_{q}^{3}y\,(c^{\ast})_{y}\circledast
(u^{\ast})_{y}(x)=\int\text{d}_{q}^{3}y\int\text{d}_{q}^{3}x^{\prime}%
\hspace{0.01in}\psi_{L}^{\ast}(x^{\prime})\circledast u_{\hspace{0.01in}%
y}(x^{\prime})\circledast(u^{\ast})_{y}(x)\nonumber\\
&  =\int\text{d}_{q}^{3}y\,\psi_{L}^{\ast}(x)\triangleleft P_{y}.
\label{DefProOrtEig2}%
\end{align}
Note that we have introduced the following operators:%
\begin{align}
P_{y}\triangleright\,\ldots\,  &  =u_{\hspace{0.01in}y}(x)\circledast
\hspace{-0.01in}\int\text{d}_{q}^{3}x^{\prime}\hspace{0.01in}(u^{\ast}%
)_{y}(x^{\prime})\circledast\,\ldots\,,\nonumber\\
\ldots\,\triangleleft P_{y}  &  =\int\text{d}_{q}^{3}x^{\prime}\hspace
{0.01in}\ldots\,\circledast u_{\hspace{0.01in}y}(x^{\prime})\circledast
(u^{\ast})_{y}(x).
\end{align}
They are projectors corresponding to the different eigenvalues of the position
operator:%
\begin{equation}
P_{y}\triangleright\psi_{R}(x)=u_{\hspace{0.01in}y}(x)\circledast
c_{\hspace{0.01in}y},\qquad\psi_{L}^{\ast}(x)\triangleleft P_{y}=(c^{\ast
})_{y}\circledast(u^{\ast})_{y}(x). \label{AntWelEigY}%
\end{equation}
Accordingly, these projectors are subject to following orthonormality
relations:\footnote{Note that the symbol "$\ldots$" stands for an expression
the projectors are acting on.}%
\begin{align}
P_{y^{\prime}}P_{y}\triangleright\,\ldots\,  &  =(P_{y^{\prime}}%
\triangleright\,\ldots\,)\circledast\frac{1}{\operatorname*{vol}}%
\hspace{0.01in}\delta_{q}^{3}((\ominus\hspace{0.01in}\kappa^{-1}y^{\prime
})\oplus y),\nonumber\\
\,\ldots\,\triangleleft P_{y}\hspace{0.01in}P_{y^{\prime}}  &  =\frac
{1}{\operatorname*{vol}}\hspace{0.01in}\delta_{q}^{3}(\hspace{0.01in}%
y^{\prime}\oplus(\ominus\hspace{0.01in}\kappa^{-1}y))\circledast
(\,\ldots\,\triangleleft P_{y^{\prime}}).
\end{align}
We can prove the above identities by calculations of the following type:%
\begin{align}
P_{y^{\prime}}P_{y}\triangleright\psi_{R}(x)  &  =P_{y^{\prime}}\triangleright
u_{\hspace{0.01in}y}(x)\circledast c_{\hspace{0.01in}y}\nonumber\\
&  =u_{\hspace{0.01in}y^{\prime}}(x)\circledast\hspace{-0.01in}\int
\text{d}_{q}^{3}x^{\prime}\,(u^{\ast})_{y^{\prime}}(x^{\prime})\circledast
u_{\hspace{0.01in}y}(x^{\prime})\circledast c_{\hspace{0.01in}y}\nonumber\\
&  =u_{\hspace{0.01in}y^{\prime}}(x)\circledast\frac{1}{\operatorname*{vol}%
}\hspace{0.01in}\delta_{q}^{3}((\ominus\hspace{0.01in}\kappa^{-1}y^{\prime
})\oplus y)\circledast c_{\hspace{0.01in}y}\nonumber\\
&  =u_{\hspace{0.01in}y^{\prime}}(x)\circledast c_{\hspace{0.01in}y^{\prime}%
}\circledast\frac{1}{\operatorname*{vol}}\hspace{0.01in}\delta_{q}%
^{3}((\ominus\hspace{0.01in}\kappa^{-1}y^{\prime})\oplus y)\nonumber\\
&  =P_{y^{\prime}}\triangleright\psi_{R}(x)\circledast\frac{1}%
{\operatorname*{vol}}\hspace{0.01in}\delta_{q}^{3}((\ominus\hspace
{0.01in}\kappa^{-1}y^{\prime})\oplus y).
\end{align}
From Eq.~(\ref{DefProOrtEig1}) and Eq.~(\ref{DefProOrtEig2}), we finally see
that the set of the projection operators is complete in some sense as the
following relations hold:%
\begin{equation}
\int\text{d}_{q}^{3}y\,P_{y}\triangleright\,\ldots\,=\int\text{d}_{q}%
^{3}y\,\ldots\,\triangleleft P_{y}=\,\ldots
\end{equation}

Identifying the wave functions $\psi_{R}$ and $\psi_{L}^{\ast}$ with their
expansion coefficients $c_{\hspace{0.01in}y}$ and $(c^{\ast})_{y}$ [cf.
Eq.~(\ref{EntKoeOrtEigFkt}) of Chap.~\ref{KapVolRelOrtEigFkt}], we can write
the expectation value of an operator $\hat{O}$ as follows:%
\begin{align}
\langle\hat{O}\rangle_{\psi}  &  =\int\text{d}_{q}^{3}x\,\psi_{L}^{\ast
}(x)\circledast\hat{O}\triangleright\psi_{R}(x)=\int\text{d}_{q}%
^{3}y\,(c^{\ast})_{y}\circledast\hat{O}\triangleright c_{\hspace{0.01in}%
y}\nonumber\\
&  =\int\text{d}_{q}^{3}y\,\psi_{L}^{\ast}(x)\triangleleft\hat{O}%
\circledast\psi_{R}(x)=\int\text{d}_{q}^{3}y\,(c^{\ast})_{y}\triangleleft
\hat{O}\circledast c_{\hspace{0.01in}y}.
\end{align}
Accordingly, we get for the expectation value of the position operator:%
\begin{equation}
\langle X^{i}\rangle_{\psi}=\int\text{d}_{q}^{3}x\,\psi_{L}^{\ast
}(x)\circledast x^{i}\circledast\psi_{R}(x)=\int\text{d}_{q}^{3}y\,(c^{\ast
})_{y}\circledast y^{\hspace{0.01in}i}\circledast c_{\hspace{0.01in}y}.
\end{equation}
In this context, it should be mentioned that the normalization condition in
Eq.~(\ref{NorBed12}) can also be written as an expectation value of the
identity operator:%
\begin{equation}
\langle1\rangle_{\psi}=\int\text{d}_{q}^{3}x\,\psi_{L}^{\ast}(x)\circledast
\psi_{R}(x)=\int\text{d}_{q}^{3}y\,(c^{\ast})_{y}\circledast c_{\hspace
{0.01in}y}=1. \label{NorBedWdh2}%
\end{equation}

If the measurement of the position coordinates of a particle yields the
eigenvalues $y^{i}$, the wave functions $\psi_{R}$ and $\psi_{L}^{\ast}$ of
the particle are reduced to the expressions given in Eq.~(\ref{AntWelEigY}).
The expectation value of the operator $P_{y}$ is then equal to the probability
density of finding the particle in a region represented by $y=(\hspace
{0.02in}y^{i})$:%
\begin{align}
\rho_{\psi}(\hspace{0.01in}y)  &  =\int\text{d}_{q}^{3}x\,\psi_{L}^{\ast
}(x)\circledast P_{y}\triangleright\psi_{R}(x)\nonumber\\
&  =\int\text{d}_{q}^{3}x\,\psi_{L}^{\ast}(x)\triangleleft P_{y}%
\circledast\psi_{R}(x)=(c^{\ast})_{y}\circledast c_{\hspace{0.01in}y}.
\end{align}
Because of the normalization condition in Eq.~(\ref{NorBedWdh2}), the
following applies:%
\begin{equation}
\int\text{d}_{q}^{3}y\,\rho_{\psi}(\hspace{0.01in}y)=\int\text{d}_{q}%
^{3}y\,(c^{\ast})_{y}\circledast c_{\hspace{0.01in}y}=1.
\end{equation}
Considering the identifications in Eq.~(\ref{EntKoeOrtEigFkt}) of
Chap.~\ref{KapVolRelOrtEigFkt}, we can state the above results in the
following way. If the normalized wave functions $\psi_{R}(x)$ and $\psi
_{L}^{\ast}(x)$ describe the state of a particle in position space, we can
write down its \textit{probability density in position space} as follows:%
\begin{equation}
\rho_{\psi}(x)=\psi_{L}^{\ast}(x)\circledast\psi_{R}(x).
\end{equation}

Recall now that we can expand a wave function in terms of momentum
eigenfunctions [cf. Eq.~(\ref{EntIm2}) and Eq.~(\ref{EntImp1}) of
Chap.~\ref{KapVolRelImpOrt}]:%
\begin{align}
\psi_{R}(x)  &  =\hspace{-0.01in}\int\text{d}_{q}^{3}p\,u_{\hspace{0.01in}%
p}(x)\circledast c_{\hspace{0.01in}p},\nonumber\\
\psi_{L}^{\ast}(x)  &  =\hspace{-0.01in}\int\text{d}_{q}^{3}p\,(c^{\ast}%
)_{p}\circledast(u^{\ast})_{p}(x). \label{EntWelImpEigWdh}%
\end{align}
The corresponding expansion coefficients can be determined by $q$-de\-formed
Fourier transformations [cf. Eq.~(\ref{CPUnt}) and Eq.~(\ref{DPUnt}) of
Chap.~\ref{KapVolRelImpOrt}]:%
\begin{equation}
c_{\hspace{0.01in}p}=\operatorname*{vol}\nolimits^{-1/2}\mathcal{F}_{R}^{\ast
}(\psi_{R})(\hspace{0.01in}p),\qquad(c^{\ast})_{p}=\operatorname*{vol}%
\nolimits^{-1/2}\mathcal{F}_{L}(\psi_{L}^{\ast})(\hspace{0.01in}p).
\end{equation}
The considerations in this chapter also apply to the expansions%
\begin{align}
\psi_{R}^{\ast}(x)  &  =\hspace{-0.01in}\int\text{d}_{q}^{3}p\,(u^{\ast}%
)^{p}(x)\circledast(c^{\ast})^{p},\nonumber\\
\psi_{L}(x)  &  =\hspace{-0.01in}\int\text{d}_{q}^{3}p\,c^{\hspace{0.01in}%
p}\circledast u^{p}(x) \label{EntWelImpEigWdh2}%
\end{align}
with the coefficients $c^{\hspace{0.01in}p}$ and $(c^{\ast})^{p}$ given by the
following expressions [cf. Eq.~(\ref{CPUnt2}) and (\ref{DPUnt2}) of
Chap.~\ref{KapVolRelImpOrt}]:%
\begin{equation}
c^{\hspace{0.01in}p}=\operatorname*{vol}\nolimits^{-1/2}\mathcal{F}_{L}^{\ast
}(\psi_{L})(\hspace{0.01in}p),\qquad(c^{\ast})^{p}=\operatorname*{vol}%
\nolimits^{-1/2}\mathcal{F}_{R}(\psi_{R}^{\ast})(\hspace{0.01in}p).
\end{equation}
We can restrict ourselves, however, to the expansions in
Eq.~(\ref{EntWelImpEigWdh}). We can do this since we obtain the results for
the expansions in Eq.~(\ref{EntWelImpEigWdh2}) by the following substitutions:%
\begin{gather}
\psi^{\ast}\leftrightarrow\hspace{0.01in}\psi,\quad\mathcal{F}\leftrightarrow
\mathcal{F}^{\hspace{0.01in}\ast},\quad\triangleright\leftrightarrow
\bar{\triangleright},\quad\triangleleft\leftrightarrow\bar{\triangleleft
},\nonumber\\
(u^{\ast})_{p}\rightarrow\hspace{0.01in}u^{p},\quad u_{p}\rightarrow(u^{\ast
})^{p},\quad(c^{\ast})_{p}\rightarrow c^{\hspace{0.01in}p},\quad
c_{\hspace{0.01in}p}\rightarrow(c^{\ast})^{p}.
\end{gather}

If we measure the values $p^{i}$ for the momentum coordinates of a particle,
the wave function of the particle will be reduced to the component
corresponding to $p=(\hspace{0.02in}p^{i})$. We can introduce operators
projecting onto these components, i.~e.%
\begin{equation}
P_{p}\triangleright\psi_{R}(x)=u_{\hspace{0.01in}p}(x)\circledast
c_{\hspace{0.01in}p},\qquad\psi_{L}^{\ast}(x)\triangleleft P_{p}=(c^{\ast
})_{p}\circledast(u^{\ast})_{p}(x)
\end{equation}
with%
\begin{align}
P_{p}\triangleright\,\ldots\,  &  =u_{\hspace{0.01in}p}(x)\circledast
\hspace{-0.01in}\int\text{d}_{q}^{3}x^{\prime}\hspace{0.01in}(u^{\ast}%
)_{p}(x^{\prime})\circledast\,\ldots\,,\nonumber\\
\ldots\,\triangleleft P_{p}  &  =\int\text{d}_{q}^{3}x^{\prime}\hspace
{0.01in}\ldots\hspace{0.01in}\circledast u_{\hspace{0.01in}p}(x^{\prime
})\circledast(u^{\ast})_{p}(x).
\end{align}
Note that the set of these projection operators is complete as we have:%
\begin{equation}
\int\text{d}_{q}^{3}p\,P_{p}\triangleright\,\ldots=\int\text{d}_{q}%
^{3}y\,\ldots\hspace{0.01in}\triangleleft P_{y}=\,\ldots
\end{equation}
Furthermore, the projection operators to the momentum eigenvalues satisfy the
following orthonormality conditions:%
\begin{align}
P_{p^{\prime}}P_{p}\triangleright\,\ldots\,  &  =(P_{p^{\prime}}%
\triangleright\,\ldots\,)\circledast\frac{1}{\operatorname*{vol}}%
\hspace{0.01in}\delta_{q}^{3}((\ominus\hspace{0.01in}\kappa^{-1}p^{\prime
})\oplus p),\nonumber\\
\,\ldots\,\triangleleft P_{p}\hspace{0.01in}P_{p^{\prime}}  &  =\frac
{1}{\operatorname*{vol}}\hspace{0.01in}\delta_{q}^{3}(\hspace{0.01in}%
p^{\prime}\oplus(\ominus\hspace{0.01in}\kappa^{-1}p))\circledast
(\,\ldots\,\triangleleft P_{p^{\prime}}).
\end{align}

The expectation value of the projector $P_{p}$ provides the density for the
probability that a momentum measurement yields $p=(\hspace{0.02in}p^{i})$:%
\begin{align}
\rho_{\psi}(\hspace{0.01in}p)  &  =\int\text{d}_{q}^{3}x\,\psi_{L}^{\ast
}(x)\circledast P_{p}\triangleright\psi_{R}(x)\nonumber\\
&  =\operatorname*{vol}\nolimits^{-1}\mathcal{F}_{L}(\psi_{L}^{\ast
})\circledast\mathcal{F}_{R}^{\hspace{0.01in}\ast}(\psi_{R})=(c^{\ast}%
)_{p}\circledast c_{\hspace{0.01in}p}. \label{DefWskImpEig1}%
\end{align}
Note that the second step is a consequence of the $q$-de\-formed
Fou\-rier-Plan\-cherel identities [cf. Eq.~(\ref{FouPla1}) and
Eq.~(\ref{FouPla2}) of Chap.~\ref{KapFourPlaIde}]. Moreover, it follows:%
\begin{align}
\langle1\rangle_{\psi}  &  =\int\text{d}_{q}^{3}x\,\psi_{L}^{\ast
}(x)\circledast\psi_{R}(x)=\frac{1}{\operatorname*{vol}}\int\text{d}_{q}%
^{3}p\,\mathcal{F}_{L}(\psi_{L}^{\ast})(\hspace{0.01in}p)\circledast
\mathcal{F}_{R}^{\hspace{0.01in}\ast}(\psi_{R})(\hspace{0.01in}p)\nonumber\\
&  =\int\text{d}_{q}^{3}p\,(c^{\ast})_{p}\circledast c_{\hspace{0.01in}p}%
=\int\text{d}_{q}^{3}p\,\rho_{\psi}(\hspace{0.01in}p)=1.
\end{align}

Next, we consider a self-adjoint version of the three-dimensional
$q$-de\-formed momentum operator, whose components are defined in complete
analogy to Eq.~(\ref{RealKoor3dimOrt}):%
\begin{align}
P^{1}  &  =\frac{\text{i}}{2}(-\hspace{0.01in}q^{-1/2}P^{+}-\hspace
{0.01in}q^{1/2}P^{-}),\nonumber\\
P^{2}  &  =\frac{1}{2}(-\hspace{0.01in}q^{-1/2}P^{+}+\hspace{0.01in}%
q^{1/2}P^{-}),\nonumber\\
P^{3}  &  =P^{3}.
\end{align}
Using the $q$-de\-formed Fourier-Plancherel identities we obtain
($i\in\{1,2,3\}$):%
\begin{align}
\langle P^{i}\rangle_{\psi}  &  =\int\text{d}_{q}^{3}x\,\psi_{L}^{\ast
}(x)\circledast\text{i}^{-1}\partial^{i}\triangleright\psi_{R}(x)=\int
\text{d}_{q}^{3}x\,\psi_{L}^{\ast}(x)\triangleleft\hspace{0.01in}\text{i}%
^{-1}\partial^{i}\circledast\psi_{R}(x)\nonumber\\
&  =\frac{1}{\operatorname*{vol}}\int\text{d}_{q}^{3}p\,\mathcal{F}_{L}%
(\psi_{L}^{\ast})(\hspace{0.01in}p)\circledast\mathcal{F}_{R}^{\hspace
{0.01in}\ast}(\text{i}^{-1}\partial_{x}^{i}\triangleright\psi_{R}%
)(\hspace{0.01in}p)\nonumber\\
&  =\frac{1}{\operatorname*{vol}}\int\text{d}_{q}^{3}p\,\mathcal{F}_{L}%
(\psi_{L}^{\ast}\triangleleft\text{i}^{-1}\partial_{x}^{i})(\hspace
{0.01in}p)\circledast\mathcal{F}_{R}^{\hspace{0.01in}\ast}(\psi_{R}%
)(\hspace{0.01in}p)\nonumber\\
&  =\int\text{d}_{q}^{3}p\,(c^{\ast})_{p}\circledast p^{\hspace{0.01in}%
i}\circledast c_{\hspace{0.01in}p}. \label{ImpErwImpRau1}%
\end{align}

Finally, we examine how the expectation value of $P^{i}$ behaves under
conjugation if we take into account the condition given in
Eq.~(\ref{KonBedWell}):%
\begin{equation}
\overline{\langle P^{i}\rangle_{\psi}}=\int\text{d}_{q}^{3}x\,\overline
{\psi_{R}}\,\bar{\triangleleft}\,\partial^{i}\text{i}^{-1}\circledast
\overline{\psi_{L}^{\ast}}=\int\text{d}_{q}^{3}x\,\psi_{L}^{\ast}%
\circledast\text{i}^{-1}\partial^{i}\,\bar{\triangleright}\,\psi_{R}.
\end{equation}
Since the actions $\partial^{i}\triangleright\psi$ and $\partial^{i}%
\,\bar{\triangleright}\,\psi$ are generally not the same, the above result
does not imply that the expectation value of $P^{i}$ is real in all cases.
Looking at the last expression of Eq.~(\ref{ImpErwImpRau1}) shows, however,
that the condition $\overline{(c^{\ast})_{p}}=c_{\hspace{0.01in}p}$ leads to a
real expectation value for $P^{i}$.

\appendix

\section{Graphical representation of morphisms\label{AppA}}

\begin{figure}
[ptb]
\begin{center}
 \includegraphics[width=0.07\textwidth]{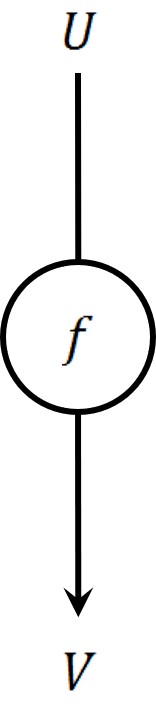}%
\caption{Graphical representation of the morphism $f:U\rightarrow V$}%
\label{Fig6}%
\end{center}
\end{figure}

The $q$-de\-formed quantum spaces we are dealing with are so-called braided
Hopf algebras \cite{Majid:1992sn}, i.~e. objects of a braided tensor category.
If $U$ and $V$ are objects of such a tensor category, a morphism
$f:U\rightarrow V$ is symbolized by a downward pointing arrow. This arrow runs
through a circle labeled by $f$ (see Fig.~\ref{Fig6}).%
In the following, we briefly discuss graphical representations of morphisms
that are important for our considerations.

\begin{figure}
[ptb]
\begin{center}
  \includegraphics[width=0.56\textwidth]{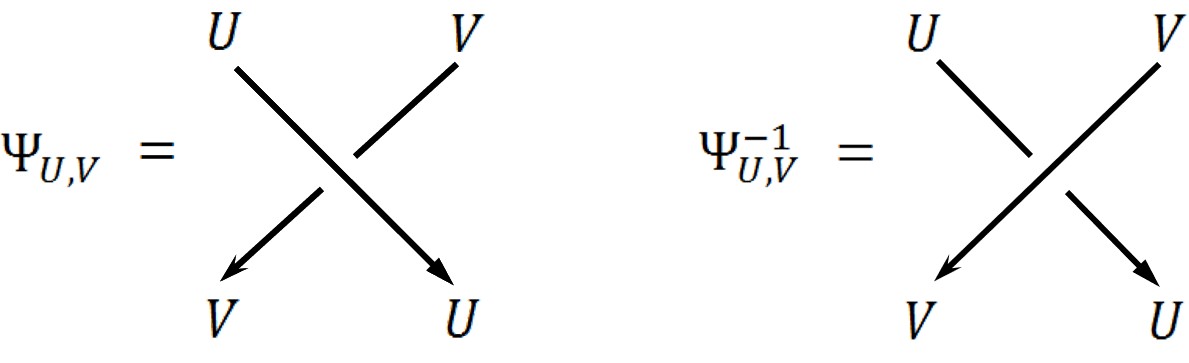}
\caption{Braiding mappings}%
\label{Fig7}%
\end{center}
\end{figure}

In Eq.~(\ref{ZopRel}) and Eq.~(\ref{InvZopRel}) of\ Chap.~\ref{KapEucQuaSpa}%
,\ we have considered isomorphisms that can be used to exchange the factors of
a tensor product. We represent these isomorphisms $\Psi_{U,V}:U\otimes
V\rightarrow V\otimes U$ and $(\Psi_{U,V})^{-1}:V\otimes U\rightarrow U\otimes
V$ known as braiding mappings by intersecting arrows (see Fig.~\ref{Fig7}).%
\textbf{\ }

\begin{figure}
[ptb]
\begin{center}
 \includegraphics[width=0.66\textwidth]{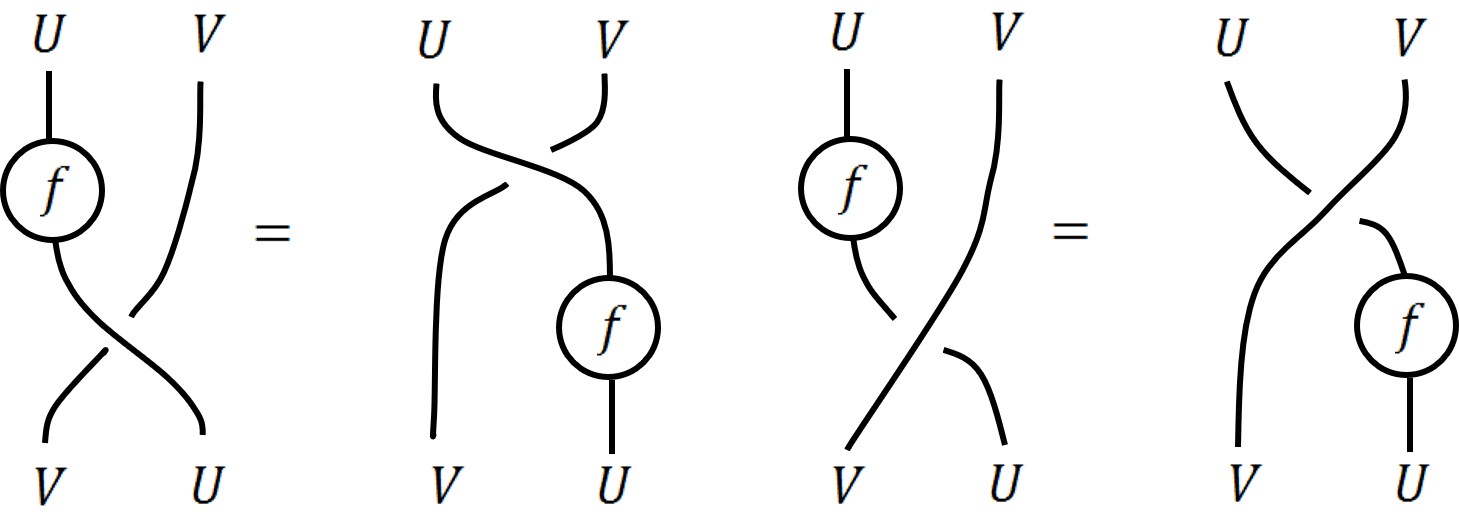}
\caption{Commuting morphisms by $\Psi$ or $\Psi^{-1}$}%
\label{Bild13}%
\end{center}
\end{figure}

\begin{figure}
[ptbptb]
\begin{center}
  \includegraphics[width=0.80\textwidth]{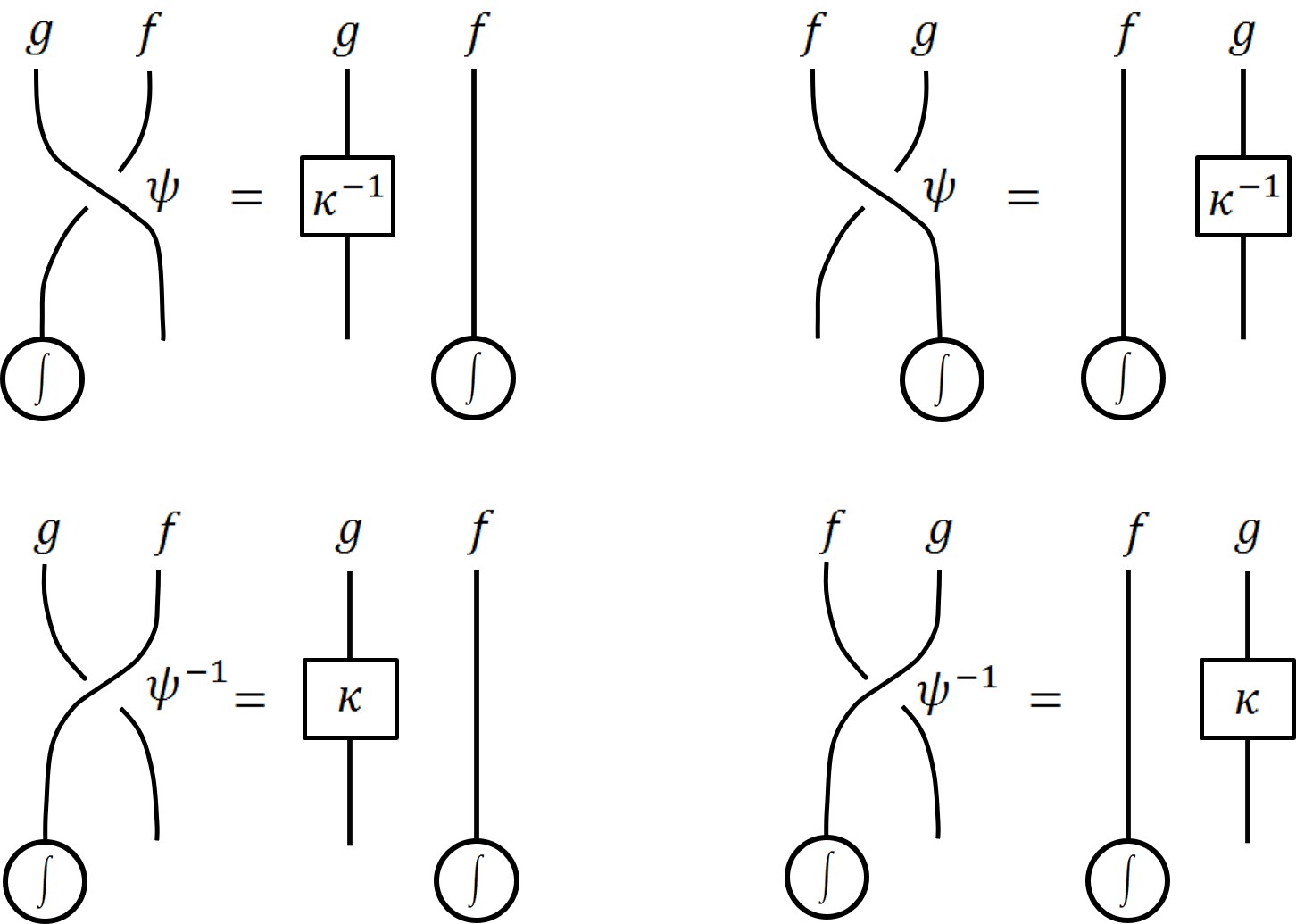}
\caption{Braiding properties of $q$-integrals}%
\label{Fig8}%
\end{center}
\end{figure}

We can draw the symbol for a morphism through a crossing that represents a
braiding mapping (cf. Fig.~\ref{Bild13}). For our $q$-in\-te\-grals
represented graphically by a free node, this statement only applies with some
restrictions. A detailed investigation shows the following. If we draw the
free node representing a $q$-in\-te\-gral through an intersection, one of the
two strands takes up a scaling (see Fig.~\ref{Fig8}). If the strand taking up
the scaling symbolizes a function, all arguments of this function are
multiplied by the same constant $\kappa$ or $\kappa^{-1}$. Note that\ in the
case of the $q$-de\-formed Euclidean quantum space we have $\kappa=q^{6}$.%

\begin{figure}
[ptb]
\begin{center}
  \includegraphics[width=0.60\textwidth]{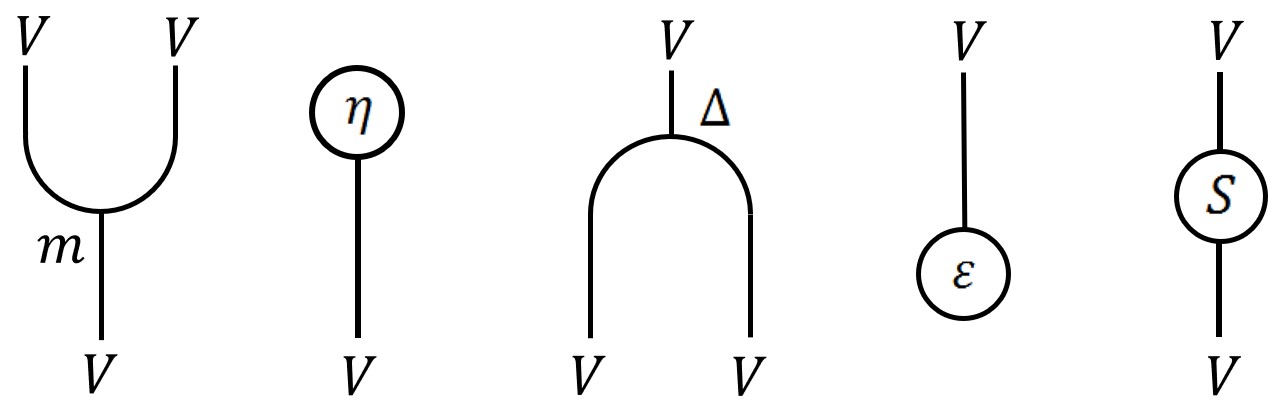}
\caption{Graphical representation of the morphisms for a Hopf algebra}%
\label{Fig9}%
\end{center}
\end{figure}

Being braided Hopf algebras, $q$-de\-formed quantum spaces have specific
morphisms. In this respect, the algebraic structure of a quantum space $V$ is
determined by a multiplication $m:V\otimes V\rightarrow V$ and a unit map
$\eta:\mathbb{K}\rightarrow V$. The dual morphisms, i.~e. the braided
co-pro\-duct $\underline{\Delta}:V\otimes V\rightarrow V$ and the (braided)
co-unit $\underline{\varepsilon}:V\rightarrow\mathbb{K}$ together with the
braided antipode $\underline{S}:V\rightarrow V$ determine the (braided) Hopf
structure of the quantum space algebra $V$. In Chap.~\ref{KapStePro} and in
Chap.~\ref{KapTra}, we have briefly explained how to realize these morphisms
on a commutative coordinate algebra as star products, $q$-trans\-la\-tions or
$q$-in\-ver\-sions. For some calculations, it does not make sense to work with
these commutative realizations. In this cases, it is much easier to perform
calculations graphically. Therefore, we have summarized the graphical
representations of the morphisms mentioned above in Fig.~\ref{Fig9} (also see
Ref.~\cite{Majid:1996kd}).%

\section{Proof for invertibility of $q$-Fourier transformations\label{AnhB}}

Graphical methods provide the simplest way to prove the identities in
Eq.~(\ref{InvFourAnf2a}) and Eq.~(\ref{InvFourAnf1a})
of\ Chap.~\ref{KapFourInv}. In this appendix, we briefly describe the most
important steps of these graphical calculations taken from
Ref.~\cite{Kempf:1994yd}. Note that we have adapted these graphical
calculations to the needs of our formalism.

\begin{figure}
[ptb]
\begin{center}
 \includegraphics[width=0.52\textwidth]{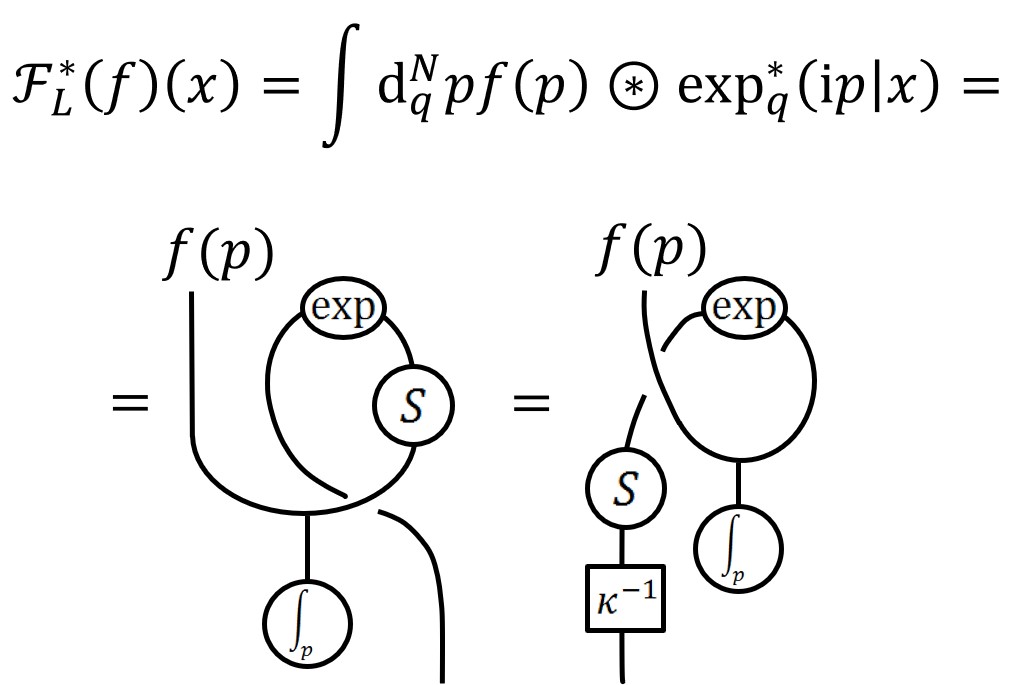}
\caption{Graphical representation of the $q$-Fourier transformation
$\mathcal{F}_{L}^{\hspace{0.01in}\ast}(f)$}%
\label{Fig10}%
\end{center}
\end{figure}

For the following considerations, we need a special graphical representation
of the Fourier transformation $\mathcal{F}_{L}^{\hspace{0.01in}\ast}(f)$ [also
see Eq.~(\ref{FTtype1}) of Chap.~\ref{KapFTDef}]. We show this representation
in Fig.~\ref{Fig10}. Note that we can obtain the right diagram in
Fig.~\ref{Fig10} from the left one by dragging the downward strand past the
node representing the $q$-in\-te\-gral and taking into account the braiding
properties of the $q$-integral (also see Fig.~\ref{Fig8}).\footnote{In
contrast to Fig.~\ref{Fig8} the scaling is given by $\kappa^{-1}$ since the
$q$-integral refers to momentum space and the downward strand to position
space.}%
\begin{figure}
[ptbptb]
\begin{center}
 \includegraphics[width=0.90\textwidth]{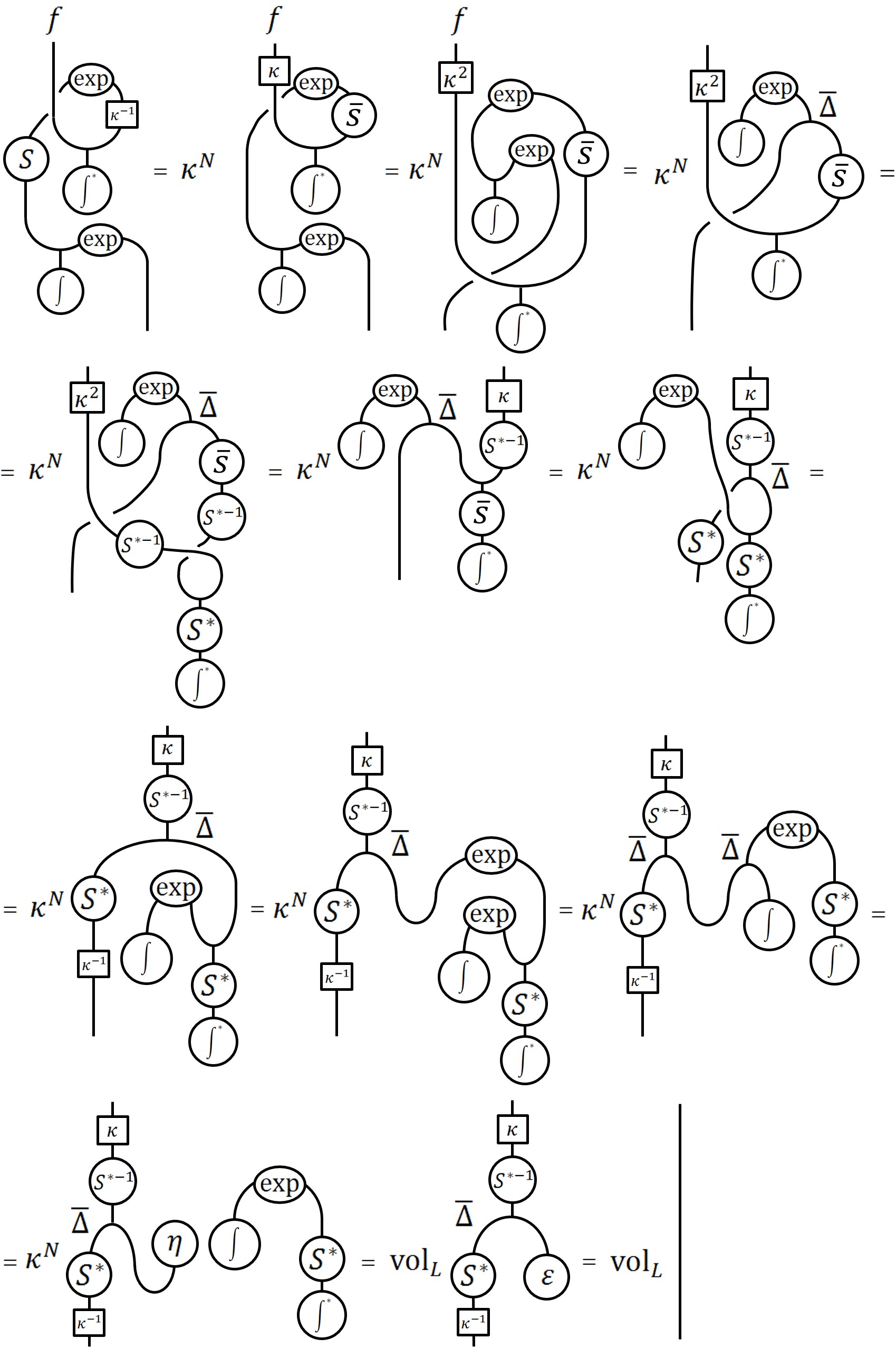}
\caption{Diagrammatic proof for the first identity of Eq.~(\ref{InvFourAnf2a}%
)}%
\label{Bild2}%
\end{center}
\end{figure}
\begin{figure}
[ptbptbptb]
\begin{center}
 \includegraphics[width=0.75\textwidth]{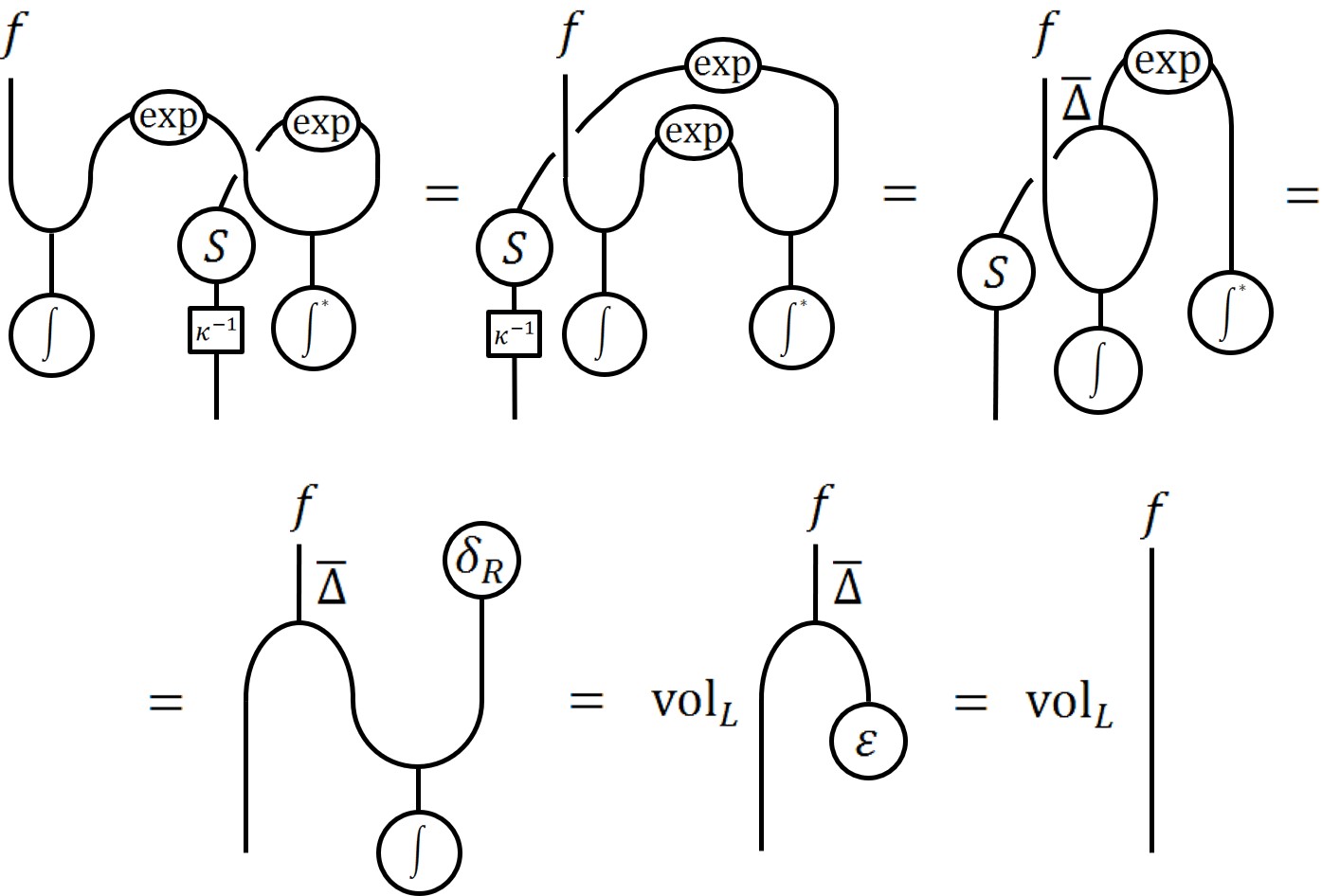}
\caption{Graphical proof for the first identity of Eq.~(\ref{InvFourAnf1a})}%
\label{Fig14}%
\end{center}
\end{figure}

Fig.~\ref{Bild2}\textbf{ }shows the graphical prove for the first identity of
Eq.~(\ref{InvFourAnf2a}). In the first step, we pull the antipode $S$ through
the crossing and apply Eq.~(\ref{InvExpAlgDefKom}) of Chap.~\ref{KapExp}. We
have labeled all morphisms referring to the dual coordinate algebra, i.~e. the
momentum algebra, with a star. Furthermore, we keep in mind that we have made
use of the scaling properties of the $q$-in\-te\-gral in the first step:%
\begin{equation}
\int\text{d}_{q}^{N}\hspace{-0.02in}x\,f(\kappa^{-1}x)\circledast
g(\kappa^{-1}x)=\kappa^{N}\int\text{d}_{q}^{N}\hspace{-0.02in}%
x\,f(x)\circledast g(x).
\end{equation}
In the second step, we pull the upward strand over the lower integral and the
lower exponential, taking into account the braiding properties of the
$q$-integral. Now, we can make use of the addition theorem for quantum space
exponentials [see Eq.~(\ref{AddTheExp}) of Chap.~\ref{KapExp}]. The fourth
step is a consequence of the following axioms of a braided Hopf
algebra:\footnote{Recall that in our convention the braiding mappings $\Psi$
and $\Psi^{-1}$ are interchanged in comparison to Ref.~\cite{Majid:1992sn}.}%
\begin{align}
m\circ(\hspace{0.01in}\underline{S}^{-1}\otimes\underline{S}^{-1})\circ\Psi &
=m\circ\Psi\circ(\hspace{0.01in}\underline{S}^{-1}\otimes\underline{S}%
^{-1})=\underline{S}^{-1}\circ m,\nonumber\\
m\circ(\hspace{0.01in}\underline{S}\otimes\underline{S}\hspace{0.01in}%
)\circ\Psi^{-1}  &  =m\circ\Psi^{-1}\circ(\hspace{0.01in}\underline{S}%
\otimes\underline{S}\hspace{0.01in})=\underline{S}\circ m.
\end{align}
In the fifth step of Fig.~\ref{Bild2}, the strand symbolizing the function $f$
is drawn to the right over the $q$-in\-te\-gral and the $q$-ex\-po\-nential,
taking into account the trivial braiding of the $q$-ex\-po\-nential as well as
the braiding properties of the $q$-integral. For the sixth step, we use a
graphical identity whose derivation we can find in Ref.~\cite{Kempf:1994yd}.
In the seventh step, the $q$-ex\-po\-nential, together with the attached
$q$-in\-te\-gral, is drawn over the left strand of the co-product. The right
strand of the co-product is then deformed in such a way that we obtain the
combination of a dual pairing (between a momentum space algebra and a position
space algebra) with a canonical element (i.~e. a $q$-ex\-po\-nential). Thus,
we can apply the addition theorem for $q$-ex\-po\-nentials again. Then, we
make use of the translation invariance of the $q$-in\-te\-gral (cf.
Fig.~\ref{Fig13} of Chap.~\ref{KapIntegral}). The penultimate step is nothing
else but the graphical representation of the following property of a dual
pairing:%
\begin{equation}
\left\langle f,1\right\rangle =\varepsilon(f).
\end{equation}
Note that in the penultimate step, we have identified the right subdiagram
with the $q$-deformed volume element (see Ref.~\cite{Kempf:1994yd} for a
detailed explanation of this identification):%
\begin{equation}
\kappa^{N}\int\nolimits_{-\infty}^{+\infty}\text{d}_{q}^{N}\hspace
{-0.02in}p\int\nolimits_{-\infty}^{+\infty}\text{d}_{q}^{N}\hspace
{-0.02in}x\,\exp_{q}(x|\ominus\text{i}p)=\kappa^{N}\int\nolimits_{-\infty
}^{+\infty}\text{d}_{q}^{N}\hspace{-0.02in}p\,\delta_{L}^{N}(\ominus
\hspace{0.01in}p)=\operatorname*{vol}\nolimits_{\hspace{0.01in}L}.
\end{equation}
Finally, the last step results from the following axiom of a braided Hopf
algebra:%
\begin{equation}
(\text{$\operatorname*{id}$}\otimes\varepsilon)\circ\Delta
=\text{$\operatorname*{id}$}=(\varepsilon\otimes\text{$\operatorname*{id}$%
})\circ\Delta. \label{KoAssExiKo}%
\end{equation}

Fig.~\ref{Fig14} shows the graphical proof for the first identity of
Eq.~(\ref{InvFourAnf1a}). Note that in the first diagram of this graphical
proof the Fourier transformation $\mathcal{F}_{L}^{\text{\hspace{0.01in}}\ast
}$ is represented again in the form we have given in Fig.~\ref{Fig10}. We
first drag the left integral over the downward strand and then apply the law
of addition for the quantum space exponentials. The third step uses the same
identity as the sixth step in Fig.~\ref{Bild2}. Now, we can identify the
expression for a $q$-del\-ta function [cf. Eq.~(\ref{DefDelLR}) of
Chap.~\ref{KapFTDef}]. The next step is nothing else but the graphical
representation of the following characteristic identity for the $q$-del\-ta
function:%
\begin{equation}
\int\nolimits_{-\infty}^{+\infty}\text{d}_{q}^{N}\hspace{-0.02in}%
x\hspace{0.01in}f(x)\circledast\delta_{L}^{N}(x)=\operatorname*{vol}%
\nolimits_{\hspace{0.01in}L}\hspace{0.01in}f(0).
\end{equation}
You can find the proof for this identity in Ref.~\cite{Kempf:1994yd}. The last
step is a consequence of the axiom in Eq.~(\ref{KoAssExiKo}) again.

{\normalsize
\bibliographystyle{abbrv}
\bibliography{book,habil}
}

\end{document}